\documentclass[11pt, oneside]{article} 
\pdfoutput=1

\usepackage{jheppub}

\usepackage{amsmath,amssymb,amsfonts,amsthm}
\usepackage{color}
\usepackage{booktabs}
\definecolor{darkblue}{rgb}{0.1,0.1,.7}
\usepackage[]{amsmath}
\usepackage[]{graphicx}
\usepackage[]{latexsym}
\usepackage{amscd}
\usepackage[all,cmtip]{xy}
\usepackage{mathrsfs}
\usepackage{bbold}
\usepackage{subfigure}
\usepackage[margin=10pt,font=small,labelfont=bf]{caption}
\usepackage{simplewick}
\usepackage{changepage}
\usepackage[]{algorithm2e}
\usepackage{booktabs,multirow}
\usepackage[OT2,OT1]{fontenc}

\newcommand{\assign}{:=}

\newcommand{\nobracket}{}

\newtheorem{theorem}{Theorem}[section]

\newtheorem{exercise}[theorem]{Exercise}
\newtheorem*{definition*}{Definition}
\theoremstyle{remark}
\newtheorem{remark}[theorem]{Remark}
\newtheorem{example}{Example}

\newcommand{\myinclude}[2][]{\raisebox{0.6ex}{\raisebox{-0.5\height}{\includegraphics[#1]{#2}}}}

\makeatletter
\def\@fpheader{\ }
\makeatother

\title{Four Lectures on the Random Field Ising Model, \\Parisi-Sourlas
	Supersymmetry,\\ and Dimensional Reduction }
\author{Slava Rychkov}
\affiliation{Institut des Hautes \'Etudes Scientifiques, 91440 Bures-sur-Yvette, France}

\abstract{Numerical evidence suggests that the Random Field Ising Model loses Parisi-Sourlas SUSY and the dimensional reduction property somewhere between 4 and 5 dimensions, while a related model of branched polymers retains these features in any $d$. These notes give a leisurely introduction to a recent theory, developed jointly with A.~Kaviraj and E.~Trevisani, which aims to explain these facts. Based on the lectures given in Cortona and at the IHES in 2022.}

\begin{document}

\maketitle

\section{History, Basics, Experiments and Simulations}\label{L1}

\subsection{History}\label{sec-history}

The Random Field Ising Model (RFIM) is the Ising model coupled to a random
$x$-dependent magnetic field. It was introduced in 1975 by Imry and Ma
{\cite{ImryMa}}, who predicted a phase transition in $d > 2$ but not in $d
\leqslant 2$, via the Imry-Ma argument discussed below. Imry and Ma also found
the upper critical dimension $d_{{\rm uc}} = 6$.

In 1976, Aharony, Imry and Ma {\cite{Aharony:1976jx}} found that, to all
orders in perturbation theory, the RFIM critical exponents in $d = 6 -
\varepsilon$ dimensions are the same as the critical exponents of the usual
Ising model in $d - 2$ dimension. This became known as the dimensional
reduction of critical exponents.

In 1979, Parisi and Sourlas {\cite{Parisi:1979ka}} found that RFIM has an
equivalent supersymmetric formulation. They showed that dimensional reduction
is a natural consequence of supersymmetry (SUSY).

These developments immediately led to a puzzle when extrapolating to the
physical dimension $d = 3$. Dimensional reduction cannot be true in this
dimension, since the $d = 1$ Ising model does not even have a phase
transition. Perturbation theory is apparently breaking down somewhere between
$d = 6$ and 3. Why?

Over the subsequent decades, there were many attacks on this problem. In
particular, Br{\'e}zin and De Dominicis {\cite{Brezin-1998}} in 1998, and
Feldman {\cite{Feldman}} in 2000 argued that dimensional reduction breaks down
for any $d < 6$, because of subtle RG effects. Tarjus and Tissier \cite{Tarjus2004} 
in 2004 found, in a nonperturbative RG calculation, that
dimensional reduction holds for $d > d_c \approx 5.1$ but breaks down at $d <
d_c$. While we do not fully agree with these authors, we will make contact
with some of their ideas below. 

The RFIM phase transition in $d = 3$ was studied both experimentally
\cite{BelangerYoung} and via numerical simulations (the only way available
in $d > 3$). For $d = 4$ and $d = 5$, the critical exponents were determined by
Fytas, Martín-Mayor, Picco and Sourlas in 2016. These studies led to an
interesting conclusion. In $d = 4$, dimensional reduction is clearly ruled out
{\cite{Picco1}}. On the other hand, the $d = 5$ model shows not only the
dimensional reduction of critical exponents {\cite{Picco2}}, but was also
found in their later work {\cite{Picco3}} (joint with Parisi) to respect SUSY
relations between correlation functions, providing a direct test of SUSY.

These lectures will explain a theory developed in collaboration with Apratim
Kaviraj and Emilio Trevisani {\cite{paper1,paper2,paper-summary,paper3}},
which started in 2018 and was sparked by a conference on disordered systems
in Rome. Our main conclusions are these:
\begin{itemize}
  \item Dimensional reduction is lost because SUSY is lost.
  
  \item SUSY is lost because some SUSY-breaking perturbations, present at the
  microscopic level, become relevant at $d < d_c$ (while they are irrelevant
  at $d > d_c$).
  
  \item From two-loop perturbation theory, we estimate $d_c \approx 4.5$. This
  is in good agreement with the above-mentioned numerical simulations, that
  dimensional reduction is lost between 4 and 5 dimensions.
\end{itemize}

\subsection{Basic facts}

Recall that the usual Ising model in $d \geqslant 2$ dimensions has a phase
transition at a critical temperature $T_c > 0$. For $T > T_c$ the model is in
the disordered phase of vanishing magnetization. For $T < T_c$ the model is in
the ordered phase, and the magnetization is nonzero: $m (T) > 0$. The
magnetization $m$ can be defined e.g.~as an infinite volume limit of the
average spin value at a point in the middle of the lattice with the $+$ boundary
conditions:
\begin{equation}
  m = \lim_{L \rightarrow \infty} \langle s_0 \rangle_+ . \label{IsingM}
\end{equation}
The phase transition is continuous: $m (T) \rightarrow 0$ as $T \rightarrow
T_c^{}$.

The usual Ising model is a model for pure uniaxial ferromagnets. Real
materials however always contain impurities, also known as disorder effects.
Superclean electronics-grade silicon has 1 impurity per $10^9$ atoms. This
means that the average distance between impurities is $O (10^3)$, which is not such a
huge number.

Consider then uniaxial ferromagnets with impurities. In these lectures we will
be interested in magnetic impurities. These can be modeled by adding a local
$x$-dependent magnetic field. The Hamiltonian of the model becomes
\begin{equation}
  \mathcal{H} [s, h] = - J \sum_{\langle x y \rangle} s_x s_y - \sum_x h_x
  s_x,
\end{equation}
where $h_x$ is the magnetic field created by impurities at site
$x$.\footnote{\label{bond}Nonmagnetic impurities instead can be modeled by
variations of the spin-spin ferromagnetic couplings $J_{} \rightarrow J +
\delta J_{x y}$.} We will be interested in the case when $h_{}$ can be
considered frozen (not in thermal equilibrium with $s_{}$). Such a form of
disorder is called quenched. For a given $h$, we can then define the partition
function
\begin{equation}
  Z [h] = {\rm Tr}_s e^{- \beta \mathcal{H} [s, h]},
\end{equation}
and correlation functions, e.g.
\begin{equation}
  \langle s_x s_y \rangle_h = Z [h]^{- 1} {\rm Tr}_s\, s_x s_y\, e^{- \beta
  \mathcal{H} [s, h]} .
\end{equation}

The next step is to replace a fixed disorder instance $h$ by a random one. The
rationale behind this is that a large system can be divided into many chunks
which are still large. Each chunk will have its own disorder instance, and
these instances can be assumed statistically independent, drawn from some
distribution $\mathcal{P} [h]$ (assuming, as we will, that faraway impurities
do not influence each other). Thus, any quantity involving an average over the
volume can be computed as an average over $\mathcal{P} [h]$. Such quantities
are called self-averaging. Examples include the free energy and
volume-averaged correlation functions, such as
\begin{equation}
  \frac{1}{V} \sum_x \langle s_x s_{x + r} \rangle_h .
\end{equation}

We will assume that $\mathcal{P} [h]$ respects lattice symmetries
(translations and rotations). We will also impose $\mathbb{Z}_2$ invariance:
$\mathcal{P} [h] =\mathcal{P} [- h]$. This means that magnetic impurities have
no preferred orientation. In particular we have (denoting averages w.r.t.
$\mathcal{P} [h]$ by an overbar)
\begin{equation}
  \overline{h_x} = 0 .
\end{equation}
As we said, the two-point correlation function of the disorder, $\overline{h_x
h_y} = f (x - y)$ is assumed short-ranged.

\begin{definition*}
  The Random Field Ising Model is the Ising model coupled to a random magnetic
  field $h$, drawn from a distribution $\mathcal{P} [h]$ which is
  $\mathbb{Z}_2$-invariant, translationally and rotationally invariant, and
  has short-range correlations. Correlation functions of the model are
  computed by first averaging over $s$, then over $h$, e.g.
  \begin{equation}
    \overline{\langle s_x s_y \rangle_h} = \overline{Z [h]^{- 1} {\rm Tr}_s
    s_x s_y e^{- \beta \mathcal{H} [s, h]}} .
  \end{equation}
\end{definition*}

We may assume for simplicity that
\begin{equation}
  \mathcal{P} [h] = \prod_x P [h_x],
\end{equation}
i.e.~that $h_x$ are i.i.d. random variables. Then we have
\begin{equation}
  \overline{h_x h_y} = 0 \quad (x \neq y), \qquad \overline{(h_x)^2} =
  \kappa^2, \label{hh}
\end{equation}
where the parameter $\kappa$ measures disorder strength. We may simplify
further and assume that the distribution $P [h_x]$ is Gaussian:
\begin{equation}
  P [h_x] = \frac{1}{\sqrt{2 \pi} \kappa} e^{- h_x^2 / (2 \kappa^2)} .
  \label{gaussian}
\end{equation}
The usual justification for all these simplifying assumptions is universality.
Normally, it is believed that models corresponding to different $P [h_x]$ will
flow to the same renormalization group fixed point describing the
transition.\footnote{Universality should definitely hold under small
deformations of $P [h_x]$. It is not guaranteed to hold under larger
deformations, which may leave the basin of attraction of a given fixed point.
We will come back to this possibility at the end of these lectures (Section
\ref{tuning}).}

\begin{figure}[h]
 \centering \resizebox{286pt}{99pt}{\includegraphics{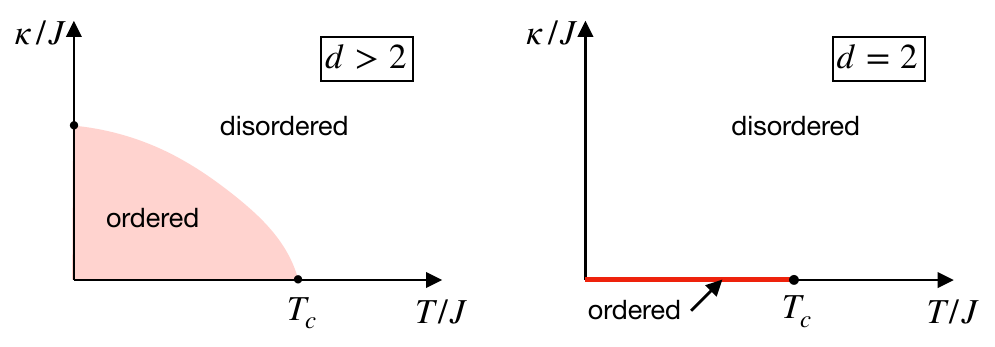}}
  \caption{\label{phase}The RFIM phase diagram for $d > 2$ (left) and $d = 2$
  (right).}
\end{figure}

The phase diagram of the model in $d > 2$ contains an ordered phase at small
$T$ and small $\kappa$, and a disordered phase at large values of these
parameters (see Fig.~\ref{phase}). The magnetization order parameter can be defined similarly to {\eqref{IsingM}}
for the pure Ising case, namely as an infinite-volume limit of a single-spin
correlation function with the $+$ boundary conditions:
\begin{equation}
	m = \lim_{L \rightarrow \infty} \overline{\langle s_0 \rangle_+}\, .
\end{equation}

On the other hand, in $d = 2$ the model is disordered for an arbitrarily small
$\kappa$ (see Fig.~\ref{phase}). The difference between $d = 2$ and $d = 3$ is
understood via the Imry-Ma argument {\cite{ImryMa}}, a kind of
Peierls' argument in presence of disorder. As in Peierls' argument, one
studies the possibility to disorder the all $+$ state by inverting droplets.
The energy cost of inverting a droplet $U$ of linear size $R$ is proportional
to the surface area $\delta E_1 \sim J | \partial U | \sim J R^{d - 1}$. In
Peierls' argument, this cost is offset by entropy effects at nonzero
temperature, in $d = 1$ (but not in $d \geqslant 2$). In the Imry-Ma argument
one works at zero temperature, and one tries to offset $\delta E_1$ by the
magnetic field contribution $\delta E_2 = 2 \sum_{x \in U} h_x$. We have
$\overline{(\delta E_2)^2} \sim \kappa^2 | U | $, i.e.~$\delta E_2$ has
typical size $\delta E^{{\rm typ}}_2 = \pm \kappa R^{d / 2}$.

We would like to find a droplet surrounding $x = 0$ such that $\delta E_1 +
\delta E_2 < 0$. We are assuming that $\kappa \ll J$ (weak disorder), so this
is unlikely for small droplets. For large droplets, the ratio $\delta
E^{{\rm typ}}_2 / \delta E_1$ decreases with $R$ for $d > 2$. We conclude
that the ordered state survives for $d > 2$ for weak disorder. For $d = 2$,
the ratio $\delta E^{{\rm typ}}_2 / \delta E_1$ stays constant with $R$. For
any $R$, there is a small finite chance to find a droplet of this size
surrounding $x = 0$ which can be flipped lowering the energy. In a
sufficiently large volume, there will be many values of $R$ to consider, and
the probability to find a good droplet tends to 1. This shows that the model
is disordered for $d = 2$.

\begin{figure}[h]
 \centering \resizebox{127pt}{94pt}{\includegraphics{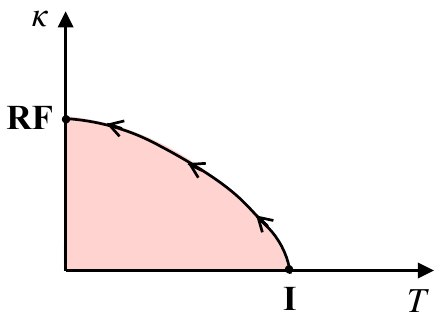}}
  \caption{\label{RGflow}RG flow along the phase separation line from the pure
  Ising fixed point I to the Random Field fixed point RF, located at zero
  temperature. }
\end{figure}

From now on, we will focus on $d > 2$. Our main interest will be the phase
transition which happens along the line separating the ordered and the
disordered phases. It is believed that this transition is continuous.
Moreover, it is believed that the whole line, excluding the point $\kappa =
0$, $T = T_c$, belongs to the same universality class. The schematic RG flow
diagram corresponding to this picture is shown in Fig.~\ref{RGflow}. We have
two fixed points. The pure Ising fixed point (I) is at $\kappa = 0$, $T =
T_c$. It is unstable with respect to disorder perturbations. All points on the
transition line flow to the Random Field fixed point (RF) located at $T = 0$,
$\kappa = \kappa_c$.

{\it Literature and further comments}

For a textbook exposition, see Cardy {\cite{Cardy-book}}.

Another way to understand that the $d = 2$ model is disordered is via RG. One
can argue {\cite{Bray1985}} that the ratio $w = \kappa / J$ satisfies in $d =
2$ an RG equation $d w / d \ell = A w^3$, $A > 0$. Thus any weak disorder
grows and becomes strong at long distances.

Rigorous mathematical work confirmed the RFIM phase diagram following from
the Imry-Ma argument. Imbrie {\cite{imbrieLCD2}} proved in 1985 that the RFIM in
$d = 3$ is ordered at weak $\kappa$ and $T = 0$. Bricmont and Kupiainen
{\cite{BK}} showed in 1988 that the order persists in $d = 3$ at weak $\kappa$
and small nonzero $T$. Recently there were some new developments. A new and
simple proof of the Imbrie and Bricmont-Kupiainen results was given in 2021
by Ding and Zhuang {\cite{Ding1}}. Further work
{\cite{Ding}} showed that order persists at all $T < T_c$ and
sufficiently small $\kappa$.

Disorder in $d = 2$ at arbitrarily small $\kappa$ was also shown to hold
rigorously, by Aizenman and Wehr {\cite{Aizenman1990}} in 1990. See also
{\cite{Ding2}} for recent rigorous developments in $d = 2$.

\subsection{Experiments}

Experimental studies of the RFIM in $d = 3$ and $d = 2$ were nicely reviewed in
1991 by Belanger and Young {\cite{BelangerYoung}} (see also
{\cite{Belanger}}). 2d experiments confirm the absence of a phase
transition. Here we will focus on the 3d case, where three experimental
platforms exist.

{\it Site-diluted uniaxial antiferromagnets in a uniform magnetic field.}
The microscopic Hamiltonian of this model is:
\begin{equation}
  \mathcal{H}= J \sum_{\langle x y \rangle} \varepsilon_x \varepsilon_y s_x
  s_y + H \sum_x \varepsilon_x s_x,
\end{equation}
where $s_x$ are the Ising spins, while $\varepsilon_x \in \{ 0, 1 \}$ are
disorder variables showing that some sites are vacant. They are chosen at
random, to satisfy a chosen concentration of vacancies. We see that this is a
disordered model, but the form of a disorder is different from RFIM. For the
zero uniform field, $H = 0$, this model on the cubic lattice is in the
universality class of the bond-disordered Ising model (footnote \ref{bond}).
For nonzero $H$, the phase transition should be in the RFIM universality
class, as was shown by Fishman and Aharony {\cite{Fishman}} and by Cardy
{\cite{Cardy-site-diluted}}. The idea is that nonzero $H$ creates a uniform
magnetization, on top of the antiferromagnetic order parameter experiencing
the critical fluctuations. The uniform magnetization couples linearly to the
order parameter, with a strength which is random since it depends on the local
dilution strength. The uniform magnetization times the random coupling
strength plays the role of a random magnetic field in RFIM. In this setup, the
effective random magnetic field strength can be tuned continuously by varying
$H$.

Experimentally, site-diluted uniaxial antiferromagnets can be realized by replacing magnetic
ions of a pure antiferromagnet by nonmagnetic ones. E.g.~replacing some Fe
ions in ${\rm Fe} \text{F}_2$ by ${\rm Zn}$ one obtains a site-diluted
antiferromagnet ${\rm Fe}_x {\rm Zn}_{1 - x}  \text{F}_2$ where $x$ is the
remaining concentration of ${\rm Fe}$ ions. Replacement happens at random
when the crystal is grown. Experiments in such 3d materials showed a
continuous phase transition {\cite{Belanger}}.

\begin{figure}[h]
 \centering \resizebox{145pt}{68pt}{\includegraphics{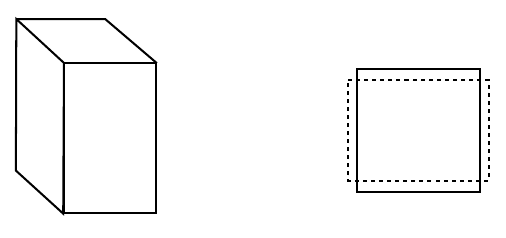}}
  \caption{\label{structural}Structural phase transition in pure ${\rm Dy} \,
  \text{V} \, \text{O}_4$. Left: The primitive cell of the tetragonal lattice
  is a square prism. Right: (View from above) At low temperatures the square
  section of the prism gets deformed to a rectangle. This can happen along the
  two axes, with the two ground states related by $\mathbb{Z}_2$.}
\end{figure}

{\it Structural phase transitions.}\footnote{See {\cite{Rong}} for a
recent review of non-disordered structural phase transitions.} In this
context, one takes a pure crystal which undergoes a transition of the Ising
universality class. One then adds impurities coupling linearly to the order
parameter. One example is the compound ${\rm Dy}\,\text{V}\,\text{O}_4$
which has tetragonal crystal lattice at high temperature, deforming to
orthorhombic at lower temperature. Deformation can happen in two orthogonal
directions, and the order parameter has $\mathbb{Z}_2$ symmetry as needed
(Fig.~\ref{structural}). This phase transition is known to be driven by the
linear coupling of electronic states of Dy and of the lattice modes (the
Jahn-Teller effect {\cite{Jahn-Teller}}). Replacing some of the V atoms by As
one gets a disordered crystal ${\rm Dy} \text{As}_x \, \text{V}_{1 - x} \,
\text{O}_4$, which realizes the RFIM. Measurements with this crystal found a
continuous phase transition {\cite{GrahamJT}}.

\begin{figure}[h]
 \centering \resizebox{124pt}{92pt}{\includegraphics{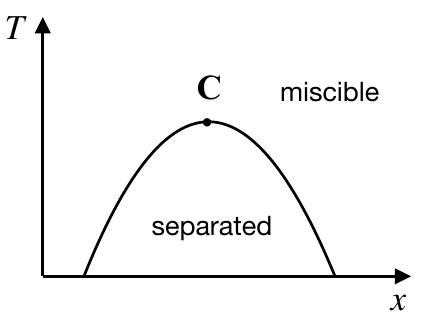}}
  \caption{\label{binary}A typical phase diagram of a binary fluid.}
\end{figure}

{\it Binary fluids in a gel.} Binary fluids are mixtures of two different
fluids, A and B, which are miscible in a range of temperatures $T$ and
concentration $x$. The typical phase diagram is in Fig.~\ref{binary}. The point
C in that diagram is a critical point, in the Ising universality class. The
fluctuating order parameter of this system is the local deviation of the
concentration from the critical value.

A gel is a rigid random network of polymer molecules. The randomness is built
when fabricating a gel. Imagine saturating a gel with a binary fluid, and
assume that the interaction between the gel and the molecules of the two fluid
components is not identical. Then, the presence of the gel will act as a
random field coupled linearly to the order parameter of the binary fluid, as
discussed by de Gennes in 1984 {\cite{DeGennes1984}}. If long-range
correlations in the gel structure can be neglected, the phase transition in
such a system should be of the RFIM universality class. Experimentally this
setup was studied in {\cite{Sinha1991}}.

\subsection{Numerical simulations}\label{sec-numsim}

Naively one might think that simulating RFIM is a hopeless task. On top of the
usual challenges of a Monte Carlo simulation at a finite temperature, arising
from the need to thermalize the system in a large volume and generate a large
number of statistically independent configurations, one seems to face the task
to repeat all of this for many instances of the magnetic field $h$. However
there is a trick which saves the day for the RFIM. Namely, the fixed point
controlling the phase transition is at zero temperature (Fig.~\ref{RGflow}).
Thus, we may simulate the model directly at $T = 0$. At $T = 0$, there is no
need to thermalize. Instead, we only need to find the ground state.

The simulation proceeds as follows. We fix the disorder distribution shape,
e.g.~the Gaussian {\eqref{gaussian}}. We choose the disorder strength
$\kappa$. For the chosen $\kappa$, we repeat many times the following two
steps: (1) generate an instance of a random field $h_x$ for every point $x$ on
the (large but finite) lattice; (2) find the ground state, i.e.~a
configuration of spins minimizing the RFIM Hamiltonian. We then vary $\kappa$
and tune it, until we find the critical value $\kappa_c$ corresponding to the
phase transition.

Note that the ground state on a finite lattice is unique for all choices of
$h$ but a set of measure zero. Indeed, suppose there are two ground states
$s_1$ and $s_2$, $\mathcal{H} [s_1, h] =\mathcal{H} [s_2, h]$. For any fixed
$s_1 \neq s_2$, the set of $h$'s solving this equation is a hyperplane, hence
of measure zero.

The problem of finding the ground state for a given $h$ is equivalent to a
classic problem of graph theory: finding a minimal cut in a graph. Let us add
two vertices $A$ and $B$ to the lattice, assigning them spins $s_A = + 1$ and
$s_B = - 1$. For vertices $x$ with $h_x > 0$ we introduce a coupling $J_{x A}
= h_x$ and for vertices $x$ with $h_x < 0$ we introduce a coupling $J_{x B} =
- h_x$. The RFIM Hamiltonian can now be rewritten equivalently as
\begin{equation}
  \mathcal{H} [s, h] = - \sum_{\langle x y \rangle} J_{x y} s_x s_y,
\end{equation}
where the sum goes over the old bonds $J_{x y} = J$ and the new bonds $J_{x
A}$ and $J_{x B}$. Crucially, all $J_{x y} > 0.$

\begin{figure}[h]
 \centering \resizebox{113pt}{101pt}{\includegraphics{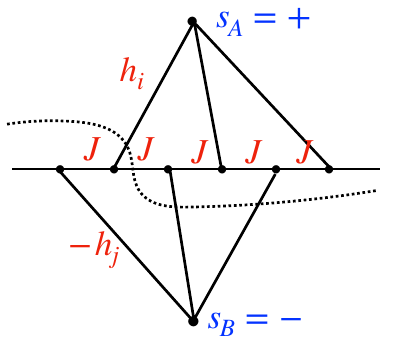}}
  \caption{\label{mincut}Finding the RFIM ground state is reduced to finding
  the minimal cut (dotted line). See the text.}
\end{figure}

To minimize the energy written in this form, we need to find a boundary
separating the $+$ spins from the $-$ spins such that the sum of all bonds
$J_{x y}$ traversing the boundary is minimal. Note that the boundary is not
empty since we assigned $s_A = + 1$ and $s_B = - 1$. Such a minimal separating
boundary is called the minimal cut between the source $A$ and the sink $B$
(Fig.~\ref{mincut}).

The minimal cut in a graph with positive weights on each link can be found by
efficient polynomial-time algorithms, such as e.g.~the push-relabel
algorithm. State-of-the-art numerical simulations of RFIM rely on such
algorithms. For an excellent review of these algorithms and many problems
in statistical physics they can be applied to (including RFIM), see
{\cite{Alava}}. To learn more about how the RFIM simulations are done, see the
excellent review {\cite{FytasReview}}.

In Table \ref{tab-exp} we report the RFIM critical exponents obtained in
numerical simulations and, for comparison, the exponents of the pure Ising
model. Note that $\nu \left( \text{4d RFIM} \right) \neq \nu \left( \text{2d
Ising} \right)$, hence no dimensional reduction in $d = 4$. On the other hand,
$\nu \left( \text{5d RFIM} \right)$ is remarkably close to $\nu \left(
\text{3d Ising} \right)$. Also $\eta \left( \text{5d RFIM} \right)$ agrees, at
$\sim 1.5 \sigma$ level, with $\eta \left( \text{3d Ising} \right)$. These
agreements suggest that the reduction may well hold in $d = 5$. A further
special property which appears to hold in $d = 5$ is $\eta = \bar{\eta}$. We
will see in Section \ref{sec-Ward} that this equality is a sign of
the Parisi-Sourlas SUSY, which is closely related to dimensional reduction.

\begin{table}[h]
	\centering
 \begin{tabular}{lllll}
    $d$ & $\nu$ & $\eta$ & $\bar{\eta}$ & Ref.\\ \toprule  
       3 & $1.38 (10)$ & 0.5153(9) & $\approx 2 \eta$ & \cite{Fytas3}\\
    4 & $0.872 (7)$ & 0.1930(13) & $\approx 2 \eta - 0.0322 (23)$ &
    \cite{Picco1}\\
    5 & $0.626 (15)$ & 0.055(15) & $\approx \eta$ & \cite{Picco2}\\
    \bottomrule
  \end{tabular}

\vspace{1em}
\begin{tabular}{llll}
    $d$ & $\nu$ & $\eta$ & Ref.\\\toprule
    1 & - & - & -\\
    2 & 1 & $1 / 4$ & \text{exact}\\
    3 & 0.629771(4) & 0.036298(2) & {\cite{Kos:2016ysd}}
    \\
    \bottomrule
  \end{tabular}
  \caption{\label{tab-exp}Critical exponents in the RFIM (above) and in the
  pure Ising model (below). Recall that $\nu$ controls the growth of the
  correlation length $\xi$ as $T \rightarrow T_c$. In the RFIM context at $T =
  0$, $\nu$ controls the growth of $\xi$ as $\kappa \rightarrow \kappa_c$:
  $\xi \sim | \kappa - \kappa_c |^{- \nu}$. The exponents $\eta$ and
  $\bar{\eta}$ will be discussed in Section \ref{PS-theory}.}
\end{table}

\section{Parisi-Sourlas theory, dimensional
reduction}\label{PS-theory}

In this lecture we will study Parisi and Sourlas's original argument
{\cite{Parisi:1979ka}} for the presence of SUSY at the RFIM critical point.
We will discuss possible caveats. We will also see why the Parisi-Sourlas SUSY
implies dimensional reduction.

\subsection{Field theory and perturbation theory}\label{2.1}

At the critical point we may replace the lattice formulation of the model by a
field theory. The appropriate field theory is a scalar field with quartic
self-interactions and coupled to a random magnetic field:
\begin{equation}
  S [\phi, h] = \int d^d x \left[ \frac{1}{2} (\partial \phi)^2 + V (\phi) - h
  \phi \right], \qquad V (\phi) = \frac{1}{2} m^2 \phi^2 + \frac{\lambda}{4!}
  \phi^4,
\end{equation}
\begin{equation}
  \overline{h (x) h (x')} = R\, \delta^{(d)} (x - x') \label{hprop},
\end{equation}
where the parameter $R$ plays the role of $\kappa^2$ in {\eqref{hh}}.
Correlation functions are computed as
\begin{equation}
  \overline{\langle A (\phi) \rangle_h} = \int \mathcal{D}h\,\mathcal{P} (h) 
  \langle A (\phi) \rangle_h, \label{Aphi}
\end{equation}
\begin{equation}
  \langle A (\phi) \rangle_h = \frac{\int \mathcal{D} \phi A (\phi) e^{- S
  [\phi, h]}}{Z [h]}
\end{equation}
where $A (\phi)$ is any function of fields, and $\mathcal{P} (h) \propto e^{-
\int d^d x \; h^2 / (2 R)}$.

We will consider perturbative expansion in this theory. Let us first discuss
how to compute $\langle A (\phi) \rangle_h$ for a fixed $h$. We will then see
what happens when we average over $h$.

For a fixed $h$, we have Feynman diagrams with the propagator $\langle \phi_p
\phi_{- p} \rangle_{\lambda = 0} = \frac{1}{p^2 + m^2}$. In addition to the quartic
interaction vertex $\lambda$, there is a vertex expressing the
linear $h \phi$ coupling:
\begin{equation}
\myinclude[height=27pt]{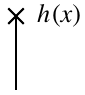} \text{(position
  space),} \qquad\myinclude[height=26pt]{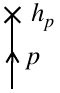}
  \text{(momentum space)} .
\end{equation}
In quantum field theory, such linear vertices are called ``tadpoles''. Since
$h (x)$ is, in general, $x$-dependent, the momentum space vertex carries
external momentum.

Due to tadpoles, $\phi$ acquires a nonzero
$h$-dependent expectation value:\footnote{We are not paying attention to the
symmetry factors here, and in many other equations in this section.}
\begin{equation}
  \langle \phi (x) \rangle_h =
\myinclude{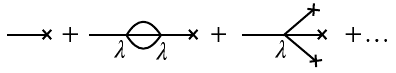}\,.
  \label{phih}
\end{equation}
To make sure that the notation is clear, we will give just once a translation
of these Feynman diagrams into an equation:
\begin{eqnarray}
  \langle \phi (x) \rangle_h & = & \int G (x - u) h (u) d^d u \nonumber\\
  & + & \lambda^2 \int G (x - y) G (y - z)^3 G (z - u) h (u) d^d y d^d z d^d
  u \nonumber\\
  & + & \lambda \int G (x - y) \prod_{i = 1}^3 G (y - u_i) h (u_i) d^d u_i
  d^d y + \ldots, 
\end{eqnarray}
where $G (x) = \int \frac{d^d p}{(2 \pi)^d} \frac{1}{p^2 + m^2} e^{i p x}$ is
the position-space propagator.

As another example, we will give the first few terms in the perturbative
expansion of the two-point functions $\langle \phi \phi \rangle$:
\begin{equation}
  \langle \phi (x) \phi (y) \rangle_h =\ 
\raisebox{-6em}{\includegraphics[scale=1.1]{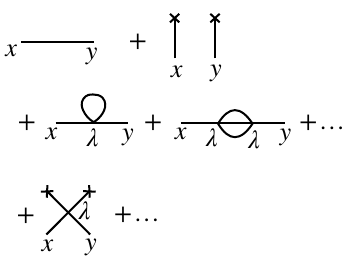}}
\end{equation}
In the first line we gave the terms present for $\lambda = 0$ (Gaussian
theory), in the second line the corrections involving $\lambda$ but not $h$,
and in the third line corrections from both $\lambda$ and $h$.

Finally let us discuss what happens when we perform average over $h$. This
average is denoted below by a dashed line. We do this average using the
propagator \eqref{hprop}, which becomes $\langle h_p h_{- p} \rangle
= R$ in momentum space. In the Feynman diagram notation, this average means
joining all crosses pairwise, in all possible combinations.

\subsection{Connected and disconnected 2pt functions}\label{sec-2pt}

The expression for $\langle \phi (x) \rangle_h$ in Eq.~{\eqref{phih}} has an
odd number of crosses which cannot be paired, hence $\overline{\langle \phi
(x) \rangle_h} = 0$, as expected from $\mathbb{Z}_2$ symmetry.

On the other hand for $\overline{\langle \phi \phi \rangle_h}$ we obtain a
nontrivial result. At $\lambda = 0$ (Gaussian theory) we get, in momentum
space
\begin{eqnarray}
  \overline{\langle \phi_p \phi_{- p} \rangle_h} & = &
 \myinclude[scale=1.1]{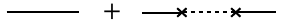} \nonumber\\
  & = & \frac{1}{p^2 + m^2} + \frac{R}{(p^2 + m^2)^2}  \qquad (\lambda = 0) .
  \label{phiphiquenched}
\end{eqnarray}
This goes as $\sim R / p^4$ for $p \rightarrow 0$ at the critical point
($m^2 = 0$). Translating to position space, we get $\sim
{\rm const} / r^{d - 4}$ behavior for $r \rightarrow \infty$. This was for
the Gaussian theory, while at the critical point of the interacting theory we
will have a correction in the power, denoted $\bar{\eta}$:
\begin{equation}
  G_{\text{disc}} (r) \assign \overline{\langle \phi (0) \phi (r) \rangle_h} =
  \frac{\text{const}}{r^{d - 4 + \bar{\eta}}} \quad (r \rightarrow \infty) .
  \label{Gdisc}
\end{equation}
This is the {\it disconnected} 2pt function.

We can also consider a slightly different 2pt function:
\begin{equation}
  \overline{\langle \phi (x) \rangle_h \langle \phi (y) \rangle_h} .
\end{equation}
This quantity measures how expectation values of $\phi$ induced by the random
field at two different points are correlated with each other. Unlike
$\overline{\langle \phi (x) \rangle_h}$, this does not vanish. The result in
the Gaussian theory is given by
\begin{equation}
  \overline{\langle \phi_p \rangle_h \langle \phi_{- p} \rangle_h} =
  \myinclude[scale=1.1]{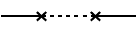} = \frac{R}{(p^2 + m^2)^2}  \qquad
  (\lambda = 0) .
\end{equation}

The {\it connected} 2pt function is defined by the difference:
\begin{equation}
  G_{\text{conn}} (r) : = \overline{\langle \phi (0) \phi (r) \rangle_h} -
  \overline{\langle \phi (0) \rangle_h \langle \phi (r) \rangle_h}\, .
  \label{Gconn}
\end{equation}
This represented the susceptibility with respect to adding a localized
non-random magnetic field:
\begin{equation}
  G_{\text{conn}} (r) = \frac{\partial}{\partial H} \overline{\langle \phi (0)
  \rangle_{h + H \delta (x - r)}} .
\end{equation}
Another rationale for the definition {\eqref{Gconn}} is that the leading $p
\rightarrow 0$ singularity cancels in the difference. Indeed in the Gaussian
theory we will have $G_{\text{conn}} (p) = \frac{1}{p^2 + m^2},$ which becomes
$1 / p^2$ at criticality, hence $\sim {\rm const} / r^{d - 2}$ in the
position space. So as usual, the connected 2pt function decays faster at
infinity than the disconnected one.

In the interacting theory we expect that the critical behavior of
$G_{\text{conn}}$ will be modified:
\begin{equation}
  G_{\text{conn}} (r) = \frac{\text{const}}{r^{d - 2 + \eta}} \quad (r
  \rightarrow \infty) .
\end{equation}
In general, $\eta$ may be different from $\bar{\eta}$ from Eq.~{\eqref{Gdisc}},
and indeed we see from Table \ref{tab-exp} that $\eta \neq \bar{\eta}$ in $d =
3, 4$. On the other hand, simulations in $d = 5$ are consistent with $\eta =
\bar{\eta}$. This, as we will see below, may be a sign of the Parisi-Sourlas
SUSY.

\begin{remark}
  Let's see the difference between the quenched disorder average
  {\eqref{Aphi}} which treats the random magnetic field $h$ frozen, i.e.~out
  of thermal equilibrium with $\phi$, and the usual average (called annealed
  disorder) which would consider $h$ as just another field in the path
  integral on par with $\phi$. In the latter case we would define correlation
  functions by
  \begin{equation}
    \langle A (\phi) \rangle_{{\rm ann}} = Z^{- 1} \int
    \mathcal{D}h\mathcal{D} \phi A (\phi) e^{- S [\phi, h] - \int h^2 / (2 R)}
    .
  \end{equation}
  Integrating out $h$, we are left with an action for the field $\phi$ alone,
  with a shifted mass: $m^2 \rightarrow m^2 - R$.
  
  Not surprisingly, the values of correlation functions are completely
  different for the two procedures. Consider e.g.~the disconnected 2pt
  function at $\lambda = 0$. In the quenched case it is given by Eq.
  {\eqref{phiphiquenched}}, while in the annealed case by
  \begin{equation}
    \langle \phi_p \phi_{- p} \rangle_{{\rm ann}} = \frac{1}{p^2 + m^2 - R} .
  \end{equation}
  That's clearly different from {\eqref{phiphiquenched}}, although agrees to
  first order in $R$ at fixed $p$, but that's not what matters. What matters
  is the behavior at $p \rightarrow 0$ when $m^2$ is fixed to the critical
  value. At criticality, the annealed 2pt function behaves as $1 / p^2$ for $p
  \rightarrow 0$, while the quenched 2pt function shows a stronger $1 / p^4$
  singularity. 
\end{remark}

\subsection{Selection of important diagrams}\label{selection}

Let us study the structure of perturbative expansion in the interacting theory
$(\lambda \neq 0)$. Consider first the pure Ising case $(h = 0)$. It is interesting to consider
the relative importance of loops vs tree level contributions. Compare these
two diagrams contributing to the 4pt function:
\begin{equation}
  \myinclude[scale=1.1]{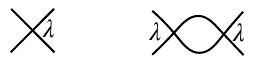}\,.
\end{equation}
We can view the second diagram, involving an integral over the external loop
momentum, as giving a correction $\delta \lambda = O (\lambda^2)$ to the
quartic coupling $\lambda$. At zero external momentum (as appropriate for
studying the long-distance behavior of the model), $\delta \lambda / \lambda$
is given by:
\begin{equation}
  \delta \lambda / \lambda = \lambda^{} \int_0^{\Lambda} \frac{d^d k}{k^4},
  \label{corr-int}
\end{equation}
where $\Lambda$ is the UV cutoff. For $d > 4$, this integral converges at the
lower limit. Thus $\delta \lambda / \lambda$ is finite. On the other hand, for
$d < 4$, the integral is IR-divergent. This is a simple way to identify the
upper critical dimension of the model: $d_{\text{uc}}( \text{pure Ising}) = 4$.

For a nonzero external momentum $p$, the integral {\eqref{corr-int}} will be
IR-convergent and, for dimensional reasons, of order $\lambda^{} p^{-
\epsilon}$, where $\epsilon = 4 - d > 0$. Further loops will introduce further
factors of $\lambda p^{- \epsilon}$. Resumming these factors modifies the
scaling behavior of correlators as $p \rightarrow 0$. Such resummation is
usually performed by renormalization group methods, which convert fixed-order
perturbative results into all-order expressions having good asymptotic scaling
behavior.

Let us now apply similar logic to the RFIM. In addition to the above loop
effects we will find new ones, more singular as $p \rightarrow 0$. Consider a
few terms in the perturbative expansion of $\langle \phi (x) \rangle_h$:
\begin{equation}
  \langle \phi (x) \rangle_h \ni \myinclude{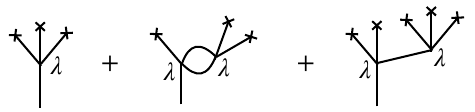} .
  \label{phi-h-corr}
\end{equation}
The second diagram loop is a correction $\delta \lambda$ of the same form as
for the pure Ising. But the last diagram, when averaging over $h$, \ gives a
new effect:
\begin{equation}
  \includegraphics{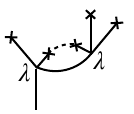} \label{new-effect}\,.
\end{equation}
We are imagining here that the crosses left free will connect to some other crosses as
when computing $\overline{\langle \phi (x) \rangle_h \langle \phi (y)
\rangle_h}$. We are focusing on just the shown subdiagram, which is of the
same form the first diagram in {\eqref{phi-h-corr}}, but with a 
coupling correction $\delta \lambda'$,
\begin{equation}
  \delta \lambda' / \lambda = \lambda^{} R \int_0^{\Lambda} \frac{d^d k}{k^6}
  .
\end{equation}
This new correction is IR-divergent for $d < 6$. Hence we conclude that the
upper critical dimension of RFIM is raised to $d_{\text{uc}}(
\text{RFIM}) = 6$.

When we compute any correlation function characterized by an external momentum
$p$, the just described effect will cause, in the $n$-th order of perturbation
theory, relative corrections of the order
\begin{equation}
  (\lambda R)^n p^{- n \varepsilon} (1 + O (p^2 / R)^{}), \qquad \varepsilon =
  6 - d, \label{nth}
\end{equation}
by dimensional analysis. Suppose we drop $O (p^2 / R)$ correction in
{\eqref{nth}}. This appears reasonable since we are interested in the $p
\rightarrow 0$ limit (but see the caveats in Section \ref{caveats} below).
This corresponds to dropping ``pure Ising'' loop corrections like in the
second diagram in {\eqref{phi-h-corr}} and keeping only the more singular loop
corrections {\eqref{new-effect}}. This, in turn, means that before doing the
average over $h$ we must keep, in each order in $\lambda$, the diagrams with
the maximal number of crosses. These are the tree-level diagrams:
\begin{equation}
  \phi_h (x) \assign \myinclude[scale=1.1]{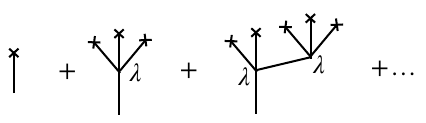},
\end{equation}
and we introduced notation $\phi_h (x)$ for their sum.

To recap, according to the above logic, when we compute a correlation function
such as $\overline{\langle \phi (x) \rangle_h \langle \phi (y) \rangle_h}$ at
criticality and for $p \rightarrow 0$, we should be allowed to replace
$\langle \phi (x) \rangle_h \rightarrow \phi_h (x)$ before doing the average
over $h$:
\begin{equation}
  \overline{\langle \phi (x) \rangle_h \langle \phi (y) \rangle_h} \sim
  \overline{\phi_h (x) \phi_h (y)} \qquad (p \rightarrow 0) .
\end{equation}
This corresponds to dropping some Feynman diagrams which go to zero faster as
$p \rightarrow 0$ than the diagrams we keep. This logic can be applied also to
any other correlator, e.g.~we can argue that
\begin{equation}
  \overline{\langle \phi (x) \nobracket \nobracket \phi (y) \rangle_h} \sim
  \overline{\phi_h (x) \phi_h (y)} \qquad (p \rightarrow 0) .
\end{equation}
This should not be surprising since we already noted that the disconnected 2pt
function is more singular at $p \rightarrow 0$ than the connected one.

\subsection{Stochastic PDE representation, Parisi-Sourlas action}

The above argument that the diagrams with the maximal number of crosses are
the most IR singular ones in every order of perturbation theory is due to
{\cite{ImryMa,Aharony:1976jx}}. Ref.~{\cite{Aharony:1976jx}} then showed that
perturbation theory based on just keeping these diagrams is the same, diagram
by diagram, as for the pure Ising in 2 dimensions lower, implying dimensional
reduction of the critical exponents. Let us see how this coincidence can be
efficiently explained in terms of SUSY (Parisi and Sourlas
{\cite{Parisi:1979ka}}).

The basic observation is that $\phi_h (x)$ defined by the sum of the tree
diagrams is a perturbative solution of the classical equation of motion
following from the action $S [\phi, h]$:
\begin{equation}
  - \partial^2 \phi + V' (\phi) - h = 0 . \label{stoPDE}
\end{equation}
Such classical equations with random sources ($h$ in our case) are called
stochastic PDEs.

We wish to compute averaged correlation functions such as e.g.
\begin{equation}
  \overline{\phi_h (x) \phi_h (y)} = \int \mathcal{D}h \; \phi_h (x) \phi_h
  (y) \, e^{- \int h^2 / (2 R)} . \label{ex-corr-av}
\end{equation}
We can rewrite this using the following crucial identity:
\begin{equation}
  \phi_h (x) \phi_h (y) = \int \mathcal{D} \varphi\, \varphi (x) \varphi (y)
  \delta [- \partial^2 \varphi + V' (\varphi) - h] \det [- \partial^2 + V''
  (\varphi)] . \label{most-imp}
\end{equation}
Here, the $\delta$-function localizes to the solution of {\eqref{stoPDE}}. It
will be convenient to call, as we did, the field appearing in this functional
integral and restricted to satisfy the classical equation of motion, by a
letter $\varphi$ to distinguish it from the field $\phi$ in the original
action. If we just had the $\delta$-function, the $\varphi$ integral would
produce a fluctuation determinant in the denominator. We don't want this
determinant, and so we introduced an explicit determinant factor to cancel it.

Plugging Eq.~{\eqref{most-imp}} into {\eqref{ex-corr-av}} and doing the $h$
integral, we obtain
\begin{equation}
  \overline{\phi_h (x) \phi_h (y)} = \int \mathcal{D} \varphi\, \varphi (x)
  \varphi (y) e^{- \int [- \partial^2 \varphi + V' (\varphi)]^2 / (2 R)} \det
  [- \partial^2 + V'' (\varphi)] \, . \label{eq-with-det}
\end{equation}
Parisi and Sourlas realized that the path integral in the r.h.s.~of this
equation has a hidden symmetry (supersymmetry). To see this, one writes this
path integral in an equivalent form by introducing additional auxiliary
fields. First, one introduces a scalar field $\omega$ to write the exponential
factor in {\eqref{eq-with-det}} as
\begin{equation}
  e^{- \int [- \partial^2 \varphi + V' (\varphi)]^2 / (2 R)} = \int
  \mathcal{D} \omega\, e^{\int \frac{R}{2} \omega^2 - \omega [- \partial^2
  \varphi + V' (\varphi)]} .
\end{equation}
\begin{remark}
  For the path integral to be convergent, the integration contour for the
  field $\omega$ has to run along the imaginary axis. However the action for
  $\omega$ is quadratic and the field $\omega$ could be eliminated by its
  classical equation of motion
  \begin{equation}
    \omega = \frac{1}{R} [- \partial^2 \varphi + V' (\varphi)],
    \label{meaning}
  \end{equation}
  which is real. The true purpose of introducing $\omega$ is that this will
  simplify the form of supersymmetry transformations. We will treat the field
  $\omega$ below as real.\footnote{This reality ``problem'' was also mentioned
  by Wegner {\cite{Wegner:2016ahw}}, p.209.}
\end{remark}

\begin{exercise}
  \label{phiomega}Using Eq.~{\eqref{meaning}}, show that, in the approximation
  of keeping the diagrams with the maximal number of crosses, the correlation
  function $\langle \varphi (x) \omega (y) \rangle$ measures the
  susceptibility, i.e.
  \begin{equation}
    \partial_H \overline{\langle \phi (x) \rangle_{h + H \delta (\cdot - y)}}
    \approx \partial_H \overline{\phi_{h + H \delta (\cdot - y)}}\,,
  \end{equation}
  where we add a small deterministic magnetic field $H$ localized at $y$ and
  measure the response of $\langle \phi (x) \rangle_h \approx \phi_h(x)$.
\end{exercise}

Second, one introduces two fermionic (i.e.~anticommuting, or Grassmann) scalar
fields $\psi, \bar{\psi}$ to represent the determinant in
{\eqref{eq-with-det}} as:
\begin{equation}
  \det [- \partial^2 + V'' (\varphi)] = \int \mathcal{D} \psi \mathcal{D}
  \bar{\psi} e^{- \int \bar{\psi} [- \partial^2 + V'' (\varphi)] \psi} .
\end{equation}
\begin{remark}
  \label{RemFP}This is analogous to how one represents the determinant appearing
  in the Faddeev-Popov quantization of nonabelian gauge fields, $\psi,
  \bar{\psi}$ playing the role of the Faddeev-Popov ghosts. Anticommuting
  scalar fields do not obey the spin-statistics relation, signaling violation of
  unitarity. For gauge fields the unitarity is restored thanks to the cancellation
  of two effects: ghosts and longitudinal gauge bosons. In the PS construction
  there is no such cancellation and the theory is truly non-unitary. Perhaps
  this is not so surprising since disordered models can be treated using the
  replica method involving the $n \rightarrow 0$ limit, as we will see in
  Section \ref{Sec3}. At any rate, we should not worry about the lack of unitarity. Indeed, unitarity (or rather its Euclidean
  counterpart reflection positivity) is not a fundamental requirement for the
  field theories of statistical physics, and many physically important models
  violate it.
\end{remark}

All in all we see that {\eqref{eq-with-det}} can be rewritten as
\begin{equation}
  \overline{\phi_h (x) \phi_h (y)} = \int \mathcal{D} \varphi \mathcal{D}
  \omega \mathcal{D} \psi \mathcal{D} \bar{\psi}\, \varphi (x) \varphi (y) e^{-
  S_{\text{PS}}},
\end{equation}
where the {{\em Parisi-Sourlas action}} $S_{\text{PS}}$ is given by
\begin{equation}
  S_{\text{PS}} = \int d^d x \left\{ - \frac{R}{2} \omega^2 + \omega [-
  \partial^2 \varphi + V' (\varphi)] + \int \bar{\psi} [- \partial^2 + V''
  (\varphi)] \psi \right\} . \label{SPS}
\end{equation}
The claim is that computations based on this action are completely equivalent
to perturbation theory described in Section \ref{selection}, i.e.~computing
$\phi_h$ via the sum of tree level diagrams and then averaging over $h$ by
joining crosses.

The quadratic part of $S_{\text{PS}}$ gives the following momentum-space
propagators:
\begin{equation}
  G_{\varphi \omega} (p) = \frac{1}{p^2 + m^2}, \quad G_{\varphi \varphi} (p)
  = \frac{R}{(p^2 + m^2)^2}, \quad G_{\omega \omega} (p) = 0, \quad
  G_{\bar{\psi} \psi} (p) = \frac{1}{p^2} . \label{PSprops}
\end{equation}
\begin{exercise}
  Check this. For the propagators involving $\varphi$ and $\omega$ you have to
  invert a matrix since these fields appear in $S_{\text{PS}}$ non-diagonally.
\end{exercise}

The interaction vertices are associated with the quartic couplings:
\begin{equation}
  \omega V' (\varphi) \rightarrow \frac{\lambda}{3!} \omega \varphi^3, \qquad
  \psi \bar{\psi} V'' (\varphi) \rightarrow \frac{\lambda}{2} \bar{\psi} \psi
  \varphi^2 . \label{quartics}
\end{equation}

Since the manipulations leading to the action $S_{\text{PS}}$ may appear somewhat
formal, let us check the equivalence with Section \ref{selection} in a simple
example. To $O (\lambda^2)$, the 2pt function $\langle \varphi (p) \varphi (-
p) \rangle$ computed from $S_{\text{PS}}$ is given by the sum of the following
diagrams:
\begin{equation}
  \myinclude[scale=1.1]{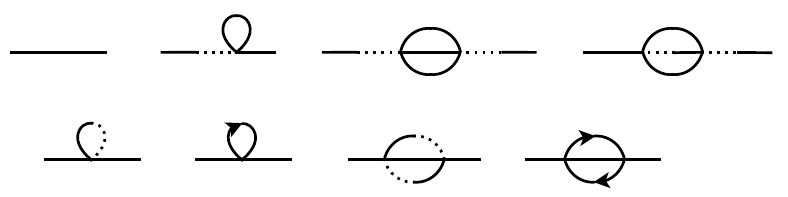} \label{five-diag} .
\end{equation}
Here $G_{\varphi \varphi}$ are solid lines, $G_{\varphi \omega}$ are
solid-dotted lines, $G_{\bar{\psi} \psi}$ are solid lines with arrows. The
vertices are read accordingly. Note that these diagrams are {\it not}
amputated.

The first line of {\eqref{five-diag}} reproduces, in the massless limit $m^2 =
0$, the four diagrams for $\overline{\phi_h (x) \phi_h (y)}$ to $O
(\lambda^2)$ computed as in Section \ref{selection}:
\begin{equation}
  \myinclude[scale=1.1]{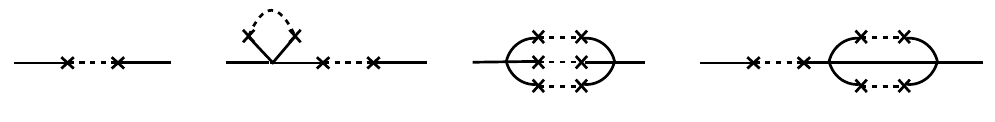} .
\end{equation}
As for the diagrams in the second line of {\eqref{five-diag}}, they cancel
pairwise thanks to the fermionic loop minus sign. We thus see the equivalence,
and the crucial role played by fields $\psi, \bar{\psi}$ in ensuring it.

For later use, we record here the free scaling dimensions of the fields $\varphi,
\omega, \psi, \bar{\psi}$, which are read off from the scaling behavior of
propagators {\eqref{PSprops}} in the massless limit:
\begin{equation}
  [\varphi] = \frac{d}{2} - 2, \quad [\psi] = [\bar{\psi}] = \frac{d}{2} - 1,
  \quad [\omega] = \frac{d}{2} . \label{freedims}
\end{equation}
For example, transforming $G_{\varphi \varphi}$ to position space we get
$G_{\varphi \varphi} (x) \sim R / x^{d - 4}$, hence $[\varphi] = \frac{d -
4}{2}$.

\begin{remark}
  The scaling dimensions determine how the propagators scale when rescaling
  distances. Parameter $R$, as a coupling constant, is kept fixed when we
  rescale $x$ to determine the scaling dimensions. The value of $R$ is of
  secondary importance since it can be changed by changing normalization of
  the fields: $\omega \rightarrow z^{- 1} \omega$, $\varphi \rightarrow z
  \varphi$ changes $R \rightarrow R / z^2$. Below we will sometimes use this
  to set $R = 2$. 
\end{remark}

\subsection{Parisi-Sourlas supersymmetry}

The action $S_{\text{PS}}$ is invariant under supersymmetry transformation.
The best way to see this is via the ``superfield formalism,'' which makes
supersymmetry manifest, analogously to how rotation invariance in field theory
is made manifest by the tensor notation.

We consider a superspace whose points $y = (x, \theta, \bar{\theta})$ are
parameterized by $x \in \mathbb{R}^d$ and two real Grassmann number coordinates
$\theta, \bar{\theta}$ satisfying $\theta^2 = 0$, $\bar{\theta}^2 = 0$,
$\theta \bar{\theta} = - \bar{\theta} \theta$. We define Grassmann parity
$[a]$ to be 0,1 for the bosonic/fermionic components of $y^a$, so that $y^a y^b =
(- 1)^{[a] [b]} y^b y^a$.

We then consider a superfield $\Phi (y) = \Phi (x, \theta, \bar{\theta})$
which is a commuting (Grassmann-even) function on the superspace. Any function
of $\theta, \bar{\theta}$, in particular $\Phi (x, \theta, \bar{\theta})$, can
be expanded as a polynomial in these variables, whose expansion stops at
$\theta \bar{\theta}$. We define $\Phi (x, \theta, \bar{\theta})$ so that the
coefficients of this expansion are identified with the fields appearing in
$S_{\text{PS}}$:
\begin{equation}
  \Phi (x, \theta, \bar{\theta}) = \varphi (x) + \theta \bar{\psi} (x) +
  \bar{\theta} \psi (x) + \theta \bar{\theta} \omega (x) . \label{Phidef}
\end{equation}
Fields $\varphi, \psi, \bar{\psi}, \omega$ are thus ``packaged'' into $\Phi
(x, \theta, \bar{\theta})$, and are referred to as components of $\Phi$.

To discuss the supersymmetry transformations, we endow the superspace with the
metric
\begin{equation}
  d s^2 = d x^2 + 2 \alpha d \bar{\theta} d \theta,
\end{equation}
where $\alpha$ is a fixed real number. We consider all transformations of the
superspace which preserve this metric. These are (super)translations $x
\rightarrow x + a$, $\theta \rightarrow \theta + \varepsilon$, $\bar{\theta}
\rightarrow \bar{\theta} + \bar{\varepsilon}$, and (super)rotations
\begin{equation}
  y^a \rightarrow y^{\prime a} = \textbf{R}^a_{\ b} y^b, \qquad y^2 = (y')^2,
  \label{superrot}
\end{equation}
where $y^2 = y^a_{} g_{a b} y^b = x^2 + 2 \alpha \bar{\theta} \theta$. The
metric tensor $g_{a b}$ has nonzero components $g_{\mu \nu} = \delta_{\mu
\nu}$ and $g_{\bar{\theta} \theta} = \alpha = - g_{\theta \bar{\theta}}$. It
is symmetric/antisymmetric in the bosonic/fermionic indices: $g^{}_{a b} = (-
1)^{[a] [b]} g^{}_{b a}$.

\begin{exercise}
  Show that the condition $(y')^2 = y^2$ can be written in matrix form as
  ${\rm\bf R}^{st} g {\rm\bf R} = g$ with the supertranspose ${\rm\bf R}^{st}$
  defined as
  \begin{equation}
    {\rm\bf R} = \left(\begin{array}{cc}
      A & B\\
      C & D
    \end{array}\right) \quad \Rightarrow \quad {\rm\bf R}^{st} =
    \left(\begin{array}{cc}
      A^t & C^t\\
      - B^t & D^t
    \end{array}\right), \label{block22}
  \end{equation}
  where we write ${\rm\bf R}$ in block-diagonal form with $A_{d \times d}$ and $D_{2
  \times 2}$ even blocks and $B_{d \times 2}$ and $C_{2 \times d}$ odd blocks,
  and $t$ is the ordinary transpose.
\end{exercise}

Superlinear transformations ${\rm\bf R}^a_{\ b}$ in {\eqref{superrot}} form a supergroup
called $\text{Osp} (d | 2 \nobracket)$ where Osp stands for
``orthosympletic,'' and $d | 2 \nobracket$ means $d$ bosonic and 2 fermionic
directions in the superspace. The maximal bosonic subgroup of this supergroup
is $\text{O} (d) \times \text{Sp} (2)$ where orthogonal $\text{O} (d)$
transformations act on $x$ preserving $x^2$ and symplectic $\text{Sp} (2)$
transformations act on $\theta, \bar{\theta}$ preserving $\theta
\bar{\theta}$. This subgroup is formed by the blocks $A$ and $D$ in
{\eqref{block22}}, with $B = C = 0$. $\text{Osp} (d | 2 \nobracket)$ also
contains transformations mixing $x$ and $\theta, \bar{\theta}$, represented by
$B$ and $C$ in {\eqref{block22}}. We will only need the infinitesimal form of
these ``superrotations'':
\begin{equation}
  \delta x^{\mu} = \alpha (\varepsilon^{\mu}_{} \bar{\theta} +
  \bar{\varepsilon}^{\mu}_{} \theta), \quad \delta \theta = -
  \varepsilon^{\mu}_{} x_{\mu}, \quad \delta \bar{\theta} =
  \bar{\varepsilon}^{\mu}_{} x_{\mu} . \label{inf-superrot}
\end{equation}

We define covariant vectors, i.e.~vectors with lower indices, by
\begin{equation}
  y_a = g_{a b} y^b, \label{covariant}
\end{equation}
so that the invariant contraction between a contravariant vector $y$ and a
covariant vector $z$ is $y^a z_a$.\footnote{Note that this is {\bf not} equal to
$y_a z^a$. We have to pay attention to how lower and upper indices of
anticommuting objects are contracted.} The vector of derivatives of a function
is a covariant vector, as is clear from the expression for the differential $d
\Phi = y^a \partial_a \Phi$.

Inverting {\eqref{covariant}}, we find
\begin{equation}
  y^a = g^{b a} y_b,
\end{equation}
where the metric with upper indices satisfies $g^{a c} g_{a b} = \delta^c_{\;
b}$. Note that $g^{a c}$ is not equal to the inverse metric $(g^{- 1})^{a c}$
which satisfies $(g^{- 1})^{a c} g_{c b} = \delta^a_{\; b}$. Rather, we have:
\begin{equation}
  g^{a c} = (- 1)^{[a] [c]} (g^{- 1})^{a c} .
\end{equation}
Specifically, we have $g_{}^{\mu \nu} = \delta^{\mu \nu}_{}$ and
$g_{}^{\bar{\theta} \theta} = \alpha^{- 1} = - g^{\theta \bar{\theta}}$. Using
the metric with upper indices, we write an invariant quadratic form on
covariant vectors:
\begin{equation}
  z^b z_b = z_a g^{a b} z_b = z_{\mu} z_{\nu} + 2 \alpha^{- 1}
  z_{\bar{\theta}} z_{\theta} . \label{invcov}
\end{equation}
We will need this expression below, for $z_a = \partial_a \Phi$.

Just like coordinate transformations induce transformations of fields in
ordinary field theory, superspace transformations $y \rightarrow y + \delta y$
induce transformations of the superfield $\Phi$:\footnote{Equation $\Phi (y)
\rightarrow \Phi (y + \delta y)$ means that $\Phi$ transforms as a scalar
field under superspace transformations.}
\begin{equation}
  \Phi (y) \rightarrow \Phi (y + \delta y) = \Phi (y) + \delta \Phi,
\end{equation}
where $\delta \Phi$ can be computed expanding $\Phi (y + \delta y)$ to first
order in $\delta y$. We call these transformations of fields supersymmetry
transformations. For example, consider the supertranslation $\theta
\rightarrow \theta + \varepsilon$. We have:
\begin{eqnarray}
  \Phi \rightarrow  \varphi + (\theta + \varepsilon) \bar{\psi} +
  \bar{\theta} \psi + (\theta + \varepsilon) \bar{\theta} \omega
 = (\varphi + \varepsilon \bar{\psi}) + \theta \bar{\psi} + \bar{\theta}
  (\psi - \varepsilon \omega) + \theta \bar{\theta} \omega, 
\end{eqnarray}
from where we infer the corresponding supersymmetry transformations of the
components:
\begin{equation}
  \delta \varphi = \varepsilon \bar{\psi}, \quad \delta \bar{\psi} = 0, \quad
  \delta \psi = - \varepsilon \omega, \quad \delta \omega = 0.
  \label{supertrans}
\end{equation}
\begin{exercise}
  \label{totder}Work out transformations of the superfield components
  corresponding to the superrotations {\eqref{inf-superrot}}. In particular
  show that $\delta \omega = - \alpha \varepsilon^{\mu}_{} \partial_{\mu}
  \bar{\psi} + \alpha \bar{\varepsilon}^{\mu}_{} \partial_{\mu} \psi$.
\end{exercise}

We have defined a class of supersymmetry transformation of fields, depending
on parameter $\alpha$. We claim that $S_{\text{PS}}$ is invariant under the
transformations we defined, provided that we choose the parameter $\alpha$
appropriately. This follows from three observations:

1. We note that $S_{\text{PS}}$ can be written as a superspace integral, as
follows:
\begin{gather}
  S_{\text{PS}} = \int d^d x\, d \bar{\theta} d \theta \mathcal{L}, \\
  \mathcal{L}= - \frac{1}{2} \Phi (\partial_{\mu} \partial^{\mu} + R
  \partial_{\bar{\theta}} \partial_{\theta}) \Phi + V (\Phi) . \label{LPS}
\end{gather}
The integral over the Grassmann directions is normalized conventionally as
$\int d \bar{\theta} d \theta\, \theta \bar{\theta} = 1$.

\begin{exercise}
  \label{checkL}Check this. Namely show that $\int d \bar{\theta} d \theta
  \mathcal{L}$ equals, up to integration by parts, the Lagrangian in
  {\eqref{SPS}}.
\end{exercise}

2. Let us fix $\alpha = 2 / R$. For this choice of $\alpha$ the second-order
differential operator in the definition of $\mathcal{L}$ is the
super-Laplacian associated with $\text{Osp} (d | 2 \nobracket)$, because it can
be written in the $\text{Osp} (d | 2 \nobracket)$ covariant form (see
{\eqref{invcov}}):
\begin{equation}
  \partial_{\mu}^{} \partial^{\mu} + 2 \alpha^{- 1} \partial_{\bar{\theta}}
  \partial_{\theta} = \partial_a g^{a b} \partial_b .
\end{equation}
It follows that the Lagrangian $\mathcal{L}$ is a superfield which transforms
under the supersymmetry transformation in the same way as $\Phi$, i.e.
$\mathcal{L} (y) \rightarrow \mathcal{L} (y + \delta y)$.

\begin{exercise}
  A radial function is an arbitrary function of $y^2$. Check that the
  super-Laplacian maps radial functions to radial functions.
\end{exercise}

3. We can expand $\mathcal{L}$ in components:
\begin{equation}
  \mathcal{L} (x) = L (x) + \theta \bar{\Psi} (x) + \bar{\theta} \Psi (x) +
  \theta \bar{\theta} \Omega (x) .
\end{equation}
We know from Exercise \ref{checkL} that $\Omega = \int d \bar{\theta} d \theta
\mathcal{L}$ equals, up to integration by parts, the Lagrangian in
{\eqref{SPS}}. The fields $L, \bar{\Psi}, \Psi$ can also be expressed in terms
of the components of $\Phi$, but we won't need their expressions. We need to
show that $S_{{\rm PS}} = \int d^d x\, \Omega$ is invariant under SUSY
transformation. Since $\mathcal{L}$ is a superfield, transformation rules for
$\Omega$ are the same as for $\omega$ up to replacement $\varphi, \psi,
\bar{\psi} \rightarrow L, \Psi, \bar{\Psi}$. In particular,
{\eqref{supertrans}} shows that $\Omega$ is supertranslation-invariant. It
also follows from Exercise \ref{totder} that $\Omega$ transforms into a total
derivative under superrotations. Hence $S_{{\rm PS}}$ is invariant. Q.E.D.

\begin{remark}
  Parisi-Sourlas supersymmetry, with its scalar fermionic directions and
  scalar supercharges and lack of unitarity (see Remark \ref{RemFP}) will look
  unfamiliar to high-energy theorists interested in unitary supersymmetries
  having spinor supercharges $Q_{\alpha}$ and spinor fermionic superspace
  directions $\theta_{\alpha}$. Nevertheless it obeys very similar rules. It's
  also much simpler - it is the simplest field-theoretic supersymmetry around.
\end{remark}

\subsection{SUSY Ward identities}\label{sec-Ward}

One immediate consequence of SUSY is that the correlation functions of the
superfield $\Phi$ have to be invariant under SUSY. For example, the 2pt
function of $\Phi$ has to be a function of the superspace distance, i.e.
\begin{equation}
  \langle \Phi (x_1, \theta_1, \bar{\theta}_1) \Phi (x_2, \theta_2,
  \bar{\theta}_2) \rangle = f [(x_1 - x_2)^2 + 2 \alpha (\bar{\theta}_1 -
  \bar{\theta}_2) (\theta_1 - \theta_2)] .
\end{equation}
Let us expand both sides of this relation in $\theta, \bar{\theta}$. In the
r.h.s.~we get
\begin{equation}
  f [(x_1 - x_2)^2] + 2 \alpha f' [(x_1 - x_2)^2] (\bar{\theta}_1 -
  \bar{\theta}_2) (\theta_1 - \theta_2), \label{exprhs}
\end{equation}
while in the l.h.s., using {\eqref{Phidef}}, we get a linear combination of
correlation functions of fields $\varphi, \omega, \psi, \bar{\psi}$. Matching
term by term we get:
\begin{eqnarray}
  \langle \varphi (x_1) \varphi (x_2) \rangle & = & f [(x_1 - x_2)^2],
  \nonumber\\
  \langle \varphi (x_1) \omega (x_2) \rangle & = & \alpha f' [(x_1 - x_2)^2],
  \\
  \langle \psi (x_1) \bar{\psi} (x_2) \rangle & = & \alpha f' [(x_1 - x_2)^2],
  \nonumber\\
  \langle \omega (x_1) \omega (x_2) \rangle & = & 0 . \nonumber
\end{eqnarray}
[The last relation follows from the fact that {\eqref{exprhs}} has no term
proportional to $\theta_1 \bar{\theta}_1 \theta_2 \bar{\theta}_2$.]

These relations among correlators, referred to as supersymmetric Ward
identities, should hold in any PS SUSY theory. In a scale invariant theory,
where $f$ is a power, they imply relations among scaling dimensions of the
fields:
\begin{equation}
  [\psi] = [\bar{\psi}] = [\varphi] + 1, \qquad [\omega] = [\varphi] + 2 .
  \label{SUSYrels}
\end{equation}
This holds in the free theory with $[\varphi] = d / 2 - 2$, see
{\eqref{freedims}}. In an interacting theory, fields will acquire anomalous
dimensions, but the integer spacings will be preserved if SUSY holds.

This has an important consequence. In Section \ref{sec-2pt}, we introduced
critical exponents $\eta$ and $\bar{\eta}$ for the connected and disconnected
2pt functions. In the SUSY theory, these 2pt functions become $\langle \varphi
\omega \rangle$ (see Exercise \ref{phiomega}) and $\langle \varphi \varphi
\rangle$. Since the scaling dimensions of $\varphi$ and $\omega$ are related
by {\eqref{SUSYrels}}, we conclude that SUSY requires $\eta = \bar{\eta}$.
Numerical simulations (Section \ref{sec-numsim}) are consistent with this
equality in $d = 5$ but not in $d = 3, 4$.

\subsection{Dimensional reduction}\label{dimred}

The most dramatic consequence of the Parisi-Sourlas SUSY is that it implies the
dimensional reduction property, which we now define. Consider any $n$-point
correlation function
\begin{equation}
  \langle \Phi (y_1) \ldots \Phi (y_n) \rangle, \label{Phin}
\end{equation}
where $y_i$ are points of the superspace. Let us split
\begin{equation}
  y = (x_{\| \nobracket}, x_{\perp}, \theta, \bar{\theta})\,,
\end{equation}
where $x_{\| \nobracket}$ parameterize the first $d - 2$ bosonic directions,
and $x_{\perp} = (x_{d - 1}, x_d)$. Suppose that all $n$ points $y_i$ lie in
the $d - 2$ dimensional bosonic subspace $\mathcal{M}_{d - 2}$ of the
superspace, i.e.~we impose
\begin{equation}
  x_{\perp i} = \theta_i = \bar{\theta}_i = 0 .
\end{equation}
Dimensional reduction says that the correlation function {\eqref{Phin}} is
then exactly equal to the correlation function of another theory which lives
in $d - 2$ dimensions. Namely we have
\begin{equation}
  \langle \Phi (y_1) \ldots \Phi (y_n)\rangle |_{\mathcal{M}_{d - 2}} = \langle
  \hat{\varphi} (x_{\| \nobracket 1}) \ldots \hat{\varphi} (x_{\| \nobracket
  n}) \rangle, \label{DRnpt}
\end{equation}
where $\hat{\varphi}$ is a $d - 2$ dimensional scalar field, whose correlators
are defined with respect to the action
\begin{equation}
  S^{}_{{\rm DR}} = \frac{4 \pi}{R} \int d^{d - 2} x_{\| \nobracket} \left(
  \frac{1}{2} (\partial_{\| \nobracket} \hat{\varphi})^2 + V (\hat{\varphi})
  \right), \label{SDR}
\end{equation}
where $V$ is the same potential which appears in the Parisi-Sourlas action
{\eqref{LPS}}. For the RFIM we are only interested in the quartic potential.
However it's useful to keep things general, since this theory also applies to
the branched polymers described by the cubic potential, as we will discuss in
Section \ref{sec-BP}. The prefactor $\frac{4 \pi}{R}$ in {\eqref{SDR}} could
have been absorbed by rescaling $\hat{\varphi}$ but it does not pay off to do
this.

There are many proofs of dimensional reduction from supersymmetry. The first
proof {\cite{Parisi:1979ka}} was perturbative. The main idea is easy to
understand. In a SUSY theory we can do perturbation theory in terms of
superpropagators. Working in position space, Feynman diagrams have to be
integrated over internal vertices. One shows that if the external vertices
belong to $\mathcal{M}_{d - 2}$, in all internal vertices the integral over
$\theta, \bar{\theta}$ exactly compensates the integral over 2 out of $d$
bosonic directions. The mechanism of the cancellation was explained in 
{\cite{Parisi:1979ka}}. For a more detailed proof along these lines see e.g.~{\cite{paper1}}, App.A.

Here we would like to explain a non-perturbative argument for dimensional
reduction due to Cardy {\cite{CARDY1983470}}. Let us write the SUSY action as,
see {\eqref{LPS}},
\begin{gather}
  S_{\text{PS}} = \int d y (\mathcal{L}_{\| \nobracket} +\mathcal{L}_{\perp}),\\
  \mathcal{L}_{\| \nobracket} = - \frac{1}{2} \Phi \partial_{\| \nobracket}^2
  \Phi + V (\Phi), \qquad \mathcal{L}_{\perp} = - \frac{1}{2} \Phi
  (\partial_{\perp}^2 + R \partial_{\bar{\theta}} \partial_{\theta}) \Phi .
\end{gather}
We then introduce an interpolating action:
\begin{equation}
  S_{\lambda, J} = \int d y \left[(\lambda + C (1 - \lambda) \delta^{(2)}
  (x_{\perp}) \delta (\theta) \delta (\bar{\theta})) \mathcal{L}_{\|
  \nobracket} +\mathcal{L}_{\perp} + J (x_{\perp}) \Phi (y) \delta^{(2)}
  (x_{\perp}) \delta (\theta) \delta (\bar{\theta})\right],
\end{equation}
which interpolates between the $d$-dimensional SUSY theory for $\lambda = 1$
and the dimensionally reduced theory for $\lambda = 0$.\footnote{The term
$\int d y\mathcal{L}_{\perp}$ produces, at $\lambda = 0$ and upon integrating
out $\omega$, the action $\int (\partial_{\perp}^2 \varphi)^2$ which is
completely decoupled from the dimensionally reduced theory and contributes an
overall constant.} The constant $C$ will be fixed so that the interpolation
preserves correlation functions. We will see that this requires $C = 4 \pi /
R$.

We consider the generating function:
\begin{equation}
  Z_{\lambda, J} = \int \mathcal{D} \Phi e^{- S_{\lambda, J}} .
\end{equation}
The idea is to show that it does not depend on $\lambda$. We have:
\begin{equation}
  \partial_{\lambda} Z_{\lambda, J} = - \int d y\, \partial_{\lambda}
  S_{\lambda, J} e^{- S_{\lambda, J}} \equiv - Z_{\lambda, J} \int d y \langle
  \partial_{\lambda} S_{\lambda, J} \rangle . \label{car1}
\end{equation}
Furthermore,
\begin{equation}
  \langle \partial_{\lambda} S_{\lambda, J} \rangle = \bigl(1 - C \delta^{(2)}
  (x_{\perp}) \delta (\theta) \delta (\bar{\theta})\bigr) \langle \mathcal{L}_{\|
  \nobracket} (x_{\| \nobracket}, x_{\perp}, \theta, \bar{\theta}_{})
  \rangle_{\lambda, J} . \label{car2}
\end{equation}
The crucial point is that
\begin{equation}
  \langle \mathcal{L}_{\| \nobracket} (x_{\| \nobracket}, x_{\perp}, \theta,
  \bar{\theta}_{}) \rangle_{\lambda, J} = F \left( x_{\| \nobracket},
  x^2_{\perp} - \frac{4}{R} \theta \bar{\theta} \right), \label{cp}
\end{equation}
i.e.~can depend on $x_{\perp}, \theta, \bar{\theta}$ only through the shown
symmetric combination - the superspace metric in the directions orthogonal to
$\mathcal{M}_{d - 2}$. This follows since all terms in the action $S_{\lambda,
J}$ are superrotation-invariant in these orthogonal directions. We also have
(omitting the $x_{\| \nobracket}$ dependence)
\begin{equation}
  \int d^2 x_{\perp} d \bar{\theta} d \theta F \left( x^2_{\perp} -
  \frac{4}{R} \theta \bar{\theta} \right) = - \frac{4}{R} \int d^2 x_{\perp}
  F' (x^2_{\perp}) = \frac{4 \pi}{R} [F (0) - F (\infty)],
\end{equation}
and $F (\infty)$ is zero since all sources are localized at $x_{\perp} = 0$.
Thus when {\eqref{cp}} is used in {\eqref{car2}} and in turn in
{\eqref{car1}}, we find that indeed $\partial_{\lambda} Z_{\lambda, J} = 0$
provided that $C = 4 \pi / R$, proving dimensional reduction.

Yet another proof of dimensional reduction was given by Zaboronsky
{\cite{Zaboronsky:1996qn}}. His proof follows the idea of supersymmetric
localization. In supersymmetric localization (for a review see e.g.~{\cite{Cremonesi:2013twh}}, Sec.~3), one chooses a supersymmetry generator $Q$
which squares to zero: $Q^2 = 0$. Correlation functions of $Q$-invariant
operators\footnote{(Local) operator $\mathcal{O} (x)$ in a Lagrangian field
theory is simply any interaction term one can write out of fields at a point $x$
and of their derivatives.} can then be computed by a path integral restricted to
$Q$-invariant field configurations ($\equiv$ localize). In the problem at
hand, one chooses $Q = M_{d - 1, \theta} + M_{d, \bar{\theta}}$ where $M_{a
b}$ are the superrotation generators. The $Q$-invariant fields are fields
$\Phi_{\text{rot}}$ invariant with respect to all superrotations around
$\mathcal{M}_{d - 2}$. The path integral over all fields $\int \mathcal{D}
\Phi$ computing the correlator of fields inserted at $\mathcal{M}_{d - 2}$
then localizes to the path integral over rotationally invariant fields $\int
\mathcal{D} \Phi_{\text{rot}}$. An extra step is then required to show that
the latter path integral equals the path integral of the dimensionally reduced
theory.

A complementary viewpoint on dimensional reduction was given by Kaviraj,
Trevisani and myself {\cite{paper1}}. We considered the $d$-dimensional SUSY
theory as a (super) conformal field theory (CFT), with local operators
classified as primaries and (super)descendants. We then exhibited a map of
operators and correlation functions from a Parisi-Sourlas supersymmetric CFT
in~$d$~dimensions to a~$(d - 2)$-dimensional ordinary CFT. One rule that CFTs
are supposed to obey is the operator product expansion (OPE), and we showed that if
the mother theory obeys it then the reduced theory also does, modulo some
operators which decouple. This construction may be called ``cohomological OPE
reduction''. Moreover, we showed that the reduced theory is local, i.e.~it has
a local conserved stress tensor operator. This $(d - 2)$-dimensional stress
tensor arises naturally as a member of the supersymmetric multiplet to which
the $d$-dimensional stress tensor belongs. Furthermore, as required by
reduction, we showed a perfect match between superconformal blocks and the
usual conformal blocks in two dimensions lower.

\subsection{Caveats}\label{caveats}

As discussed in Section \ref{L1}, dimensional reduction definitely does not
hold for the RFIM in $d = 3$ and $d = 4$, so something must yield. With all
the proofs and tests described in Section \ref{dimred}, the implication
``Parisi-Sourlas SUSY $\Rightarrow$ dimensional reduction'' appears on solid
ground.

On the other hand, the argument for the Parisi-Sourlas SUSY itself had one
key assumption, which may be questioned: dropping all diagrams but those with
the maximal number of crosses. Quoting Parisi {\cite{ParisiLH}}, Sec.3: ``The
argument is good near 6 dimensions. Decreasing the dimensions, in particular
near 4 dimensions, the other diagrams start to be infrared divergent. A
careful analysis of the anomalous dimensions of $\varphi^4$-like operators is
needed to decide if these extra diagrams have the effect of changing the
critical exponents and destroying dimensional reduction. This problem maybe
solved in the $\varepsilon$-expansion (for $\varepsilon \sim 2$!) or in the
loop expansion using the standard techniques for computing anomalous
dimensions of composite operators.''

Previous attempts to carry out this program {\cite{Brezin-1998,Feldman}} were
not fully satisfactory. In particular they concluded, for various reasons (different for {\cite{Brezin-1998} and {\cite{Feldman}),
that SUSY should be lost arbitrarily close to 6 dimensions. In Section
\ref{Sec3} we will describe a new and more systematic approach. It leads to a
result in agreement with the numerical simulations, which suggest that SUSY is
lost between $d = 4$ and 5.

\subsection{Literature and further comments}

See Secs. 1-4 of Parisi's Les Houches lectures {\cite{ParisiLH}} for an
insightful introduction to the RFIM.

How did Parisi and Sourlas manage to follow so many steps from the stochastic
PDE {\eqref{stoPDE}} to the action {\eqref{SPS}}, to noticing that this action
is supersymmetric, to realizing that supersymmetry implies dimensional
reduction? In fact, they worked backwards! They started with the action
{\eqref{LPS}} so that dimensional reduction holds by construction, and only
then, having expanded the action in fields, discovered the connection to RFIM
(see Parisi's remarks in {\cite{ParisiRome}}, 1:07:00).

\section{Replicas, Cardy transform, leaders, loss of
SUSY}\label{Sec3}

\subsection{Replicas}

To investigate systematically the hints about what may go wrong with the
derivation of Parisi-Sourlas SUSY (Section \ref{caveats}), it will be
convenient to change the formalism. In this lecture we will use the method of
replicas, a time-honored approach to the physics of disordered systems.
Instead of talking about important and unimportant diagrams, we will be able
to use the renormalization group intuition and talk about relevant and irrelevant
interactions.

We will need a variant of the method of replicas adapted to the study of
correlation functions. Suppose we want to compute a correlation function
{\eqref{Aphi}}. The idea is to insert $1 = Z [h]_{}^{n - 1} / Z [h]_{}^{n -
1}$ under the $h$ integral. We call $\phi = \phi_1$ and introduce $n - 1$
independent fields $\phi_2, \ldots, \phi_n$ to represent $Z [h]^{n - 1}$ in
the numerator
\begin{equation}
  Z [h]^{n - 1} = \prod_{i = 2}^n \int \mathcal{D} \phi_i e^{- S [\phi_i, h]}
  .
\end{equation}
So we have $n$ independent fields, called replicas, all coupled to the same
random magnetic field $h$. In the denominator we now have $Z [h]^n$. We
imagine analytically continuing the expression to complex $n$ and taking the
$n \rightarrow 0$ limit. In this limit the denominator $Z [h]^n \rightarrow 1$
and we obtain:
\begin{equation}
  \overline{\langle A (\phi) \rangle_h} = \lim_{n \rightarrow 0}  \int
  \mathcal{D}h \hspace{0.17em} \mathcal{P} (h)  \int \mathcal{D} \vec{\phi} A
  (\phi_1) e^{- \sum_{i = 1}^n S [\phi_i, h]} .
\end{equation}
Let us now perform the $h$ integral. As usual, assume that the measure is
Gaussian, $\mathcal{P} (h) \propto e^{- \int h^2 / (2 R)}$. The important term
in the action is $h \left( \sum \phi_i \right)$. Performing the Gaussian
integral
\begin{equation}
  \int \mathcal{D}h e^{- \int d^d x h^2 / (2 R) + h \left( \sum \phi_i
  \right)} = {\rm const} . \times e^{- \int d^d x \frac{R}{2} \left( \sum
  \phi_i \right)^2},
\end{equation}
we thus eliminate $h$ and we are left with the effective action for $n$
replicas including a quadratic term coupling them:
\begin{equation}
  \mathcal{S}_n = \int d^d x \left[ \frac{1}{2}  \sum_{i = 1}^n (\partial
  \phi_i)^2 + \sum_{i = 1}^n V (\phi_i) - \text{{\small $\frac{1}{2}$}} R
  \left( \sum_{i = 1}^n \phi_i \right)^2 \right] . \label{Sn}
\end{equation}
The action has $\mathrm{} S_n \times \mathbb{Z}_2$ global symmetry, where
$S_n$ is the permutation group.

In terms of this action, our original correlator is computed simply as
\begin{equation}
  \overline{\langle A (\phi) \rangle_h} = \lim_{n \rightarrow 0} \int
  \mathcal{D} \vec{\phi} A (\phi_1) e^{-\mathcal{S}_n [\vec{\phi}]} .
\end{equation}
That's the main equation of the replica method.

The method also works for more complicated correlators such as
$\overline{\langle A (\phi) \rangle_h \langle B (\phi) \rangle_h}$. When we
write this similarly to {\eqref{Aphi}} we will have the product of two $\int
\mathcal{D} \phi$ integrals in the numerator and $Z [h]^2$ in the denominator.
We introduce fields $\phi_1, \phi_2$ for the numerator, and additional fields
$\phi_3, \ldots, \phi_n$. In the $n \rightarrow 0$ limit we have $- 2$
additional fields, which reproduce the denominator. So we get
\begin{equation}
  \overline{\langle A (\phi) \rangle_h \langle B (\phi) \rangle_h} = \lim_{n
  \rightarrow 0} \int \mathcal{D} \vec{\phi} A (\phi_1) B (\phi_2)
  e^{-\mathcal{S}_n [\vec{\phi}]} .
\end{equation}

Thus, our main task becomes to analyze theory {\eqref{Sn}} in the $n
\rightarrow 0$ limit. Gathering all quadratic terms (kinetic terms, the mass
terms from $V (\phi_i)$, and the term coupling the replicas), we can compute
the propagator.

\begin{exercise}
  Show that the momentum-space propagator is given by
  \begin{equation}
    G_{i j} (p) = \frac{\delta_{i j}}{p^2 + m^2} + \frac{R M_{i j}}{(p^2 +
    m^2) (p^2 + m^2 - n R)}, \label{prop}
  \end{equation}
  where $M$ is the matrix with $M_{i j} = 1$ for all $i, j$.
\end{exercise}

Using this result, in the free case ($\lambda = 0$) we obtain at criticality
($m^2 = 0$) and in the $n \rightarrow 0$ limit:
\begin{equation}
  \overline{\langle \phi \phi \rangle_h} = \langle \phi_1 \phi_1 \rangle =
  \frac{1}{p^2} + \frac{R}{p^4},
\end{equation}
\begin{equation}
  \overline{\langle \phi \rangle_h \langle \phi \rangle_h} = \langle \phi_1
  \phi_2 \rangle = \frac{R}{p^4} .
\end{equation}
We have thus reproduced the results from the previous lecture (Section
\ref{sec-2pt}).

We can now introduce the quartic interaction $\lambda \sum \phi_i^4$ and build
perturbation theory. If we keep the most IR-singular diagrams, we will again
reproduce the conclusions of the previous lecture, but it will be hard to see
what is wrong. But now we have a more attractive strategy. Instead of
selecting diagrams, we can try to understand the scaling dimension of
interaction terms, and study which ones are relevant and which are irrelevant.

\begin{remark}
  Note that because of the formal $n \rightarrow 0$ limit, our approach still
  remains perturbative in nature. There is another famous model with $S_n$
  symmetry, the Potts model, for which the $n \rightarrow 0$ limit can be
  defined non-perturbatively and it described percolation. Is there an
  analogue of the percolation picture for RFIM? It's an interesting open
  question.
\end{remark}

\subsection{Upper critical dimension}\label{UCD}

Here we will present a ``quick and dirty'' approach to scaling dimensions. It
will be improved in the next section. Suppose we want to compute the scaling
dimension of the quartic interaction term
\begin{equation}
  \mathcal{O}= \frac{1}{4!} \sum_{i = 1}^n \phi_i^4 .
\end{equation}
We are interested in this scaling dimension in the Gaussian theory, i.e.~with
$V$ set to zero in {\eqref{Sn}}.

Usually, the scaling dimension $\Delta$ of an operator can be extracted from its
2pt function
\begin{equation}
  \langle \mathcal{O} (0) \mathcal{O} (x) \rangle \sim \frac{1}{x^{2 \Delta}}
  .
\end{equation}
But this approach will not work for our operator $\mathcal{O}$. Indeed, its
2pt function vanishes in the $n \rightarrow 0$ limit. Explicitly we have,
applying Wick's theorem:
\begin{equation}
  \langle \mathcal{O} (0) \mathcal{O} (x) \rangle = \frac{1}{4!} \sum_{i =
  1}^n \sum_{i = 1}^n [G_{i j} (x)]^4 .
\end{equation}
Plugging in {\eqref{prop}} we find that this is $O (n)$ because of $\sum_i
\sum_j \delta_{i j} = \sum_i 1 = n$.

\begin{remark}
  More generally, it is true that any correlation function of an arbitrary
  number of $S_n$ invariant operators vanishes in the $n \rightarrow 0$ limit.
  This property is closely related to the fact that the partition function of
  the replicated theory is exactly 1 in the $n \rightarrow 0$ limit.
\end{remark}

In the absence of the 2pt function to look at, we can extract the scaling
dimension of an operator from the operator product expansion (OPE). Here we
will only need rudimentary understanding of the OPE in the free theory, see e.g.~Cardy
{\cite{cardy_1996}}, Sec.5.1. In general an operator $\mathcal{O}_k (0)$
appears in the OPE of two operators $\mathcal{O}_i (x) \mathcal{O}_j (0)$, $x
\rightarrow 0$, with a coefficient ({\cite{cardy_1996}}, Eq.(5.6))
\begin{equation}
  C_{i j k} (x) = \frac{f_{i j k}}{| x |^{\Delta_i + \Delta_j - \Delta_k}},
  \label{Cijk}
\end{equation}
where $f_{i j k}$ are pure numbers and $\Delta_i, \Delta_j, \Delta_k$ are the
scaling dimension of the three operators. We will apply this equation for
$\mathcal{O}_i =\mathcal{O}_j =\mathcal{O}_k =\mathcal{O}$, in which case we
should have $C(x) \propto 1/|x|^{\Delta_\mathcal{O}}$.

We will need the propagator {\eqref{prop}} in position space, in the $n
\rightarrow 0$ limit and at criticality $m^2 = 0$:
\begin{equation}
  G_{i j} (x) = \langle \phi_i (0) \phi_j (x) \rangle =\# \frac{\delta_{i
  j}}{x^{d - 2}} +\# \frac{R}{x^{d - 4}} ; \label{prop-pos}
\end{equation}
with some nonzero coefficients $\#$ which will not be important (all numerical proportionality factors will be set to one in the rest of the argument). Consider
first the OPE $\phi_i^4 (x) \phi_j^4 (0)$. Focusing on the quartic operators
in the OPE, the important part of this OPE is:
\begin{equation}
  \phi_i^4 (x) \phi_j^4 (0) \supset B \langle \phi_i (x) \phi_j (0) \rangle^2
  \phi^2_i (x) \phi_j^2 (0) = K_{i j} (x) \phi^2_i (0) \phi_j^2 (0) + \ldots\,,
  \label{OPE-rel}
\end{equation}
where $B$ is a combinatorial coefficient, $K_{i j} (x) = B \langle \phi_i (x)
\phi_j (0) \rangle^2$ and $\ldots$ stands for the less singular terms with
derivatives of $\phi$, appearing when $\phi^2_i (x)$ is Taylor-expanded around
$x = 0$. Up to constants, $K_{i j}$ is a sum of three terms (see
{\eqref{prop-pos}}):
\begin{equation}
  K_{i j} (x) = \frac{\delta_{i j}}{x^{2 d - 4}} + \frac{\delta_{i j} R}{x^{2
  d - 6}} + \frac{R^2}{x^{2 d - 8}} .
\end{equation}
Plugging this into {\eqref{OPE-rel}} and summing over $i, j$ we have:
\begin{equation}
  \mathcal{O} (x) \mathcal{O} (0) \supset \left( \frac{1}{x^{2 d - 4}} +
  \frac{R}{x^{2 d - 6}} \right) \mathcal{O} (0) + \frac{R^2}{x^{2 d - 8}}
  \tilde{\mathcal{O}} (0), \label{OxO}
\end{equation}
where $\tilde{\mathcal{O}} = \left( \sum_{i = 1}^n \phi_i^2 \right)^2$.

Let us compare this with {\eqref{Cijk}}. The second term involving
$\tilde{\mathcal{O}} (0)$ does not concern us here; we focus on the first term
which involves $\mathcal{O} (0)$. Its coefficient is not a pure power but a
sum of two powers. This means that $\mathcal{O}$ does not have a single
scaling dimension but is a sum of several operators with different scaling
dimensions. The smallest of these is $\Delta = 2 d - 6$, determined by the
least singular term which is $R / x^{2 d - 6}$. In other words we obtain
\begin{equation}
  \mathcal{O}=\mathcal{O}_{2 d - 6} + \text{higher dimension operators} .
  \label{Odec}
\end{equation}
The leading operator, $\mathcal{O}_{2 d - 6}$, will become relevant for $d <
6$. Thus we reproduce in this language the result that the upper critical
dimension of the RFIM model (quartic interaction) equals 6.

Note that already the propagator {\eqref{prop-pos}} has two different powers,
suggesting that the multiplet $\phi_i$ hides inside itself fields of unequal
scaling dimensions. When we construct composite operators, we should not be
surprised that those are also in general sums of operators of different
scaling dimensions, as we found for $\mathcal{O}$. However, computing scaling
dimensions using the OPE is bound to become tedious when many operators need
to be considered. In the next section we will present a much more efficient
approach to the scaling dimensions in the free theory, and to the anomalous dimensions in
the interacting theory.

\subsection{Cardy transform: second argument for the PS SUSY}\label{sec-Cardy}

We will now describe a field transform due to Cardy {\cite{CARDY1985123}}. I
am not sure how Cardy originally arrived at his transform, but here's how one
can motivate it and look for it systematically. As mentioned, the form of the
propagator {\eqref{prop}}, {\eqref{prop-pos}} hints that there must be fields
of unequal scaling dimensions inside $\phi_i$. This suggests an analogy with
the Parisi-Sourlas Lagrangian {\eqref{SPS}}, where $\omega, \varphi, \psi,
\bar{\psi}$ had different dimensions {\eqref{freedims}}. The quadratic part of
the PS Lagrangian has the form

\begin{equation}
  \partial \varphi \partial \omega - \frac{R}{2} \omega^2 + \partial
  \bar{\psi} \partial \psi \label{PSquad},
\end{equation}
while the quadratic part of the replicated Lagrangian {\eqref{Sn}}, setting
$m^2 = 0$, is
\begin{equation}
  \sum_{i = 1}^n \frac{1}{2} (\partial \phi_i)^2 - \frac{R}{2} \left( \sum_{i
  = 1}^n \phi_i \right)^2 . \label{Repquad}
\end{equation}
It would be great if we could map {\eqref{Repquad}} to {\eqref{PSquad}} via a
field transformation of $\phi_i$. This is of course impossible verbatim, since
$\psi, \bar{\psi}$ are fermionic, and all fields in {\eqref{Repquad}} are
bosonic. But note that, at the quadratic level, 2 fermions are equivalent to
$- 2$ bosons, as integrating these fields out gives the same functional
determinant raised to the same power. So perhaps we may find a transform
$(\phi_i)_{i = 1}^n \rightarrow \varphi, \omega, \chi_i$ which maps
{\eqref{Repquad}} to
\begin{equation}
  \partial \varphi \partial \omega - \frac{R}{2} \omega^2 + \frac{1}{2}
  \sum_{}^{} (\partial \chi_i)^2, \label{aim}
\end{equation}
up to some terms vanishing as $n \rightarrow 0$. In the limit $n \rightarrow
0$ we will have $- 2$ fields $\chi_i$. If we replace them by two fermions
$\bar{\psi}, \psi$, we will land on the PS Lagrangian. It turns out that the
following transform does the job (Cardy {\cite{CARDY1985123}}):
\begin{eqnarray}
  \phi_1 & = & \varphi + \omega / 2 \nonumber\\
  \phi_i & = & \varphi - \omega / 2 + \chi_i \quad (i = 2, \ldots, n), \quad
  \sum_{i = 2}^{n^{}} \chi_i = 0 .  \label{eq-Cardy}
\end{eqnarray}
Following Cardy, we introduced $n - 1$ fields $\chi_i$ satisfying the
constraint $\sum_{i = 2}^{n^{}} \chi_i = 0$. So effectively we have $n - 2$
fields $\chi$, which becomes $- 2$ as $n \rightarrow 0$, just as we need.

The inverse transformations are $\varphi = \frac{1}{2} (\phi_1 + \rho)$,
$\omega = \phi_1 - \rho$, $\chi_i = \phi_i - \rho$, where $\rho= \frac{1}{n -
1} (\phi_2 + \cdots + \phi_n)$.

\begin{exercise}
  Plug {\eqref{eq-Cardy}} into {\eqref{Repquad}} and see that this indeed
  reproduces {\eqref{aim}}, up to terms which are $O (n)$.
\end{exercise}

From the quadratic Lagrangian {\eqref{aim}}, we can compute the scaling dimension
of the fields:
\begin{equation}
  [\varphi] = d / 2 - 2, \quad [\chi_i] = d / 2 - 1, \quad  [\omega] = d / 2.
  \label{dims}
\end{equation}
This is also consistent with the momentum dependence of the propagators
computed from {\eqref{aim}}:
\begin{equation}
  \langle \varphi \omega \rangle = \frac{1}{p^2}, \quad \langle \varphi
  \varphi \rangle = \frac{R}{p^4}, \quad \langle \chi_i \chi_j \rangle =
  \frac{\delta_{i j} - \frac{1}{n - 1} M_{i j}}{p^2}\,. \label{Cardyprop}
\end{equation}
Thus, the Cardy fields $\varphi, \chi_i, \omega$ have, unlike $\phi_i$,
well-defined scaling dimensions. This will be a crucial simplification in what
follows.

\begin{remark}
  This simplification was not used much before our work {\cite{paper2}}, and
  one may ask why. One reason might be that after the Cardy transform the full $S_n$
  invariance of the theory is not manifest, only $S_{n - 1}$ is. The Cardy fields
  are nice from the renormalization group perspective, but they somewhat
  confound the symmetry structure. It appears to be a feature of our problem
  that one cannot have both the manifest $S_n$ invariance, and good scaling. In
  our work we found that good scaling is crucial, so we will use the Cardy fields.
  $S_n$ invariance is of course also very important, and it will play a role.
  As we will see, there is a way around the fact that it is not realized
  manifestly.
  
  We stress that, although $S_n$ invariance is not manifest, it is still
  present after the transformation to the Cardy fields. In
  particular it's not spontaneously broken.
\end{remark}

\begin{exercise}
  Spontaneous breaking of $S_n$ invariance could be detected in the lack of
  $S_n$ invariance of the 2-point function $\langle \phi_i \phi_j \rangle$,
  $i, j = 1 \ldots n$. Use {\eqref{eq-Cardy}} and the propagators
  {\eqref{Cardyprop}} to compute the propagators $\langle \phi_1 \phi_1
  \rangle$, $\langle \phi_1 \phi_i \rangle$ and $\langle \phi_i \phi_j
  \rangle$, $i, j = 2 \ldots n$. Show that these are consistent with the $S_n$
  invariant expression {\eqref{prop}} (for $m^2 = 0$).
\end{exercise}

With the Cardy fields at our disposal, we can now transform any composite operator
to this field basis in order to reveal it scaling dimension content. Let us
start with the interaction term $\sum_i V (\phi_i)$ in {\eqref{Sn}}. We have
\begin{equation}
  \sum_{i = 1}^n V (\phi_i) = V \left( \varphi + \frac{\omega}{2} \right) +
  \sum\nolimits' V \left( \varphi - \frac{\omega}{2} + \chi_i \right), \label{masterV}
\end{equation}
where we introduced the notation $\sum\nolimits' \equiv \sum_{i = 2}^n$.

We are interested in polynomial interactions $V (\phi) = \phi^z$ where $z = 2$
(mass term), $z = 4$ (quartic interaction, RFIM), or $z = 3$ (cubic
interaction, branched polymers, Section \ref{sec-BP} below). Since the scaling
dimension of $\varphi$ is lower than of $\chi_i$ and $\omega$, we are
supposed to expand {\eqref{masterV}} in $\chi_i$ and $\omega$, and higher
powers of these fields will give us fields of higher and higher dimension. The
zeroth order term
\begin{equation}
  V (\varphi) + \sum\nolimits' V (\varphi) = n V (\varphi)
\end{equation}
vanishes for $n \rightarrow 0$. At the first order, we find terms
\begin{equation}
  V' (\varphi) \frac{\omega}{2} + V' (\varphi) \sum\nolimits' \left( - \frac{\omega}{2}
  + \chi_i \right) = V' (\varphi) \omega \left( \frac{1}{2} - \frac{n - 1}{2}
  \right) = \omega V' (\varphi) + O (n),
\end{equation}
At the second order, we find, using $\sum\nolimits' \chi_i = 0$,
\begin{equation}
  V'' (\varphi) \frac{1}{2} \left[ \left( \frac{\omega}{2} \right)^2 + \sum\nolimits'
  \left( - \frac{\omega}{2} + \chi_i \right)^2 \right] = \frac{1}{2} \sum\nolimits'
  \chi^2_i V'' (\varphi) + O (n) .
\end{equation}
Dropping $O (n)$ terms, we obtain
\begin{equation}
  \sum_i V (\phi_i) = \omega V' (\varphi) + \frac{1}{2} \sum\nolimits' \chi^2_i V''
  (\varphi) \label{quarticL} + \ldots,
\end{equation}
The first two terms here have the same scaling dimension (recall that $V =
\phi^z$):
\begin{equation}
  (z - 1) [\varphi] + [\omega] = (z - 2) [\varphi] + 2 [\chi_i] = z (d / 2 -
  2) + 2.
\end{equation}
For $z = 2$ this becomes $d - 2$, i.e.~the mass term is always relevant. For
$z = 4$ we get $2 d - 6$, which is the same answer as using the OPE method in
Section \ref{UCD}. We thus reproduce yet again the upper critical dimension 6
of the RFIM.

Terms $\ldots$ in {\eqref{quarticL}}, arising from the higher derivatives of
$V$, contain higher powers of $\chi_i$ and/or $\omega$ than for the shown
terms, and have a higher scaling dimension. For example for the third
derivative we will get terms like
\begin{equation}
  V''' (\varphi) \times \left\{ \sum\nolimits' \chi_i^3, \sum\nolimits' \chi_i^2 \omega,
  \omega^3 \right\} . \label{quartic-foll}
\end{equation}
Their scaling dimensions are at least 1 unit higher than in {\eqref{quarticL}}. The terms
from the fourth derivative are at least 1 more unit higher.

Suppose we stay close to the upper critical dimension, so that terms
{\eqref{quarticL}} are weakly relevant. Then terms from higher derivatives are
irrelevant and can be dropped.\footnote{This observation was first made in
{\cite{paper1}}, while Cardy {\cite{CARDY1985123}} used a different argument
for dropping these terms.} We are then left with the action:
\begin{equation}
  S_{\text{pre-SUSY}} = \int d^d x \left\{ \partial \varphi \partial \omega -
  \frac{R}{2} \omega^2 + \frac{1}{2} \sum\nolimits' (\partial \chi_i)^2 + \omega
  V' (\varphi) + \frac{1}{2} \sum\nolimits' \chi^2_i V'' (\varphi) \right\} .
  \label{cardy-action}
\end{equation}
Note that the fields $\chi_i$, $- 2$ in number, enter quadratically. Thus we
can still replace them by two fermions. It is convenient to choose
normalization so that $\sum \chi_i^2 \rightarrow 2 \bar{\psi} \psi$. After
this replacement, {\eqref{cardy-action}} maps precisely onto the full PS SUSY
action {\eqref{SPS}}, including the interaction terms. We will therefore call
{\eqref{cardy-action}} the pre-SUSY action.

\subsection{Testing the second argument}\label{testing}

In the first argument for the emergence of PS SUSY (Section \ref{PS-theory})
we dropped a class of diagrams, while in the second argument we had to drop
some interaction terms. The advantage of the new argument is that it's easier
to test for consistency. Indeed, there are no general rules for dropping
diagrams, but there is one for interaction terms: the dropped terms must be
irrelevant in the renormalization group sense. As mentioned in Section
\ref{sec-Cardy}, the dropped terms are indeed irrelevant close to the upper
critical dimension. We now need to see if all dropped terms stay irrelevant in
lower $d$. If any term becomes relevant in lower $d$, this opens the door to
the loss of SUSY. This test was carried out in {\cite{paper2,paper-summary}},
and we will explain the key points and results in subsequent sections.

First of all, it will turn out that terms {\eqref{quartic-foll}} that we
dropped do stay irrelevant even in lower $d$. In Section \ref{sec-leaders}
these interactions will be classified as ``followers'' of the ``leader''
interaction described by {\eqref{quarticL}}, and we will argue that the
followers can always be dropped.

However, we should extend the test to more interaction terms, which we forgot
to write so far. It's now time to bring them up. The point is that since
Lagrangian {\eqref{Sn}} has $S_n \times \mathbb{Z}_2$ symmetry, we are
supposed to consider all possible $S_n \times \mathbb{Z}_2$ invariant
interactions (which will be called {\bf singlets}). Indeed, in a theory
with a short-distance cutoff, all interactions allowed by symmetry will be
generated by the renormalization group flow. And so far we only considered a
small subclass of $S_n \times \mathbb{Z}_2$ interactions of the form $\sum_i V
(\phi_i)$.

The need to consider more general interactions was pointed out by Br{\'e}zin
and De Dominicis {\cite{Brezin-1998}}. In addition to the above symmetry
considerations, they showed their appearance in a microscopic model. They
considered the RFIM spin model on the lattice, using the replica method.
Mapping this model to a continuous field theory via the Hubbard-Stratonovich
transformation, they found a host of additional $S_n \times \mathbb{Z}_2$
invariant interactions. Denoting
\begin{equation}
  \sigma_k = \sum_{i = 1}^n \phi_i^k,
\end{equation}
they observed interactions such as $\sigma_2, \sigma_4, \sigma_1^2$ (these are
present in {\eqref{Sn}}), but in addition terms like $\sigma_6,
\sigma_{2^{}}^2, \sigma_2 \sigma_4$ etc.

\begin{remark}
  \label{Brezin}While we agree with Ref.~{\cite{Brezin-1998}} about the need
  to deal with these extra singlet interactions, we disagree in {\it how}
  one should deal with them. Ref.~{\cite{Brezin-1998}} considered a joint
  beta-function involving the usual quartic interaction $\sigma_4$ and all the
  other quartic interactions $\sigma_1 \sigma_3$, $\sigma^2_2$, $\sigma_1^2
  \sigma_2$, $\sigma^4_{1^{}}$, as if all these operators were marginally
  relevant near $6$ dimensions. Based on such considerations they concluded
  that the SUSY fixed point is unstable with respect to adding these extra
  couplings, arbitrarily close to 6d (see also the book
  {\cite{de_dominicis_giardina_2006}}, Sec.~2.7).
  
  However, the starting point of their calculation appears to be incorrect:
  the extra interactions are {\bf not} close to marginality near 6d. This is easy to
  check, by transforming to the Cardy field basis or by the OPE method: the
  interactions $\sigma_1 \sigma_3$, $\sigma^2_2$, $\sigma_1^2 \sigma_2$,
  $\sigma^4_{1^{}}$ are all {\bf strongly irrelevant} near $d = 6$. We
  have (see {\cite{paper2}}, Table 1, Sec.~8.1)
  \begin{eqnarray}
    \sigma_1 \sigma_3, \sigma_2^2 & = & \text{dim. $8 - 2 \varepsilon$ +
    higher}, \nonumber\\
    \sigma_1^2 \sigma_2 & = & \text{dim. $10 - 2 \varepsilon$ + higher}, 
    \label{dims-quartics}\\
    \sigma_1^4 & = & \text{dim. $12 - 2 \varepsilon$ + higher} . \nonumber
  \end{eqnarray}
  Thus close to 6d there cannot be perturbative mixing between $\sigma_4$ and
  these operators, or in fact any other singlets: perturbative instability
  reported in {\cite{Brezin-1998}} is not an option.
  
  Instead, what we believe may well happen is that some singlet interaction,
  while strongly irrelevant in $d = 6 - \varepsilon$ dimensions, gets a
  negative anomalous dimension and becomes relevant at some $d_c < 6$. SUSY
  will then be lost for $d < d_c$, and not arbitrarily close to 6d. We will
  see below that this indeed does seem to happen for some specific
  interactions, with $d_c$ between 4 and 5.
\end{remark}

\begin{exercise}
  Reproduce the leading scaling dimensions in {\eqref{dims-quartics}}, by
  transforming these interactions into the Cardy field basis.
\end{exercise}

\subsection{Analogies}\label{sec-analogies}

\begin{example}
  \label{ex-pure-Ising}As an example, recall that in the pure Ising model
  context, one studies the field theory with the Lagrangian $(\partial
  \varphi)^2 + \mu \varphi^2 + \lambda \varphi^4$, having $\mathbb{Z}_2$
  invariance. Although the higher even powers of $\varphi$, as well as other
  $\mathbb{Z}_2$ invariant interactions, are generated by the Wilsonian RG
  flow, it turns out consistent not to worry about them in this case, because
  they remain irrelevant in any $d < 4$ (most operators have positive
  anomalous dimensions in $d = 4 - \varepsilon$). But if any of such terms
  became relevant, the Ising fixed point would have been destabilized and the
  perturbative analysis leading to it would be invalid below some $d_c$. This
  does not happen for the pure Ising, but we should check if something like
  that perhaps happens for the RFIM.
\end{example}

\begin{example}
  \label{ex-cubic}As an example of the model where something like this does
  happen, consider the cubic model, which is a theory of 3 scalar fields
  $\varphi_1, \varphi_2, \varphi_3$ with the quartic potential
  \begin{equation}
    u (\varphi_1^2 + \varphi_2^2 + \varphi_3^2)^2 + v (\varphi_1^4 +
    \varphi_2^4 + \varphi_3^4) .
  \end{equation}
  The coupling $u$ preserves the full $O (3)$ invariance, while the coupling
  $v$ preserves only the discrete subgroup $G = S_3 \ltimes (\mathbb{Z}_2)^3$,
  which permutes the fields and flips their signs.\footnote{$G$ is called the
  cubic group because it's the symmetry group of the cube.} It turns out that in
  $d = 4 - \varepsilon$ dimensions the cubic interaction is irrelevant at long
  distances. Thus even if $v$ is nonzero at short distances, it flows to zero
  in the IR and the fixed point will have an emergent $O (3)$ invariance.
  However for $d < d_c$ the cubic interaction becomes relevant, and the model
  flows to another fixed point having only the cubic symmetry. This is
  analogous to how Parisi-Sourlas SUSY could emerge for the RFIM for $d > d_c$
  and yet be broken for $d < d_c$.
  
  For a review of the cubic model, and its generalization to $N$ fields, see
  {\cite{Pelissetto:2000ek}}, Sec.~11.3. For recent proofs of the cubic
  interaction becoming relevant in $d = 3$, see
  {\cite{Chester:2020iyt,Hasenbusch:2022zur}},
\end{example}

There is actually an interesting difference between these two examples. In
Example \ref{ex-pure-Ising}, the interactions which we worried could become
relevant (e.g.~$\varphi^6$) all had the same symmetry as the
interactions already present in the action (e.g.~$\varphi^4$). Interactions
having the same symmetry mix, and their scaling dimensions are not expected to
cross.\footnote{This is analogous to how, in quantum mechanics, energy levels
having the same symmetry do not cross, without finetuning.} In Example
\ref{ex-cubic}, the interactions multiplying couplings $u$ and $v$ have
different symmetry. So there is no mixing between them, and there is no reason
to prevent crossing.

In the RFIM problem, we suspect that some singlet interaction which is strongly
irrelevant near 6d, becomes relevant at $d < d_c$. At the same time there is
at least one singlet interaction which is irrelevant in any $d$ - it's the
quartic singlet $\sigma_4$ which drives the flow from the Gaussian theory in
the UV to the nontrivial fixed point in the IR. For some other interaction to
become relevant, there should be crossing between that interaction and
$\sigma_4$. And crossing requires that this other interaction should have a
different symmetry from $\sigma_4$. Thus we expect that there should be a
finer classification of interactions by symmetry, rather than them being just
singlets of $S_n \times \mathbb{Z}_2$. This finer classification indeed
exists, see Section \ref{sec-class} below.

\subsection{Leaders}\label{sec-leaders}

As explained in Section \ref{testing}, we have to keep an eye not only on the
terms considered in Section \ref{sec-Cardy}, but on all singlets (i.e.~$S_n
\times \mathbb{Z}_2$ invariant interactions). We have to see if any of these
may become relevant as $d$ is lowered. The total number of singlets is
infinite, so the task is potentially arduous.

Let us see what happens to singlets when we apply the Cardy transform. We have the
following master formula, which we already discussed in Section
\ref{sec-Cardy} with $A (\varphi) = V (\varphi)$:\footnote{One can generalize
this formula to the case when $A$ also depends on derivatives of $\varphi$. It
suffices to replaces derivatives of $A$ by functional derivatives.}
\begin{eqnarray}
  \sum_{i = 1}^n A (\phi_i) & = & A \left( \varphi + \frac{\omega}{2} \right)
  + \sum\nolimits' A \left( \varphi - \frac{\omega}{2} + \chi_i \right) \nonumber\\
  & = & A' (\varphi) \omega + \frac{1}{2} A'' (\varphi)  \sum\nolimits' \chi_i^2
  \nonumber\\
  &  & + \sum_{k = 3}^{\infty} \frac{1}{k!} A^{(k)} (\varphi)  \left[ \left(
  \frac{\omega}{2} \right)^k + \sum\nolimits' \left( - \frac{\omega}{2} + \chi_i
  \right)^k \right] .  \label{ACardy}
\end{eqnarray}
Let us write in full the result when we apply this formula to $\sigma_4 = \sum
\phi_i^4$. We get, in the $n \rightarrow 0$ limit (if there is no confusion we
will sometimes drop $\sum\nolimits'$ and write e.g.~$\chi_i^2$ instead of $\sum\nolimits'
\chi_i^2$)
\begin{eqnarray}
  \sigma_4 & = & [4 \omega \varphi^3 + 6 \chi_i^2 \varphi^2]_{\Delta = 6}
  \nonumber\\
  &  & + [4 \varphi \chi_i^3]_{\Delta = 7} + [\chi_i^4 - 6 \varphi \omega
  \chi_i^2]_{\Delta = 8} \nonumber\\
  &  & - [2 \omega \chi_i^3]_{\Delta = 9} + \Bigl[ \frac{3}{2} \omega^2
  \chi_i^2 + \varphi \omega^3 \Bigr]_{\Delta = 10},  \label{s4}
\end{eqnarray} where we grouped terms according to their scaling dimension in
$d = 6$. We will call the lowest dimension part of any singlet interaction its
{\it leader}, and the rest the {\it followers}. E.g.~the leader of
$\sigma_4$ is $4 \omega \varphi^3 + 6 \chi_i^2 \varphi^2$, and the rest of the
terms in {\eqref{s4}} are the followers.

The difference in scaling dimensions between the leader and its followers is
due to the different scaling dimensions of $\varphi, \chi_i, \omega$ in the
SUSY theory. As long as the theory stays SUSY, the relations
\begin{equation}
  [\chi] = [\varphi] + 1, \qquad [\omega] = [\chi] + 1,
\end{equation}
will be preserved, although $[\varphi]$ may get an anomalous dimension, see
Section \ref{sec-Ward}.

Combination of SUSY and $S_n$ invariance implies that scaling dimension
splitting between the leader and followers is preserved in presence of
interactions.\footnote{Recall that a Wilsonian RG step consists of two
substeps - integrating out a momentum shell and rescaling momenta. The integrating-out substep respects $S_n$ and will renormalize the coefficients of the leader
and of the followers in in the same way. In the rescaling substep the fields
are rescaled according to their scaling dimensions. This breaks $S_n$, but in
a controlled way, and creates integer spacings between the scaling dimensions
of the leader and the followers. These spacings are not renormalized - they are
the same as in free theory. See {\cite{paper2}}, Sec.~7.1, for an example.}
This means that, as the RG flow progresses, the importance of followers keeps
decreasing with respect to that of the leader. Hence, we arrive at a very
important conclusion {\cite{paper2}}: {\bf we may set the followers to
zero, and study only the scaling dimension of the leaders.} This is a huge
simplification since the leader is a small part of the interaction.

Let us discuss the dropped terms {\eqref{quartic-foll}} from this perspective.
These terms belong to the follower part of the $\sigma_4$ singlet. The leader
of $\sigma_4$ is relevant in $d < 6$ dimensions in the free theory (i.e.~at
short distances). As usual, at long distances, once the theory flows to an IR
fixed point, the interaction driving the flow (the leader of $\sigma_4$ in our
case) becomes irrelevant. The followers have dimension equal to the leader
dimension plus an integer, both in the UV and IR. Since the leader is
irrelevant in the IR, the follower is even more irrelevant in the IR than the
leader. This shows that it was indeed consistent to drop the terms
{\eqref{quartic-foll}}, in any $d$, and not only in $d = 6 - \varepsilon$.

It is interesting to discuss how the full $S_n$ invariant interaction can be
reconstructed from its leader. Note that leaders are generally not $S_n$
invariant by themselves.\footnote{For a leader to be $S_n$ invariant, all
followers need to vanish. This happens for the interactions at most quadratic or
linear in the fields, and products of such interactions. E.g.~$\sigma_2$ or
$\sigma_1^2$ have no followers.} They are symmetric under the subgroup
$S_{n - 1}$ permuting the $\chi_i$'s. These are the same transformations which
permute $\phi_i$ for $i = 2, \ldots, n$. However, they are not in general
invariant under the transformation $P_k$ which permutes $\phi_1$ and $\phi_k$.

\begin{exercise}
  Show that the transformation $P_k$ acts on the fields $\varphi, \chi,
  \omega$ (in the $n \rightarrow 0$ limit) as {\cite{paper2}}
  \begin{equation}
    \varphi \rightarrow \varphi + (\chi_k - \omega), \quad \omega \rightarrow
    \omega, \quad \chi_k \rightarrow 2 \omega - \chi_k, \quad \chi_i
    \rightarrow \chi_i - \chi_k + \omega\quad (i \neq k) .
  \end{equation}
\end{exercise}

If $L$ is a leader, we can reconstruct the full $S_n$ invariant interaction
$\mathcal{O}$ symmetrizing over all permutations $P_k$ {\cite{paper-summary}}:
\begin{equation}
  \mathcal{O}= L + \sum_{k = 2}^n P_k L . \label{sym}
\end{equation}
\begin{exercise}
  Show that this expression is fully $S_n$ invariant.
\end{exercise}

Note that some interactions are not leaders of any singlet interaction. For
example, $\varphi^2$, $\varphi \omega$, $\omega \sum\nolimits' \chi_i^3$ are not
leaders of any singlet. [On the other hand $\varphi \omega + \frac{1}{2}
\sum\nolimits' \chi_i^2$ is a leader, of $\sigma_2$.]

\begin{exercise}
  If $L$ is not a leader, then formula {\eqref{sym}} will still give a singlet
  interaction $\mathcal{O}$, but its leader will not be equal to $L$. Compute
  what this formula gives for $L = \varphi^2$, $\varphi \omega$, $\omega \sum\nolimits'
  \chi_i^3$.
\end{exercise}

\subsection{Classification of leaders}\label{sec-class}

In what follows it will be very important to classify leaders into three types:

{\it Susy-writable leaders} are those which involve $\chi_i$ in $O (n -
2)$ invariant combinations such as $\chi_i^2$, $(\partial \chi_i)^2$, etc.
This preserved subgroup is an accidental enhancement of the subgroup $S_{n -
1}$ preserved by the Cardy transform. Examples of singlets with susy-writable
leaders are:
\begin{eqnarray}
  \sigma_4 & = & [4 \omega \varphi^3 + 6 \chi_i^2 \varphi^2]_{\Delta = 6} +
  \ldots, \nonumber\\
  \sigma_1 \sigma_3 & = & [3 \varphi^2 \omega^2 + 3 \varphi \omega
  \chi_i^2]_{\Delta = 8} + \ldots, \nonumber\\
  \sigma_1^2 \sigma_2 & = & [4 \varphi^2 \omega^2 + 4 \varphi \omega \chi_i^2
  + (\chi_i^2)^2]_{\Delta = 8}, \nonumber\\
  \sigma_1^4 & = & [\omega^4]_{\Delta = 12} . 
\end{eqnarray}
We can map these to the SUSY variables by $\chi_i^2 \rightarrow 2 \bar{\psi}
\psi$, $(\partial \chi_i^{}) \rightarrow 2 \partial \bar{\psi} \partial \psi$,
etc. The susy-writable leaders are the simplest interactions to study.

{\it Susy-null leaders} still involve $\chi_i$ in $O (n - 2)$ invariant
combinations, so they can also be mapped to the SUSY variables. Their defining
property is that they vanish after such a map, because of the Grassmann
conditions $\psi^2 = \bar{\psi}^2 = 0$.

The lowest-dimension susy-null {\it interaction} is $\left( \sum\nolimits'
\chi_i^2 \right)^2$ which maps to $(2 \bar{\psi} \psi)^2 = 0$. Is this a
{\it leader}? It turns out that yes, the corresponding singlet
interaction being
\begin{equation}
  \sigma_2^2 - \frac{4}{3} \sigma_1 \sigma_3 = [(\chi_i^2)^2]_{\Delta = 8} -
  \frac{4}{3} [\omega \chi_i^3]_{\Delta = 9} + \cdots
\end{equation}
What happens is that $\sigma_2^2$ and $\sigma_1 \sigma_3$ have susy-writable
leaders which are the same, up to a proportionality factor. Taking their
appropriate linear combination we can cancel the susy-writable part and we are
left with a susy-null leader. We will see in a second a more systematic way to
guess this special linear combination.

There are more examples like this. A generic singlet has a susy-writable
leader, shown in {\eqref{ACardy}}, 2nd line. Sometimes this susy-writable part
cancels in a linear combination of several different singlets, giving rise to
a leader which is susy-null, or non-susy-writable, see below.

Should we care at all about susy-null leaders, if they vanish after mapping to
susy-variables? It's an interesting subtle question, to which we will come
back later, in Section \ref{tuning}. The answer is yes, we should potentially worry about them.

{\it Non-susy-writable leaders}. These leaders involve $\chi_i$ in
combinations such as $\sum_{}' \chi_i^r$ with a power $r \geqslant 3$, which
only preserve exactly the $S_{n - 1}$ subgroup permuting $\chi_i$. Therefore
in this case $S_{n - 1}$ is not enhanced to $O (n - 2)$ as it is for the
previous two classes of leaders. Clearly there are many non-susy-writable
{\it interactions}, but it's not a priori obvious that there are any such
leaders. To see that they do exist, consider the following singlet
interaction:
\begin{equation}
  \mathcal{F}_k = \sum_{i = 1}^n \sum_{j = 1}^n (\phi_i - \phi_j)^k .
\end{equation}
where $k$ is an even integer. These interactions were first considered by
Feldman {\cite{Feldman}}, long before the notion of leaders was introduced in
{\cite{paper2}}. On the one hand, we can express $\mathcal{F}_k$ as a linear
combination of products of $\sigma_k$'s ({\bf Exercise}):
\begin{equation}
  \mathcal{F}_k = \sum_{l = 1}^{k - 1} (- 1)^l \binom{k}{l} \sigma_l \sigma_{k
  - l} .
\end{equation}
On the other hand, we can write $\mathcal{F}_k$ in terms of Cardy fields as
\begin{eqnarray}
  \mathcal{F}_k & = & \sum_{i = 2}^n \sum_{j = 2}^n (\phi_i - \phi_j)^k + 2
  \sum_{j = 2}^n (\phi_1 - \phi_j)^k \nonumber\\
  & = & \sum\nolimits' \sum\nolimits' (\chi_i - \chi_j)^k + 2 \sum\nolimits' (\omega_{} - \chi_j)^k . 
\end{eqnarray}
We see in particular that there is no $\varphi$ in this expression, as it
cancels in the differences $\phi_i - \phi_j$. What remains are $\chi$'s and
$\omega$'s. For $k \geqslant 4$, the leader will come from $\chi$'s and is
given by ({\bf Exercise}):
\begin{equation}
  [\mathcal{F}_k]_L = \sum_{l = 2}^{k - 2} (- 1)^l \binom{k}{l} \left( \sum\nolimits'
  \chi^l_i \right) \left( \sum\nolimits' \chi^{k - l}_i \right) .
\end{equation}
[For $k = 2$ the part consisting of $\chi$'s vanishes and we have
$\mathcal{F}_2 \propto \omega^2$.]

For $k = 4$ this construction gives the susy-null leader $[\mathcal{F}_4]_L =
6 (\chi_i^2)^2$ considered above. For $k = 6$ we get a non-susy-writable
leader:
\begin{equation}
  [\mathcal{F}_6]_L = [30 (\chi_i^2) (\chi_i^4) - 20 (\chi_i^3)^2]_{\Delta =
  12} . \label{eq-F6}
\end{equation}

In {\cite{paper2}}, App.~D, we have worked out leaders of all singlet interactions
with scaling dimension $\Delta \leqslant 12$ in 6d. This census shows that
{\eqref{eq-F6}} is actually the lowest-dimension non-susy-writable leader.

\subsection{Renormalization group fixed point and its
stability}\label{RG-stab}

We now have to start realizing our plan to find out if any of the many singlet
interactions, while irrelevant in $d = 6 - \varepsilon$, may become relevant
at smaller $d$. Thus we have to compute scaling dimensions of the leaders of these
interactions. The scaling dimension of interest is not the scaling dimension at
short distances, at the Gaussian fixed point - the so-called classical scaling
dimension, but the scaling dimension at long distances, at the nontrivial IR
fixed point, which is a sum of the classical scaling dimension and the
anomalous dimension.

Let us first discuss the IR fixed point. We consider the SUSY theory
described either by the SUSY action $S_{\text{PS}}$ {\eqref{SPS}} with the fields
$\varphi, \omega, \psi, \bar{\psi}$ or, equivalently, the pre-SUSY action
{\eqref{cardy-action}} with the Cardy fields $\varphi, \omega, \chi_i$ and in the
$n \rightarrow 0$ limit. We consider the quartic potential
\begin{equation}
  V (\phi) = \frac{1}{2} m^2 \phi^2 + \frac{\lambda}{4!} \phi^4 .
\end{equation}
This theory flows in the IR to a SUSY fixed point, which in $d = 6 -
\varepsilon$ dimensions is perturbative, the fixed point coupling $\lambda$
being $O (\varepsilon)$. The mass term $m^2$ has to be finetuned to reach the
fixed point. In practice it's best to work in dimensional regularization,
where one simply sets $m^2 = 0$. Computations leading to the fixed point are
standard. SUSY guarantees that the renormalized Lagrangian has the same form
as the bare one, up to renormalizing the quartic coupling
$\lambda$.\footnote{In particular the relative coefficients of the two quartic
interaction terms, one proportional to $\omega \varphi^3$, and another
proportional to $\varphi^2 \bar{\psi} \psi$, see {\eqref{quartics}}, are fixed
by SUSY.} The beta-function for this coupling $\lambda$ has the form.
\begin{equation}
  \beta (\lambda) = \frac{d \lambda}{d \log a} = \varepsilon \lambda - B
  \lambda^2 + \text{higher order terms }. \label{beta-function}
\end{equation}
The first term in the beta function, $\varepsilon \lambda$, just tells us that
the coupling $\lambda$ is relevant in the free theory, the interaction having
the dimension $d - \varepsilon$. The second term comes from the one-loop
correction effect. Coefficient $B$, which is positive, can be found by a
one-loop computation. The fixed-point coupling is found by solving $\beta
(\lambda_{\ast}) = 0$ which gives $\lambda_{\ast} = \varepsilon / B + O
(\varepsilon^2)$. See {\cite{paper2}} for details. Because of
dimensional reduction, the beta-function and the fixed point for the SUSY theory in $6 -
\varepsilon$ dimensions turns out to be simply related to the beta-function and the fixed-point for the
non-supersymmetric $\varphi^4$ theory in $4 - \varepsilon$ dimension (the
Wilson-Fisher fixed point).

We next have to study the scaling dimensions of additional (i.e.~not included in
the original Lagrangian) singlet interactions around this SUSY IR fixed point.
If any additional interaction becomes relevant for $d < d_c$, this signals
instability of the SUSY IR fixed point with respect to this perturbation. At
short distances, any singlet perturbation is expected to be present. If any
perturbation becomes relevant, the RG flow will not be able to reach the SUSY
IR fixed point but it will be deviated away from it, to some other fixed
point or to a massive phase.

\begin{remark}
  Here we should recall that the fixed point we are discussing is supposed to
  describe the phase transition of the RFIM. In a real experiment or in a
  numerical simulation, this phase transition is obtained by tuning exactly
  one parameter, which may be the temperature or the magnetic disorder
  strength (see Fig.~\ref{phase} left). This means that the fixed point should
  have exactly one relevant singlet perturbation. The SUSY fixed point has
  {\it one} such relevant perturbation - the mass term interaction
  $\sigma_2 = \sum_{i = 1}^n \phi_i^2 = 2 \varphi \omega + \sum\nolimits' \chi_i^2 = 2
  \varphi \omega + 2 \bar{\psi} \psi$. What we are saying that it should not
  have {\it more} than one. If it does, it will become unstable, and will
  not be able to describe the phase transition. At least, not without
  additional tuning - see Section \ref{tuning}.
\end{remark}

We next discuss the scaling dimensions. One way to think about them is as
follows. We perturb the SUSY fixed point action by a singlet perturbation
times an infinitesimal coefficient. As explained in Section \ref{sec-leaders},
it's enough to perturb by the leader, not by the full perturbation. Then we do
an RG step and see how the coefficient gets rescaled. From the rescaling
factor we infer the scaling dimension of the leader. In particular, irrelevant
leaders get their coefficients suppressed, and relevant ones get them
enhanced.

The previous paragraph is overly simplistic in one aspect. Namely, when we do
an RG step, we do not only get the same leader with a rescaled coefficient,
but we may also generate other leaders.\footnote{That leaders should generate
under RG only leaders, as opposed to general interactions terms, follows from
the logic of Section \ref{sec-leaders}. See however the discussion in Section
\ref{sym-meaning} below.} \ This is referred to as operator mixing. In
perturbation theory and in dimensional regularization, only leaders of the
same classical scaling dimension can mix. For every classical scaling dimension
$\Delta$, this gives us a mixing matrix $\Gamma_{i j}$ of size $n \times n$, where
$n$ is the number of leaders having classical scaling dimension $\Delta$.
Anomalous dimensions are eigenvalues of this matrix.

In addition, there are selection rules for the three classes of leaders
described in Section \ref{sec-leaders}:
\begin{eqnarray}
  \text{sn} & \rightarrow & \text{sn}, \nonumber\\
  \text{sw} & \rightarrow & \text{sw} + \text{ sn},  \label{eq-selection}\\
  \text{nsw} & \rightarrow & \text{nsw + } \text{sw} + \text{ sn} . \nonumber
\end{eqnarray}
Let us explain this pattern. Perturbations by susy-null (sn) leaders have no
effect on the SUSY theory. This implies that after an RG step, susy-null
leaders may generate only susy-null leaders - the first line of
{\eqref{eq-selection}}. Susy-writable (sw) leaders respect $O (n - 2)$
invariance when written in the Cardy variables, which is also the symmetry of the
Cardy action. After an RG step only operators respecting this symmetry may be
produced - that's what the second line says. Finally, the third line says that
non-susy-writable (nsw) leaders may generate anything under an RG step.

The selection rules imply a block-lower-triangular structure for the mixing matrix $\Gamma$:
\begin{eqnarray}
  &  & \begin{array}{lll}
    \text{\quad sn} & \text{sw} & \text{nsw}
  \end{array} \nonumber\\
  & \begin{array}{l}
    \text{sn}\\
    \text{sw}\\
    \text{nsw}
  \end{array} & \left( \begin{array}{lll}
    \ast \quad & 0 \quad & 0\\
    \ast & \ast & 0\\
    \ast & \ast & \ast
  \end{array} \right) . 
\end{eqnarray}
To find the eigenvalues, the off-diagonal blocks are not needed. It's enough
to diagonalize separately the blocks which describe mixings within each class
of operators: sn-sn, sw-sw and nsw-nsw. This simplifies the computations.

See {\cite{paper2}} for technical details about the computations of anomalous
dimensions. Here I will just comment about their physical meaning, and how
they arise in field theory in general. Take a scalar operator $\mathcal{O}$ in
a massless Gaussian theory, and consider its two-point correlator. If it
decays as a power of the distance: $\langle \mathcal{O} (r) \mathcal{O} (0)
\rangle = \text{const.} r^{- 2 \Delta}$, we say that $\mathcal{O}$ has scaling
dimension $\Delta$ - that's the classical scaling dimension. Now let us
compute the same correlator in the Gaussian theory perturbed by an interaction
term. For example, in the $\varphi^4$ model we have to consider the average:
\begin{equation}
  \left\langle \mathcal{O} (r) \mathcal{O} (0) \exp^{} \left( - g \int d^d x
  \varphi^4 (x) \right) \right\rangle .
\end{equation}
Expanding the exponential gives the perturbative series for the correlator
$\langle \mathcal{O} (r) \mathcal{O} (0) \rangle$ in the interacting theory,
whose terms are integrals of correlators $\langle \mathcal{O} (r) \mathcal{O}
(0) \varphi^4 (x_1) \ldots \varphi^4 (x_n) \rangle$ of the Gaussian theory.
Anomalous dimensions appear from the parts of those integrals where $\varphi^4
(x_{})$ approach the $\mathcal{O} (r)$ and $\mathcal{O} (0)$. When $\varphi^4
(x)$ sits near $\mathcal{O} (0)$, viewed from far away this will look like $C
(x) \mathcal{O} (0)$ where $C (x)$ is a function of $x$ which is singular as
$x \rightarrow 0$ (this is the OPE already encountered in Section \ref{UCD}).
When we integrate $C (x)$ over $x$, we get $\mathcal{O} (0)$ times a rescaling
factor $F$. Since $C (x)$ is singular, we have to regulate the integral,
cutting it off at $x \sim a$ where $a$ is a short-distance cutoff. Hence the
rescaling factor depends on $a$. The fact that the operator $\mathcal{O} (0)$
in the interacting theory looks like $F (a) \mathcal{O} (0)$ means that the
scaling dimension changed. It now equals $\Delta + \gamma$ where $\gamma$ is
the anomalous dimension computable from the dependence of $F$ on $a$.

What we just described is the basis of the so-called ``OPE method'' to compute
the anomalous dimensions. Formulated in the $x$-space, it is the fastest
method at the leading order in the coupling constant, but it becomes awkward
to use beyond the leading order. At higher orders in the coupling constant,
one usually uses another method - computing Feynman diagrams in momentum space,
and regulating the theory via dimensional regularization.

The above discussion expresses the anomalous dimensions in terms of the
coupling constant $g$, which then has to be set equal to the fixed point value
$g_{\ast}$. For models with quartic interactions, like $\varphi^4$ or the RFIM,
near the upper critical dimension we typically have $g_{\ast} = O
(\varepsilon)$, see {\eqref{beta-function}}. The anomalous dimensions become
power series in $\varepsilon$. Most operators have anomalous dimensions
starting at $O (\varepsilon)$, or sometimes at $O (\varepsilon^2)$.

In the $\varphi^4$ model in $d = 4 - \varepsilon$ dimensions, anomalous
dimensions of many operators have been thus computed. Anomalous dimensions of
operators $\varphi$, $\varphi^2$, $\varphi^4$ are known in this theory up to
very high order $\varepsilon^7$ or even $\varepsilon^8$. These series are then
resummed and extrapolated to $\varepsilon = 1$ to obtain predictions of
critical exponents of the pure Ising model in $d = 3$ dimensions, in excellent
agreement with experiments, numerical simulations and the conformal bootstrap.
Anomalous dimensions of more complicated operators are also known, albeit to a
smaller order in $\varepsilon$. The O$(N)$ model - generalization of the
$\varphi^4$ model to $N$ fields - is a similar success story.

Our theory - the SUSY fixed point described at the beginning of this section
- is a bit more complicated that the $\varphi^4$ model or the O$(N)$ model:
there are several fields with different classical scaling dimensions, and two
quartic coupling constants. In our work {\cite{paper2}}, we went only to the
second order in the $\varepsilon$-expansion, although in principle our results
should be extendable to higher orders by standard techniques.

We examined the susy-writable, susy-null and non-susy-writable leader
perturbations of the SUSY fixed point. Anomalous dimensions of susy-writable
leaders are computed in the $\varphi, \omega, \psi, \bar{\psi}$ variables.
Dimensional reduction relates them to the anomalous dimensions of composite
operators of the $\varphi^4$ model in $d = 4 - \varepsilon$ dimensions, many
of which are already known, and often we were able to simply borrow these
known results. In contrast, anomalous dimensions of susy-null and
non-susy-writable leaders cannot be inferred from dimensional reduction and
should be computed from scratch. These computations are done in the $\varphi,
\omega, \chi_i$ variables.

We surveyed all leaders with classical scaling dimension $\Delta \leqslant 12$
in $d = 6$, and some leaders with even higher classical dimensions. We looked
for any operator which may turn relevant as we lower $d$. In the susy-writable
sector all was good: apart from the mass term operator $\sigma_2$ which is
relevant for any $d$, we did not find any operator which may become relevant.
On the other hand, in the susy-null and the non-susy-writable sectors we did
identify two operators with negative anomalous dimensions at order
$\varepsilon^2$.\footnote{It is not common to encounter negative anomalous
dimensions around the upper critical dimension. In the ordinary $\varphi^4$
model in $d = 4 - \varepsilon$ dimensions, most operators get positive
anomalous dimensions. Negative anomalous dimensions in our model may be
related to the lack of unitarity.} These are precisely the leaders of the
singlets $\mathcal{F}_4$ and $\mathcal{F}_6$, see Section \ref{sec-class}.
Their scaling dimensions are given by:
\begin{eqnarray}
  &  & \Delta_{(\mathcal{F}_4)_L} = (8 - 2 \varepsilon)_{\text{class}} -
  \frac{8}{27} \varepsilon^2 + \ldots, \nonumber\\
  &  & \Delta_{(\mathcal{F}_6)_L} = (12 - 3 \varepsilon)_{\text{class}} -
  \frac{7}{9} \varepsilon^2 + \ldots,  \label{eq-res}
\end{eqnarray}
where we show separately the classical dimension and the anomalous one,
arising at order $\varepsilon^2$ (there is no order-$\varepsilon$ anomalous
contribution).

In Fig.~\ref{fig-results} we plot the results {\eqref{eq-res}} neglecting the
unknown higher order terms. There we see that both these scaling dimensions
crossed marginality line $\Delta = d$ for $d$ between 4 and 5:
\begin{eqnarray}
  \Delta_{(\mathcal{F}_4)_L} = d &  & \quad {\rm at} \quad d = d_{c 1}
  \approx 4.6, \nonumber\\
  \Delta_{(\mathcal{F}_6)_L} = d &  & \quad {\rm at} \quad d = d_{c 2}
  \approx 4.2.  \label{eq-dc}
\end{eqnarray}
\begin{figure}[h]
 \centering \resizebox{!}{200pt}{\includegraphics{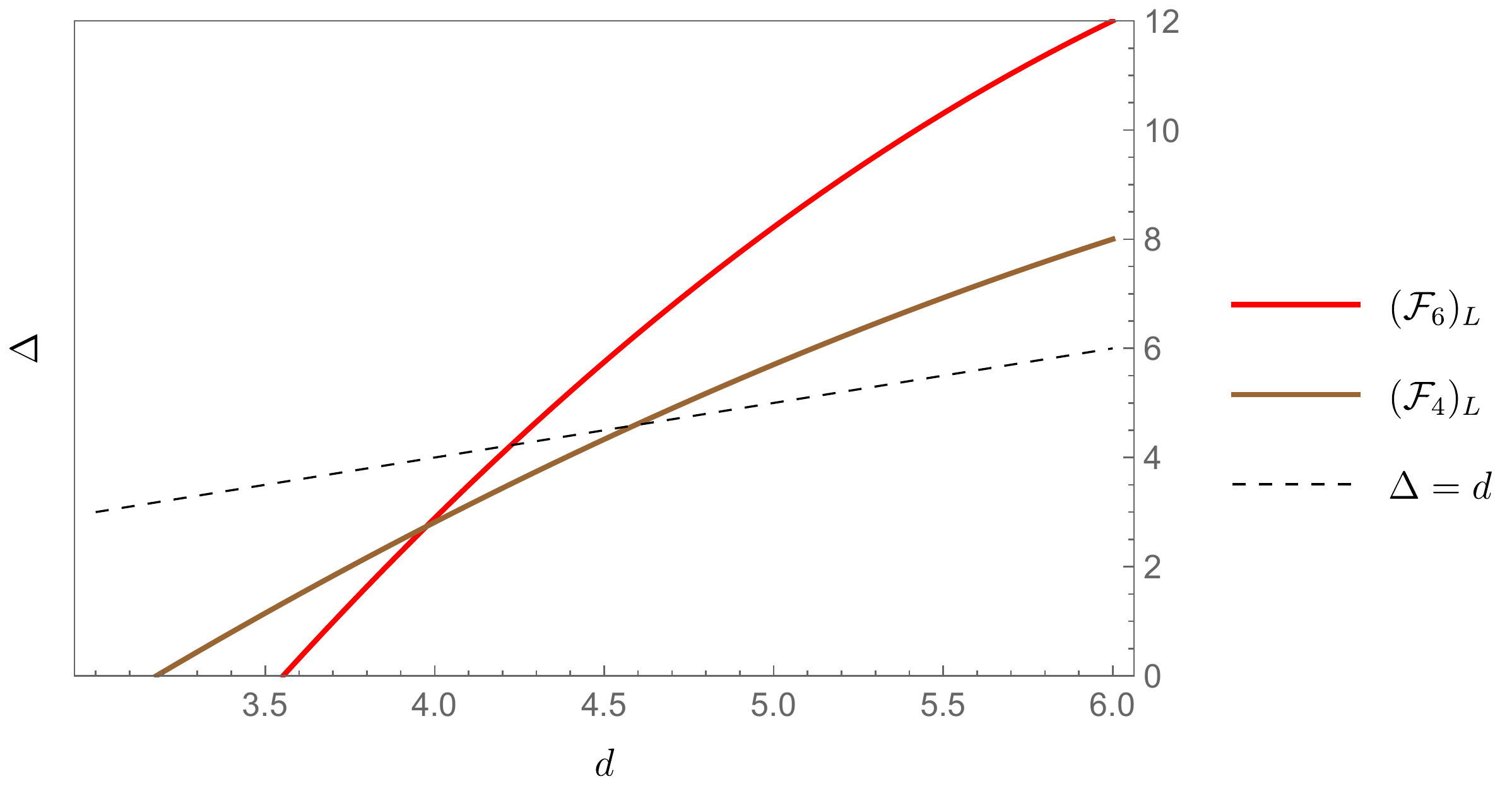}}
  \caption{Scaling dimensions {\eqref{eq-res}} plotted as a function of $d = 6
  - \varepsilon$. \label{fig-results}}
\end{figure}
We may estimate the effect of the unknown higher order terms in
{\eqref{eq-res}} by using Pad{\'e}-resummed versions of the same equations.
Let us use $\mathrm{\text{Pad{\'e}}}_{[1, 1]}$, which means that we look for
rational functions $(a_0 + a_1 \varepsilon) / (b_0 + b_1 \varepsilon)$ whose
expansions coincide to $O (\varepsilon^2)$ with {\eqref{eq-res}}. Proceeding
this way, we find that $\Delta_{(\mathcal{F}_4)_L}$ crosses marginality at
$d_{c 1} \approx 4.7$, while $\Delta_{(\mathcal{F}_6)_L}$ at $d_{c 2} \approx
4.5$ ({\bf Exercise}). That these numbers do not differ too much from
{\eqref{eq-dc}} gives us hope that the conclusion is robust -
$(\mathcal{F}_4)_L$ and $(\mathcal{F}_6)_L$ do become relevant somewhere
between $d = 4$ and 5. This will destabilize the SUSY fixed point, leading to
the loss of SUSY and with it to the loss of dimensional reduction.

Possible tests, predictions, and open problems will be discussed in the next
lecture.

\subsection{Relation to previous work}

Ideas from the previous work by Br{\'e}zin and De Dominicis
{\cite{Brezin-1998}} and by Feldman {\cite{Feldman}} were important for us, up
to some important disagreements in both cases.

We relied on the observation of {\cite{Brezin-1998}} about the need to
consider all singlet interactions. On the other hand we disagree with
{\cite{Brezin-1998}} when they treated some of these extra interactions as
though they were marginal in $d = 6$, while they are strongly irrelevant. For
this reason our conclusion ended up different from {\cite{Brezin-1998}}, see
Remark \ref{Brezin}.

As to Ref.~{\cite{Feldman}}, it identified the class of operators
$\mathcal{F}_k$ whose importance was confirmed in our work. In addition,
{\cite{Feldman}} computed their scaling dimensions finding {\eqref{eq-res}}
and, more generally:
\begin{equation}
  \Delta_k \assign \Delta_{(\mathcal{F}_k)_L} = \left( 2 k - \frac{k}{2}
  \varepsilon \right)_{\text{class}} - \frac{k (3 k - 4)}{108} \varepsilon^2 +
  \ldots \text{} \label{FkLgen}
\end{equation}
Ref.~{\cite{Feldman}} worked in the replicated action formalism, without
using the Cardy fields nor separating interactions into leaders and followers.
Ref.~{\cite{Feldman}} computed the leading scaling dimension inside
$\mathcal{F}_k$, which agrees with the scaling dimension of
$(\mathcal{F}_k)_L$. We reproduced {\eqref{FkLgen}} using our formalism. This
is a nice check of the equivalence of the two formalisms. We do believe
that for more systematic computations, and for pushing to higher orders our
formalism is preferred.

Now to the disagreement. Let us focus on $\mathcal{F}_k$ with $k$ even which
preserve $\mathbb{Z}_2$ symmetry. Feldman observed that the classical
part of $\Delta_k$ grows linearly with $k$, while the negative anomalous part
grows quadratically. Hence, he said, $\Delta_k$ will cross marginality at
$\varepsilon \propto 1 / \sqrt{k}$, which becomes closer and closer to $d = 6$
as $k$ grows. He thus concluded that SUSY should be lost arbitrarily close to
$d = 6$. From the numerical simulations described in Section \ref{sec-numsim}
(which, to be fair, appeared after {\cite{Feldman}}), this does not seem to be
correct.

We see two loopholes in Feldman's reasoning:
\begin{itemize}
  \item All operators $\mathcal{F}_k$, $k \geqslant 6$, have non-susy-writable
  leaders. If we blindly apply {\eqref{FkLgen}}, their scaling dimensions
  would cross (see Fig.~\ref{fig8}). But we don't actually expect them to
  cross due to nonperturbative mixing effects, which should repel scaling
  dimensions of operators belonging to the same symmetry class, when they come
  close to each other. These mixing effects were already mentioned in Section
  \ref{sec-analogies}.
  
  Furthermore, in {\cite{paper-summary}} we exhibited a series of operators
  $\mathcal{G}_k$ with non-susy-writable leaders and \emph{positive} anomalous
  dimensions, the first of which $\mathcal{G}_8$ is located between $\mathcal{F}_6$ and
  $\mathcal{F}_8$ for $d = 6$. Its scaling dimension to $O (\varepsilon)$ is
  given by ({\cite{paper-summary}}, Eqs. (18),(19),(21)):
  \begin{equation}
    \Delta_{(\mathcal{G}_8)_L} = (14 - 4 \varepsilon)_{\text{class}} +
    \frac{13}{3} \varepsilon + \ldots, \label{eq-G}
  \end{equation}
  plotted in Fig.~\ref{fig8}. This operator acts as a sort of ``roof'' for
  $\mathcal{F}_6$, protecting it from the influence of $\mathcal{F}_k$ with $k
  \geqslant 8$. These latter operators will mix with $\mathcal{G}_8$, and with
  higher operators of this series, when their scaling dimensions become close
  to each other, and should repel.
  
  Although there is no straightforward way to compute the nonperturbative
  mixing effects,\footnote{Functional renormalization group may be one tool.}
  it appears plausible that level repulsion will help prevent $\mathcal{F}_k$
  with $k \geqslant 8$ from becoming relevant in any $d$.
  
  \item The second reason is even simpler. Fixed-order perturbation theory
  breaks down for composite operators in the limit of many fields
  {\cite{Badel:2019oxl}}. We have a small parameter $\varepsilon$, but also a
  large parameter $k$, and one would have to first resum $k \varepsilon$
  effects before making any claims about the behavior of $\Delta_k$ for $k \gg
  1$.
  
  \ 
\end{itemize}
\begin{figure}[h]
 \centering \resizebox{293pt}{190pt}{\includegraphics{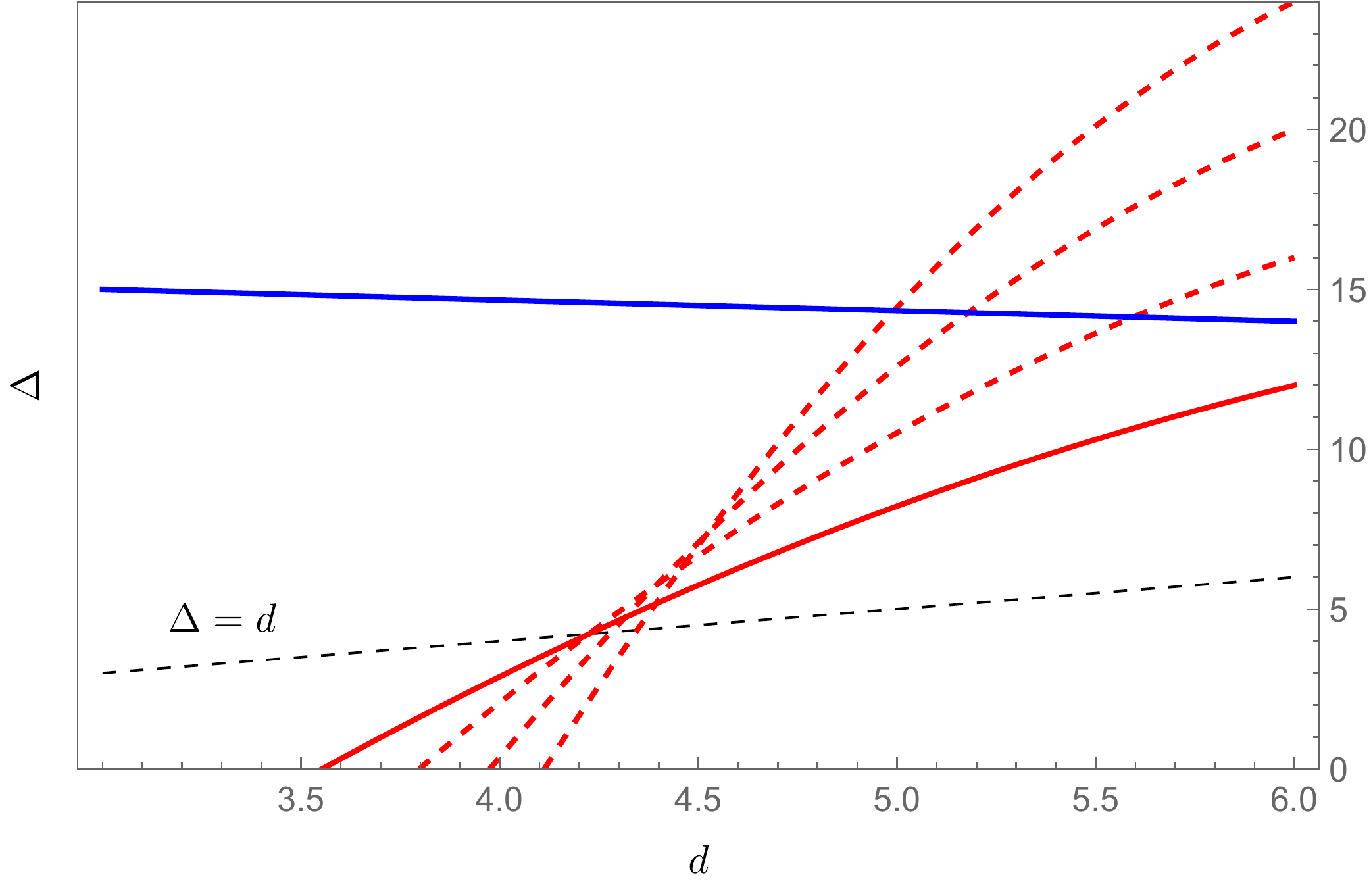}}
  \caption{\label{fig8}Scaling dimensions of $(\mathcal{F}_6)_L$ (red solid),
  $(\mathcal{F}_k)_L$ with $k = 8, 10, 12$ (red dashed) and $\mathcal{G}_L$
  (blue) according to {\eqref{FkLgen}}, {\eqref{eq-G}}.}
\end{figure}

\subsection{Literature and further comments}

We had no time to touch upon several important subtleties, whose discussion
may be found in {\cite{paper2}}. Here are several examples:
\begin{itemize}
  \item {\it Cutoff issue.} Although the pre-SUSY theory
  {\eqref{cardy-action}} is formally equivalent to the SUSY theory
  {\eqref{SPS}} at the level of the classical action, SUSY is broken by cutoff
  effects. However one can argue that SUSY does emerge in the IR, up to
  renormalization of $R$, if no further interactions become relevant. See
  {\cite{paper2}}, Section 3.2.
  
  \item {\it General rules for mapping susy-writable leaders to SUSY
  variables.} We only used the mapping for simple operators like $\sum\nolimits'
  \chi_i^2$ or $\sum\nolimits' (\partial \chi_i^{})^2$. However any $O (n - 2)$
  invariant operator can be mapped to SUSY variables. See {\cite{paper2}},
  App.C.
  
  \item {\it The meaning of interactions which are not leaders.} We
  argued that instead of perturbing the action by a full $S_n$ invariant
  interaction, it's enough to perturb by the leader, while the followers are
  always suppressed. What if we perturb the action by an interaction which is
  not a leader? As explained in {\cite{paper2}}, this corresponds to breaking
  $S_n$ invariance, which we don't want to do.
  
  \item What if we keep $n_{}$ small but finite, instead of taking the strict
  $n \rightarrow 0$ limit? Then the model develops a finite correlation
  length, which goes to infinity as $n \rightarrow 0$. This is different from
  the bond-disordered Ising model case, where a fixed point with infinite
  correlation length is believed to exist for any $n$. See {\cite{paper2}},
  Sec.6.
\end{itemize}

There is no shortage of good books explaining how to compute anomalous
dimensions in perturbative field theory for the $\varphi^4$ and related model
in $d = 4 - \varepsilon$ dimensions. For computations based on Feynman diagrams and dimensional
regularization, see Peskin and Schroeder {\cite{Peskin:1995ev}}, Vasiliev
{\cite{vasil2020field}}, Zinn-Justin {\cite{Zinn-Justin:2002ecy}}, Kleinert
and Schulte-Frohlinde {\cite{Kleinert:2001ax}}. For the OPE method, see Cardy
{\cite{Cardy-book}}, Ch.5.

For the nonperturbative level repulsion among operator dimensions, see
{\cite{Korchemsky:2015cyx,Behan:2017mwi,Henriksson:2022gpa}}.

\section{Open problems and future directions}

Here we will discuss several disconnected subjects, many of which are not
fully understood. These are great topics for future research.

\subsection{Symmetry meaning of leaders}\label{sym-meaning}

Recall our logic in the previous lecture. We introduced the Cardy field
transform {\eqref{eq-Cardy}} and used it to map the replicated action
{\eqref{Sn}} to the pre-SUSY action {\eqref{cardy-action}}, plus some
irrelevant terms. Action {\eqref{cardy-action}} is called pre-SUSY because it is
equivalent to the SUSY action {\eqref{SPS}} after mapping $- 2$ scalars
$\chi_i$ to 2 fermions $\psi, \bar{\psi}$.

The symmetry of the replicated action {\eqref{Sn}} is clear - it is $S_n$
(times $\mathbb{Z}_2$). The symmetry of the SUSY action {\eqref{SPS}} is also
clear - it is the Parisi-Sourlas supersymmetry. But what is the symmetry of
the pre-SUSY action? This question is not fully understood.

It can't be $S_n$, because we dropped follower parts of the interaction
terms. There are reasons to believe that the pre-SUSY action should have some
sort of ``pre-SUSY'' symmetry, which becomes SUSY after mapping to
{\eqref{SPS}}. But I do not know how to express this putative symmetry in
terms of fields $\varphi, \omega, \chi_i$. I only know how to write it in
terms of $\varphi, \omega, \psi, \bar{\psi}$.

Why should we care? Recall that in the previous lecture we introduced a
classification of leaders. The properties of susy-writable leaders could be
studied by mapping them to the SUSY fields, but properties of susy-null and
non-susy-writable leaders could only be studied in terms of the Cardy fields
$\varphi, \omega, \chi_i$. So all calculations for these leaders are done in
terms of the pre-SUSY action, and of course when working with an action one
would like to know its symmetry.

We argued in the previous lecture that it's enough to perturb the pre-SUSY
action by the leader parts of $S_n$ singlet interactions, and compute
anomalous dimensions of these leaders. We argued that
renormalization of leader perturbations should produce only leaders. These
arguments used the existence of the $S_n$ invariant replicated action from
which the pre-SUSY action originates. All our concrete calculations
{\cite{paper2}} confirmed this conclusion - leaders only generate leaders
under renormalization.

Normally, when there is a class of operators closed under renormalization,
this is explained by these operators being preserved by a symmetry of the
action. This begs the question: what is the symmetry which singles out leaders
out of all possible interaction terms perturbing the pre-SUSY action?

There is one case when we were able to clarify the symmetry meaning of the
leaders - namely for the susy-writable leaders. When mapping them to the SUSY
fields, they become a subset of possible interactions of the SUSY action. Which
subgroup of SUSY singles out these interactions? The answer is interesting -
it is the supertranslation invariance which acts on the fields as:
\begin{equation}
  \delta_{{\rm st}} \varphi = \bar{\varepsilon} \psi + \varepsilon \bar{\psi},
  \quad \delta_{{\rm st}}  \bar{\psi} = \bar{\varepsilon} \omega, \quad
  \delta_{{\rm st}} \psi = - \varepsilon \omega, \quad \delta_{{\rm st}}
  \omega = 0 \hspace{0.17em} .
\end{equation}
In addition, one has to impose ${\rm Sp} (2)$ invariance which rotates the
fermions $\psi, \bar{\psi}$ into each other (this just means that $\psi$ and
$\bar{\psi}$ should enter symmetrically like $\bar{\psi} \psi$, $\partial
\bar{\psi} \partial \psi$ etc.). Supertranslation invariance fixes relative
coefficients in the interaction. E.g.~$\varphi \omega + \bar{\psi} \psi$ and
$\omega \varphi^3 + 3 \bar{\psi} \psi \varphi^2$ are supertranslation
invariant, and indeed these and only these linear combinations are, up to a
constant, leaders of singlet interactions, $\sigma_2$ and $\sigma_4$. In full
generality this characterization of susy-writable leaders is proven in
{\cite{paper-summary}}, App.~B. 

It would be very interesting to get a similar
characterization of susy-null and non-susy-writable leaders, in terms of a
subset of the still elusive symmetry of the pre-SUSY action. For a step in this direction see \cite{paper-summary}, Eq.~(C2).

\subsection{Branched polymers}\label{sec-BP}

A (linear) polymer is a chain of monomers, which you can think of as rods of
fixed length joined at vertices and free to rotate, up to a self-repulsion
constraint which does not allow different vertices to come close and prevents
collapse (Fig.~\ref{fig-poly}(a)). In the limit of many monomers, polymers
show critical behavior which can be studied via the $n \rightarrow 0$ limit of
the O$(n)$ model ({\cite{Cardy-book}}, Ch.9). This universality class is also
called the self-avoiding random walk (SAW).

\begin{figure}[h]\centering
\resizebox{!}{100pt}{\includegraphics{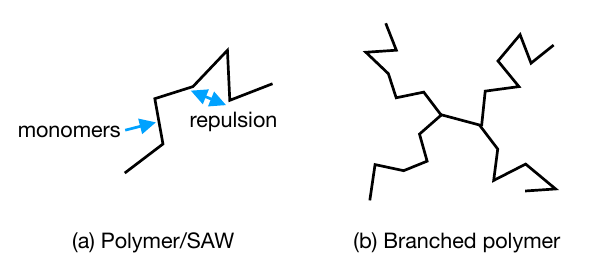}}
  \caption{\label{fig-poly}Polymers and branched polymers.}
\end{figure}

Our interest here will be the problem of {\it branched polymers}.
Branched polymers consist of several linear polymer chains arbitrarily joined
at vertices (Fig.~\ref{fig-poly}(b)). They also show critical behavior in the
limit of many monomers, which is different from SAW. Two commonly considered
critical exponents are $\theta$ and $\nu$ which are related to the total
statistical weight and to the mean size of branched polymers consisting of $N$
monomers:
\begin{eqnarray}
  Z_N & \sim & \mu^N N^{- \theta}, \\
  R_N & \sim & N^{\nu} . 
\end{eqnarray}
The constant $\mu$ depends on the details of the microscopic model, but
$\theta$ and $\nu$ are universal. On the lattice, branched polymers can be
modeled as arbitrary connected configurations of $N$ bonds, assigning to each
configuration weight 1. One may impose or not the tree topology (it turns out that the
universality class does not depend on that). Such configurations are also
called lattice animals. $Z_N$ is then the total number of lattice animals made
of $N$ bonds. Lattice animals may be efficiently simulated on the lattice, and
high quality determinations of $\theta, \nu$ are available in any $2\leqslant d \leqslant 8$ (which turns out to be the upper critical dimension) {\cite{Hsu2005}}.

We are interested in branched polymers (BP), because field theory for their
statistics turns out to be closely related to RFIM. This is surprising since
the problem is clearly quite different - it is purely geometric and there is
no disorder. The link was found by Parisi and Sourlas
{\cite{PhysRevLett.46.871}}, following the work by Lubensky and Isaacson
{\cite{PhysRevA.20.2130}}. They argued that BP can be studied via the $n
\rightarrow 0$ limit of the replicated action $\mathcal{S}_n$ {\eqref{Sn}}
where $V (\phi)$ is not the quartic potential as for the RFIM, but the cubic
potential with an imaginary coupling:
\begin{equation}
  V (\phi) = \frac{1}{2} m^2 \phi^2 + i \frac{\lambda}{3!} \phi^3 .
  \label{potential-cubic}
\end{equation}
Then, using PS SUSY, they argued for the dimensional reduction - that the critical
exponents of BP in $d$ dimensions should be the same as those for the Lee-Yang
(LY) universality class in $\hat{d} = d - 2$ dimensions. Recall that the LY
universality class is realized by the Ising model in a critical imaginary
magnetic field, and is described field-theoretically by the $i \varphi^3$
model - a single scalar field $\varphi$ with the cubic potential
{\eqref{potential-cubic}}. LY critical exponents are well known due to the
$\varepsilon$-expansion in $\hat{d} = 6 - \varepsilon$ dimensions, as well as
exact results in $\hat{d} = 0, 1, 2$. In this case it turns out that
dimensional reduction works perfectly - BP$_d$ and LY$_{\hat{d}}$ exponents do
agree in any $d$ {\cite{Hsu2005}}.

Why is the situation here so different from RFIM? Unfortunately, concerning
this question there is some confusion in the literature, related to the
rigorous work by Brydges and Imbrie {\cite{zbMATH02068689}}. This work
considered a particular type of branched polymers where the distance $r$
between vertices joined by a monomer is not fixed but is distributed according
to a weight $Q (r)$. Another weight $P (r) = e^{- V (r)}$, where $V (r)$ is
the repulsing potential, does not allow vertices not joined by a monomer to
come close to each other. A physically reasonable model requires $P (r)$
growing monotonically from $0$ at \ $r \lesssim 1$ to $1$ at $r \gtrsim 1$,
and $Q (r)$ peaked at $r \sim 1$. In a general model of this kind $P$ and $Q$
would be independent, but {\cite{zbMATH02068689}} imposes on them the special
relation
\begin{equation}
  Q (r) = P' (r), \label{BP-susy}
\end{equation}
which is consistent with physical requirements. The main observation of
{\cite{zbMATH02068689}} is that BP satisfying this relation have exact SUSY -
their partition function allows an exact fermionic-integral representation.
Furthermore, this SUSY representation can be shown to reduce to a
grand-canonical partition function of a $\hat{d}$-dimensional gas of particles
with pairwise potential $V (r)$, at negative fugacity. The latter model has,
by physics arguments, a critical point in the same universality class as LY.

The result of {\cite{zbMATH02068689}} is sometimes quoted by saying that they
proved dimensional reduction for BP, but this is misleading. They proved
dimensional reduction for a particular class of models having built-in SUSY.
This should not be so surprising, since we know that dimensional reduction is
a consequence of SUSY. The main challenge is to show how SUSY may emerge in a
model of BP which has no SUSY built in from the start, like for a model of
lattice animals. However, the theorem of Brydges-Imbrie does not answer this
question. For example, is their result stable with respect to small
perturbations violating the relation {\eqref{BP-susy}}? From the mathematical
point of view, this is an open problem.

From the physics perspective, stability of the SUSY fixed point of BP can be
studied in the same way as for the RFIM. Namely, one does the Cardy transform
of the replicated action, goes to the fixed point of the pre-SUSY action, and
studies the anomalous dimensions of leaders of singlet perturbations. This was
carried out by Kaviraj and Trevisani {\cite{paper3}}. Unlike for the RFIM,
they did not find any leader interactions with negative anomalous dimensions -
all additional interactions for BP seem to remain irrelevant for any $d$.
This, then, is an explanation of why dimensional reduction is valid for BP in
any $d$, and not only for $d > d_c$ as for RFIM .

Another confusion about BP is related to the question to what extent the theory is
conformally invariant. This goes back to the observation of Miller and De'Bell
{\cite{Miller:1992uv}} that the critical point of BP in 2d cannot have
Virasoro symmetry. Ref.~{\cite{Miller:1992uv}} looked at the free Parisi-Sourlas
action, which integrating out $\omega$ has a higher-derivative in $\varphi$
piece
\begin{equation}
  \int (\partial^2 \varphi)^2 d^d x. \label{higher-der}
\end{equation}
It is indeed well known that such a Gaussian theory is not Virasoro invariant.
However it is invariant under global conformal transformations (M{\"o}bius
transformations).\footnote{Virasoro invariance in 2d requires the stress
tensor to be traceless $T_{\mu}^{\hspace{1em} \mu} = 0$ or
$T_{\mu}^{\hspace{1em} \mu} = \partial^2 L$ in which case it can be improved
to be traceless. However, theory {\eqref{higher-der}} has
$T_{\mu}^{\hspace{1em} \mu} = \partial_{\mu} \partial_{\nu} L^{\mu \nu}$ which
is enough for the global conformal invariance but not for Virasoro
{\cite{Polchinski:1987dy}}.} This is nontrivial and should be kept in mind. In
general, the BP fixed point has global conformal invariance in any $d \geqslant
2$.

\subsection{Nonperturbative RG studies of RFIM}\label{sec-FRG}

Tarjus, Tissier and their collaborators published since 2004
{\cite{Tarjus2004}} many papers devoted to applying non-perturbative RG
techniques to the RFIM. Their method is also known as ``functional RG'' (FRG) and
is based on the Wetterich equation. For the $\varphi^4$ model (pure Ising) in
$2 \leqslant d \leqslant 4$, modern FRG calculations give reliable results in agreement with the $\varepsilon$-expansion and the conformal
bootstrap. It is thus interesting to know what they give for the RFIM, and to
compare with our results.

Ref.~{\cite{Tarjus2004}} applied FRG to the replicated action and reported a
change of the regime of the fixed point for $d < d_c \approx 5.1$. They
associated this change with the loss of analyticity of the effective action
(``cusps''), and observed that $\eta \neq \bar{\eta}$ for $d <
d_{c\text{{\it }}}$, as a sign that SUSY gets broken.

It is premature to discuss the numerical difference between our and their
estimate of $d_c$. Our scheme may be improved by including more perturbative
orders, while their scheme may be improved by including more terms in the
derivative expansion of the effective action. One would have to see first how
$d_c$ changes after such improvements.

It is interesting to compare what happens at and near $d_c$. FRG studies find,
similarly to us, two interactions heading towards the marginality threshold as
$d \rightarrow d^+_c$ (see {\cite{Balog:2020sre}}, Fig.1). There is however a
qualitative difference in the predicted behavior near $d_c$. Ref.
{\cite{Balog:2020sre}} found a square-root behavior $\sim \sqrt{d - d_c}$ in
the scaling dimensions.\footnote{In addition, they saw a small discontinuity
at $d = d_c$. This appears very strange. We don't understand how a fixed point
can possibly disappear before an interaction becomes marginal. Hopefully this
puzzling feature is an artifact of their approximation which will go away in a
more precise treatment.} Such a square root behavior is a hallmark of
fixed-point annihilation (see e.g.~{\cite{Gorbenko:2018ncu}} for a recent
discussion). To cite Ref.~{\cite{Balog:2020sre}}: ``What is found within the
NP-FRG is that the SUSY/dimensional-reduction fixed point that controls the
critical behavior of the RFIM below the upper dimension $d = 6$ annihilates
with another (unstable) SUSY/dimensional-reduction fixed point when $d =
d_{\textit{DR}}$.''

We find this scenario problematic from several points of view. First, there
is no known candidate for the unstable SUSY/dimensional-reduction fixed point.
This fixed point would dimensionally reduce to an unstable non-SUSY fixed
point in $3.1 < d < 4$ having $\mathbb{Z}_2$ global symmetry and there is no
known fixed point with such properties.

Second, when fixed points annihilate, they usually go to the complex plane,
and no real fixed points are left {\cite{Gorbenko:2018ncu}}. Here instead at $d
< 5.1$ a SUSY-breaking fixed point is supposed to exist.

In our scenario, these problems are avoided because the interactions becoming
marginal, $\mathcal{F}_4$ and $\mathcal{F}_6$, have susy-null and
non-susy-writable leaders, i.e.~belong to different symmetry classes from the
susy-writable interactions present in the SUSY actions. There is no fixed
point annihilation, which only happens when the marginal interaction has all
symmetries of the fixed point action (i.e.~is a full singlet). Instead, we
have a more conventional exchange of stability between the two fixed points -
SUSY for $d > d_c$ and non-SUSY for $d < d_c$.

There should be a way to distinguish the two scenarios by numerical
simulations in $d = 4$ dimensions. Tarjus, Tissier et al predict that the SUSY
fixed point does not exist in $d = 4$, having disappeared via annihilation. We
predict instead that the SUSY fixed point does exist even in $d = 4$, although
it is unstable. Unstable fixed points may still be realized via additional
tuning, as we discuss in the next section.

Further aspects of the work of Tarjus, Tissier et al are discussed in
{\cite{paper2}}, Apps.~A.7, A.8.

\subsection{Tuning to the SUSY fixed point}\label{tuning}

Let's suppose that our theory is true: for $d$ below a critical value, the interactions $\mathcal{O}_1 =
(\mathcal{F}_4)_L$ and $\mathcal{O}_2 = (\mathcal{F}_6)_L$ become relevant.
Consider the effective action including these terms:
\begin{equation}
  S_{\text{eff}} = S_{\text{pre-SUSY}} + g_1 \mathcal{O}_1 + g_2 \mathcal{O}_2
  . \label{g1g2}
\end{equation}
Suppose that we somehow managed to tune the couplings $g_1$, $g_2$ to zero.
Then, we will be left with $S_{\text{pre-SUSY}}$. In $d = 4$, the pre-SUSY
action flows for a particular value of the mass to a SUSY fixed point, which
should dimensionally reduce to the critical 2d Ising model. In $d = 3$, the
pre-SUSY action flows to a gapped phase for any value of the mass (this is
related to the fact that the $d = 1$ $\varphi^4$ theory is gapped for any
value of the mass). Thus, we may hope to realize the SUSY fixed point in $d =
4$ (but not in $d = 3$) by tuning. This was proposed in \cite{paper2,paper-summary} as a smoking gun for our scenario, allowing to distinguish it from other scenarios such as the work of Tarjus, Tissier et al, Section \ref{sec-FRG}.

Let us first discuss the pattern of renormalization group flows and fixed
points which arises when couplings $g_1$ and $g_2$ become relevant. Then, we
will discuss various scenarios of what may happen when the microscopic theory
is tuned. We will see that it is actually sufficient to tune one of the two
couplings to zero to observe SUSY.

We will be very schematic. We may describe the RG flow of $g_1, g_2$ by the
following approximate beta-functions:
\begin{eqnarray}
  \frac{d g_1}{d \log a} & = & A_1 (d_{c 1} - d) g_1 - B_1 g_1^2, \\
  \frac{d g_2}{d \log a} & = & A_2 (d_{c 2} - d) g_2 - B_2 g_2^2 .
  \nonumber
\end{eqnarray}
The coefficients $A_i$ are positive, so that $g_i$ is relevant (grows towards
IR) for $d < d_{c i}$, $i = 1, 2$. We will continue to assume that $d_{c
1}$ is slightly larger than $d_{c 2}$, as Section \ref{RG-stab} indicates. The
coefficient $B_i$ is proportional to the coefficient with $\mathcal{O}_i$
appears in the OPE $\mathcal{O}_i \times \mathcal{O}_i$. A computation in $d =
6$ using the expression for $\mathcal{O}_i$ from Section \ref{sec-class}
gives $B_1, B_2 > 0$. We will assume that the sign of $B_i$ does not change
between $d = 6$ and $d = d_{c i}$. We could add a term $C g_2$ to the first
beta-function, showing that $g_2$ can generate $g_1$ (while the opposite
process is forbidden by the selection rules). This term does not influence much the
discussion below,\footnote{It would change the positions of the fixed points III and IV slightly, in Fig.~\ref{fig-g1g2}.} and we will set $C= 0$.

For $d_{c 2} < d < d_{c 1}$, the coupling $g_2$ is irrelevant and $g_1$ is
relevant. For negative initial values, $g_1$ flows towards $- \infty$. If this
situation is indeed realized, then SUSY may be broken already at $d = d_{c
1}$.\footnote{In the problem of interface disorder {\cite{Wiese:2021qpk}} this
is what happens apparently, see {\cite{paper2}}, App.~A.6.} This is because a
large susy-null coupling may give large corrections to anomalous dimensions of
non-susy-writable couplings, which may become relevant and break SUSY. This answers the question from Section \ref{sec-class} why one should worry about susy-null interactions.

On the other hand, when $g_1$ has a
positive initial value, it flows to a stable fixed point $g^{\ast}_1 = A_1
(d_{c 1} - d) / B_1$. Since $\mathcal{O}_1$ is a susy-null interaction, all
susy-writable observables are not modified. We do expect corrections to
non-susy-writable observables, e.g.~dimensions of nsw interactions. These
corrections will be small since $d$ is close to $d_{c 1}$ and $g^{\ast}_1$ is
small. It is very hard to distinguish this fixed point from the one having
$g_1 = 0$. Both fixed points are SUSY and have the same values of the main
critical exponents $\nu, \eta, \bar{\eta}$. Thus, in this case, SUSY does not
get broken at $d = d_{c 1}$.

As we lower $d$ further, $d < d_{c 2}$, coupling $g_2$ also becomes relevant.
The SUSY fixed point at $g_2 = 0$ becomes unstable, and a stable non-SUSY
fixed point appears at $g^{\ast}_2 = A_2 (d_{c 2} - d) / B_2 > 0$. For the
initial $g_2$ positive, the flow will be attracted to the stable fixed point.
For the initial $g_2$ negative, the flow goes towards large
negative couplings and its fate is uncertain. It may lead to a first order
phase transition or to another non-SUSY fixed point not visible in this
approximation.

We note in passing that for $d > d_{c 2}$ the beta-functions predict
a stable fixed point at $g_2 = 0$ and an unstable fixed point at $g_2 < 0$.
The status of the latter fixed point is unclear - it may be invalidated by
non-perturbative effects similarly to how the Wilson-Fisher fixed point at $d
> 4$ is destroyed by non-perturbative instability of the $\lambda \varphi^4$
potential with negative coupling.

Analogously for $d > d_{c 1}$ the beta-functions predict
a stable SUSY fixed point at $g_1 = 0, g_2=0$ and an unstable SUSY fixed point at $g_1 < 0, g_2=0$.
From the susy-writable sector perspective, it's the same fixed point. That's why this discussion does not contradict what we said in Section \ref{sec-FRG} - that there is no known second SUSY fixed point at $d>d_c$, \textit{substantially (i.e.~in the SUSY sector) different from the one we are discussing}, and which could annihilate with it.

Coming back to $d<d_{c2}$, the full RG flow diagram is shown in Fig \ref{fig-g1g2}. We see
two SUSY fixed points I and II, and two non-SUSY fixed points III and IV. The
fixed points I and II are, as discussed, identical as far as the susy-writable observables are concerned; they have the same $\nu$ and $\eta = \bar{\eta}$ and have
dimensional reduction. The fixed points III and IV have $\nu$, $\eta,
\bar{\eta}$ different from each other and from I,II; they are not expected to obey
$\eta = \bar{\eta}$ nor dimensional reduction. The fixed point IV is the most
stable one, the first quadrant being its basin of attraction. It is the prime
candidate for describing the RFIM phase transition in $d = 4$.

\begin{figure}[h]
	\centering
	\resizebox{138pt}{103pt}{\includegraphics{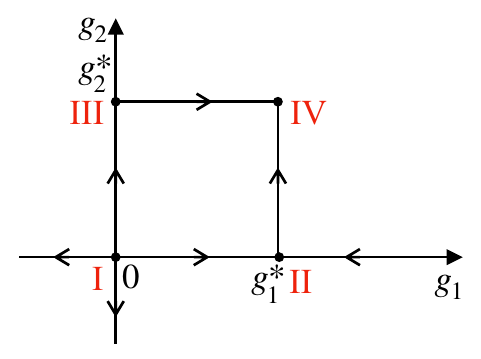}}
	\caption{\label{fig-g1g2}Pattern of fixed points and RG flows for $d < d_{c
			2}$.}
\end{figure}

What does this imply about the prospects of observing SUSY in a tuned
numerical simulation? In numerical simulations we first choose the shape of
the random magnetic field distribution $P [h]$. We then pick a parameter
$\kappa > 0$ playing the role of the strength of the random magnetic field. We
sample the rescaled distribution
\begin{equation}
	P_{\kappa} [h] = \frac{1}{\kappa} P [h / \kappa].
\end{equation}
For some $\kappa = \kappa_c$, depending on $P [h]$, a phase transition is
realized.

In Section \ref{sec-numsim} we fixed $P [h]$ to be a Gaussian, but this does
not have to be the case. When we vary $P [h]$ in the microscopic theory, this
translates into varying couplings $g_1, g_2$ in the effective field theory
description {\eqref{g1g2}} of the phase transition. For example, we may
consider a one-parameter family interpolating between the Gaussian
distribution with variance 1, and the bimodal distribution $\frac{1}{2}
[\delta (h - 1) + \delta (h + 1)]$.

Any such family will trace a curve of initial conditions in the RG flow
diagram of Fig.~\ref{fig-g1g2}. We don't a priori know how this curve will
pass. Let us consider a couple of examples. In Fig.~\ref{fig-g1g2-A} we have a
favorable situation for the observation of SUSY. The curve of initial
conditions, parameterized by $t \in [0, 1]$, exits the basin of attraction of
fixed point IV at $t = t_0$ crossing the axis $g_2 = 0$ at $g_1 > 0$, which
gets attracted to the SUSY fixed point II. For $t > t_0$ we have $g_2$ flowing
to negative values, which may lead to a first-order transition or another
fixed point not visible in our treatment.

\begin{figure}[h]
		\centering
	\resizebox{134pt}{104pt}{\includegraphics{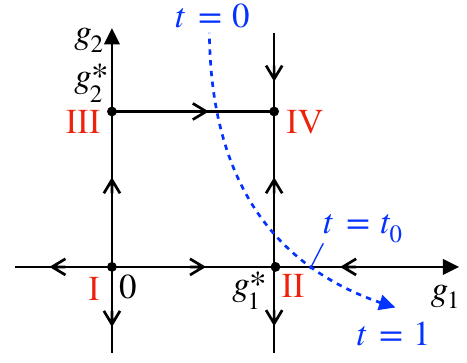}}
	\caption{\label{fig-g1g2-A}One parameter family of initial conditions where
		SUSY may be observed at $t = t_0$.}
\end{figure}

In Fig.~\ref{fig-g1g2-B} the situation is not so rosy - SUSY does not reveal
itself. We will observe the non-SUSY fixed point IV for $t < t_0$, and another
fixed point III at $t = t_0$ which is also not SUSY. For $t > t_0$ the flow
leads to large negative values of $g_1$ and/or $g_2$ and is not under control.
The $g_2 = 0$ line is crossed at $g_1 < 0$, which does not get attracted to a
SUSY fixed point.

\begin{figure}[h]
		\centering
	\resizebox{159pt}{108pt}{\includegraphics{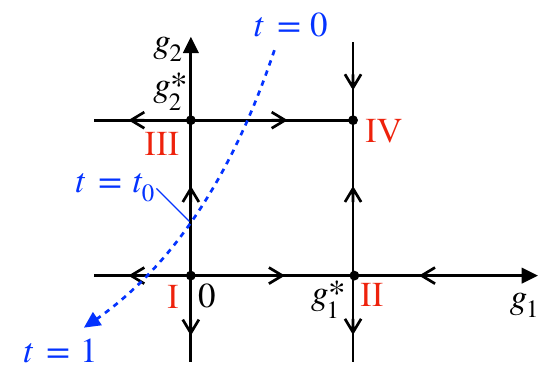}}
	\caption{\label{fig-g1g2-B}One parameter family of initial conditions where
		SUSY will not be observed.}
\end{figure}

The punchline of this discussion is that one may hope to see a SUSY fixed
point even for a one-parameter family of initial conditions, but one needs a
bit of luck and/or experimentation to find a family crossing $g_2 = 0$ line at
$g_1 > 0$. 

We note that Refs. {\cite{Fytas3,Picco1}} did already consider varying the
random magnetic field distribution. In $d = 3$, Ref.~{\cite{Fytas3}}
considered distributions along the Gaussian-bimodal interpolating family
mentioned above, as well as the Poissonian distribution. They did not see any
evidence for the SUSY fixed points - all distributions led to the same,
non-SUSY, critical exponents (albeit the distributions close to the bimodal
one had large corrections to scaling). This is consistent with our remarks
above, that a SUSY fixed point is expected to exist in $d = 4$ but not in $d =
3$.

In $d = 4$, Ref.~{\cite{Picco1}} considered only the Gaussian and the
Poissonian distributions, which led to the same, non-SUSY, critical
exponents. It would be very interesting to carry out a $d = 4$ study for more
distributions and look for changes in exponents. In the meantime, motivated by \cite{paper2}, an ongoing $d=4$ study by the authors of \cite{Picco3} along the Gaussian-bimodal family is showing intriguing preliminary results: a new fixed point close to the bimodal distribution (where the $d=3$ study saw large corrections to scaling).\footnote{I am grateful to Marco Picco and Giorgio Parisi for communicating to me their preliminary results. This Section was inspired in part by trying to foresee how their study may further develop.} Hopefully, it will be soon clarified if this fixed point obeys dimensional reduction and supersymmetry.

\section*{Acknowledgements}

I am grateful to Apratim Kaviraj and Emilio Trevisani for the collaboration and for the valuable suggestions concerning these notes. These notes are mostly based on our papers
{\cite{paper1,paper2,paper-summary,paper3}}, as well as on various lectures and talks I gave: at Cortona
{\cite{Cortona}}, in Montreal {\cite{Montreal}}, and at the IHES
{\cite{IHES}}. I am grateful to the organizers of the Cortona school and of the Montreal meeting for the invitation. Sections \ref{UCD} and \ref{tuning} contain previously
unpublished material from {\cite{Montreal,IHES}}. The work was supported by
the Simons Foundation grant 733758 (Simons Bootstrap Collaboration).

\small
\bibliographystyle{utphys}
\bibliography{mybib.bib}

\providecommand{\href}[2]{#2}\begingroup\raggedright\begin{thebibliography}{10}

\bibitem{ImryMa}
Y.~Imry and S.-k. Ma, ``{Random-Field Instability of the Ordered State of
  Continuous Symmetry},''
  \href{http://dx.doi.org/10.1103/PhysRevLett.35.1399}{{\em Phys. Rev. Lett.}
  {\bfseries 35} (1975) 1399--1401}.

\bibitem{Aharony:1976jx}
A.~Aharony, Y.~Imry, and S.~K. Ma, ``{Lowering of Dimensionality in Phase
  Transitions with Random Fields},''
\href{http://dx.doi.org/10.1103/PhysRevLett.37.1364}{{\em Phys. Rev. Lett.}
  {\bfseries 37} (1976) 1364--1367}.

\bibitem{Parisi:1979ka}
G.~Parisi and N.~Sourlas, ``{Random Magnetic Fields, Supersymmetry and Negative
  Dimensions},''
\href{http://dx.doi.org/10.1103/PhysRevLett.43.744}{{\em Phys. Rev. Lett.}
  {\bfseries 43} (1979) 744}.

\bibitem{Brezin-1998}
E.~Br{\'e}zin and C.~De~Dominicis, ``{New phenomena in the random field Ising
  model},'' \href{http://dx.doi.org/10.1209/epl/i1998-00428-0}{{\em Europhys
  Lett.} {\bfseries 44} no.~1, (1998) 13--19},
  \href{http://arxiv.org/abs/cond-mat/9804266}{{\ttfamily cond-mat/9804266}}.

\bibitem{Feldman}
D.~E. Feldman, ``{Critical Exponents of the Random-Field $O(N)$ Model},''
  \href{http://dx.doi.org/10.1103/PhysRevLett.88.177202}{{\em Phys. Rev. Lett.}
  {\bfseries 88} (2002) 177202},
  \href{http://arxiv.org/abs/cond-mat/0010012}{{\ttfamily
  arXiv:cond-mat/0010012 [cond-mat.dis-nn]}}.

\bibitem{Tarjus2004}
G.~Tarjus and M.~Tissier, ``{Nonperturbative Functional Renormalization Group
  for Random-Field Models: The Way Out of Dimensional Reduction},''
  \href{http://dx.doi.org/10.1103/physrevlett.93.267008}{{\em Phys. Rev. Lett.}
  {\bfseries 93} no.~26, (2004) 267008},
  \href{http://arxiv.org/abs/cond-mat/0410118}{{\ttfamily
  arXiv:cond-mat/0410118 [cond-mat.dis-nn]}}.

\bibitem{BelangerYoung}
D.~Belanger and A.~Young, ``The random field ising model,''
  \href{http://dx.doi.org/10.1016/0304-8853(91)90825-U}{{\em J. Magn. Magn.
  Mater.} {\bfseries 100} no.~1, (1991) 272 -- 291}.

\bibitem{Picco1}
N.~G. Fytas, V.~Martín-Mayor, M.~Picco, and N.~Sourlas, ``{Phase Transitions
  in Disordered Systems: The Example of the Random-Field Ising Model in Four
  Dimensions},'' \href{http://dx.doi.org/10.1103/PhysRevLett.116.227201}{{\em
  Phys. Rev. Lett.} {\bfseries 116} (2016) 227201},
  \href{http://arxiv.org/abs/1605.05072}{{\ttfamily arXiv:1605.05072
  [cond-mat.dis-nn]}}.

\bibitem{Picco2}
N.~G. Fytas, V.~Martín-Mayor, M.~Picco, and N.~Sourlas, ``{Restoration of
  dimensional reduction in the random-field Ising model at five dimensions},''
  \href{http://dx.doi.org/10.1103/PhysRevE.95.042117}{{\em Phys. Rev. E}
  {\bfseries 95} (2017) 042117},
  \href{http://arxiv.org/abs/1612.06156}{{\ttfamily arXiv:1612.06156
  [cond-mat.dis-nn]}}.

\bibitem{Picco3}
N.~G. Fytas, V.~Martín-Mayor, G.~Parisi, M.~Picco, and N.~Sourlas, ``{Evidence
  for Supersymmetry in the Random-Field Ising Model at $D=5$},''
  \href{http://dx.doi.org/10.1103/PhysRevLett.122.240603}{{\em Phys. Rev.
  Lett.} {\bfseries 122} (2019) 240603},
  \href{http://arxiv.org/abs/1901.08473}{{\ttfamily arXiv:1901.08473
  [cond-mat.stat-mech]}}.

\bibitem{paper1}
A.~Kaviraj, S.~Rychkov, and E.~Trevisani, ``{Random Field Ising Model and
  Parisi-Sourlas supersymmetry. Part I. Supersymmetric CFT},''
  \href{http://dx.doi.org/10.1007/JHEP04(2020)090}{{\em JHEP} {\bfseries 04}
  (2020) 090}, \href{http://arxiv.org/abs/1912.01617}{{\ttfamily
  arXiv:1912.01617 [hep-th]}}.

\bibitem{paper2}
A.~Kaviraj, S.~Rychkov, and E.~Trevisani, ``{Random field Ising model and
  Parisi-Sourlas supersymmetry. Part II. Renormalization group},''
  \href{http://dx.doi.org/10.1007/JHEP03(2021)219}{{\em JHEP} {\bfseries 03}
  (2021) 219}, \href{http://arxiv.org/abs/2009.10087}{{\ttfamily
  arXiv:2009.10087 [cond-mat.stat-mech]}}.

\bibitem{paper-summary}
A.~Kaviraj, S.~Rychkov, and E.~Trevisani, ``{Parisi-Sourlas Supersymmetry in
  Random Field Models},''
  \href{http://dx.doi.org/10.1103/PhysRevLett.129.045701}{{\em Phys. Rev.
  Lett.} {\bfseries 129} no.~4, (2022) 045701},
  \href{http://arxiv.org/abs/2112.06942}{{\ttfamily arXiv:2112.06942
  [cond-mat.stat-mech]}}.

\bibitem{paper3}
A.~Kaviraj and E.~Trevisani, ``{Random field ${\phi}^{3}$ model and
  Parisi-Sourlas supersymmetry},''
  \href{http://dx.doi.org/10.1007/JHEP08(2022)290}{{\em JHEP} {\bfseries 08}
  (2022) 290}, \href{http://arxiv.org/abs/2203.12629}{{\ttfamily
  arXiv:2203.12629 [hep-th]}}.

\bibitem{Cardy-book}
J.~L. Cardy, {\em {Scaling and renormalization in statistical physics}}.
\newblock Cambridge, UK: Univ. Pr., 238 p.,
1996.
\newblock

\bibitem{Bray1985}
A.~J. Bray and M.~A. Moore, ``{Scaling theory of the random-field Ising
  model},'' \href{http://dx.doi.org/10.1088/0022-3719/18/28/006}{{\em {J. of
  Phys. C}} {\bfseries 18} no.~28, (1985) L927--L933}.

\bibitem{imbrieLCD2}
J.~Z. Imbrie, ``The ground state of the three-dimensional random-field ising
  model,'' \href{http://dx.doi.org/10.1007/BF01220505}{{\em Comm. Math. Phys.}
  {\bfseries 98} no.~2, (1985) 145--176}.

\bibitem{BK}
J.~{Bricmont} and A.~{Kupiainen}, ``{Phase transition in the 3d random field
  Ising model},'' \href{http://dx.doi.org/10.1007/BF01224901}{{\em Comm. Math.
  Phys.} {\bfseries 116} no.~4, (1988) 539--572}.

\bibitem{Ding1}
J.~{Ding} and Z.~{Zhuang}, ``{Long range order for random field Ising and Potts
  models},'' \href{http://arxiv.org/abs/2110.04531}{{\ttfamily arXiv:2110.04531
  [math.PR]}}.

\bibitem{Ding}
J.~{Ding}, Y.~{Liu}, and A.~{Xia}, ``{Long range order for three-dimensional
  random field Ising model throughout the entire low temperature regime},''
  \href{http://arxiv.org/abs/2209.13998}{{\ttfamily arXiv:2209.13998
  [math.PR]}}.

\bibitem{Aizenman1990}
M.~Aizenman and J.~Wehr, ``Rounding effects of quenched randomness on
  first-order phase transitions,''
  \href{http://dx.doi.org/10.1007/BF02096933}{{\em Comm. Math. Phys.}
  {\bfseries 130} no.~3, (1990) 489--528}.

\bibitem{Ding2}
J.~{Ding} and J.~{Xia}, ``{Exponential decay of correlations in the
  two-dimensional random field Ising model},''
  \href{http://arxiv.org/abs/1905.05651}{{\ttfamily arXiv:1905.05651
  [math.PR]}}.

\bibitem{Belanger}
D.~P. {Belanger},
  \href{http://dx.doi.org/10.1142/9789812819437_0008}{``{Experiments on the
  Random Field Ising Model},''} in {\em Spin Glasses And Random Fields}, {A. P.
  Young}, ed., pp.~251--275.
\newblock World Scientific, 1997.
\newblock \href{http://arxiv.org/abs/cond-mat/9706042}{{\ttfamily
  arXiv:cond-mat/9706042 [cond-mat.dis-nn]}}.

\bibitem{Fishman}
S.~Fishman and A.~Aharony, ``Random field effects in disordered anisotropic
  antiferromagnets,'' \href{http://dx.doi.org/10.1088/0022-3719/12/18/006}{{\em
  J. of Phys. C} {\bfseries 12} no.~18, (1979) L729}.

\bibitem{Cardy-site-diluted}
J.~L. Cardy, ``Random-field effects in site-disordered ising
  antiferromagnets,'' \href{http://dx.doi.org/10.1103/PhysRevB.29.505}{{\em
  Phys. Rev. B} {\bfseries 29} (1984) 505--507}.

\bibitem{Rong}
J.~Rong, ``{Scalar CFTs from structural phase transitions},''
  \href{http://arxiv.org/abs/2303.12028}{{\ttfamily arXiv:2303.12028
  [hep-th]}}.

\bibitem{Jahn-Teller}
G.~A. Gehring and K.~A. Gehring, ``Co-operative jahn-teller effects,''
  \href{http://dx.doi.org/10.1088/0034-4885/38/1/001}{{\em Reports on Progress
  in Physics} {\bfseries 38} no.~1, (1975) 1}.

\bibitem{GrahamJT}
J.~T. Graham, M.~Maliepaard, J.~H. Page, S.~R.~P. Smith, and D.~R. Taylor,
  ``{Random-field effects on Ising Jahn-Teller phase transitions},''
  \href{http://dx.doi.org/10.1103/PhysRevB.35.2098}{{\em Phys. Rev. B}
  {\bfseries 35} (1987) 2098--2101}.

\bibitem{DeGennes1984}
P.~G. de~Gennes, ``Liquid-liquid demixing inside a rigid network. qualitative
  features,'' \href{http://dx.doi.org/10.1021/j150670a004}{{\em J. Phys. Chem.}
  {\bfseries 88} no.~26, (1984) 6469--6472}.

\bibitem{Sinha1991}
S.~K. Sinha, J.~Huang, and S.~K. Satija,
  \href{http://dx.doi.org/10.1007/978-1-4757-1402-9_12}{``Binary fluid phase
  separation in gels: A neutron scattering study,''} in {\em Scaling Phenomena
  in Disordered Systems}, R.~Pynn and A.~Skjeltorp, eds., pp.~157--162.
\newblock Springer (Boston), 1991.

\bibitem{Alava}
M.~Alava, P.~Duxbury, C.~Moukarzel, and H.~Rieger,
  \href{http://dx.doi.org/https://doi.org/10.1016/S1062-7901(01)80009-4}{``{Exact
  combinatorial algorithms: Ground states of disordered systems},''} in {\em
  Phase Transitions and Critical Phenomena, vol. 18}, C.~Domb and J.~Lebowitz,
  eds., pp.~143--317.
\newblock Academic Press, 2001.

\bibitem{FytasReview}
N.~G. {Fytas} and V.~{Mart{\'\i}n-Mayor}, ``{Efficient numerical methods for
  the random-field Ising model: Finite-size scaling, reweighting extrapolation,
  and computation of response functions},''
  \href{http://dx.doi.org/10.1103/PhysRevE.93.063308}{{\em Phys. Rev. E}
  {\bfseries 93} no.~6, (2016) 063308},
  \href{http://arxiv.org/abs/1512.06571}{{\ttfamily arXiv:1512.06571
  [cond-mat.dis-nn]}}.

\bibitem{Fytas3}
N.~G. {Fytas} and V.~{Mart{\'\i}n-Mayor}, ``{Universality in the
  Three-Dimensional Random-Field Ising Model},''
  \href{http://dx.doi.org/10.1103/PhysRevLett.110.227201}{{\em Phys. Rev.
  Lett.} {\bfseries 110} no.~22, (2013) 227201},
  \href{http://arxiv.org/abs/1304.0318}{{\ttfamily arXiv:1304.0318
  [cond-mat.dis-nn]}}.

\bibitem{Kos:2016ysd}
F.~Kos, D.~Poland, D.~Simmons-Duffin, and A.~Vichi, ``{Precision Islands in the
  Ising and $O(N)$ Models},''
  \href{http://dx.doi.org/10.1007/JHEP08(2016)036}{{\em JHEP} {\bfseries 08}
  (2016) 036}, \href{http://arxiv.org/abs/1603.04436}{{\ttfamily
  arXiv:1603.04436 [hep-th]}}.

\bibitem{Wegner:2016ahw}
F.~Wegner, \href{http://dx.doi.org/10.1007/978-3-662-49170-6}{{\em
  {Supermathematics and its Applications in Statistical Physics}: {Grassmann
  Variables and the Method of Supersymmetry}}}, vol.~920.
\newblock Springer, 2016.

\bibitem{CARDY1983470}
J.~L. Cardy, ``{Nonperturbative effects in a scalar supersymmetric theory},''
  \href{http://dx.doi.org/https://doi.org/10.1016/0370-2693(83)91328-X}{{\em
  Physics Letters B} {\bfseries 125} no.~6, (1983) 470 -- 472}.

\bibitem{Zaboronsky:1996qn}
O.~V. Zaboronski, ``{Dimensional reduction in supersymmetric field theories},''
  \href{http://dx.doi.org/10.1088/0305-4470/35/26/312}{{\em Journal of Physics
  A: Mathematical and General} {\bfseries 35} no.~26, (2002) 5511--5519},
  \href{http://arxiv.org/abs/hep-th/9611157}{{\ttfamily arXiv:hep-th/9611157
  [hep-th]}}.

\bibitem{Cremonesi:2013twh}
S.~Cremonesi, ``{An Introduction to Localisation and Supersymmetry in Curved
  Space},'' \href{http://dx.doi.org/10.22323/1.201.0002}{{\em PoS} {\bfseries
  Modave2013} (2013) 002}.

\bibitem{ParisiLH}
G.~Parisi, ``{An introduction to the statistical mechanics of amorphous
  systems},'' in {\em {Recent Advances in Field Theory and Statistical
  Mechanics, Proceedings of Les Houches 1982, Session XXXIX}}, J.~B. Zuber and
  R.~Stora, eds., p.~473.
\newblock North-Holland, Amsterdam, 1984.
\newblock {Reprinted in G. Parisi, ``Field Theory, Disorder and Simulations'',
  World Scientific, 1992}.

\bibitem{ParisiRome}
S.~Rychkov, ``{Random magnetic fields, supersymmetry, and negative
  dimensions}.''.
  \href{https://drive.google.com/file/d/1SF0A7BFp8aE2d-yoc024M0veAi7VLukw/view}{Talk
  in the series ``The interdisciplinary contribution of Giorgio Parisi to
  theoretical physics", La Sapienza University, Rome, Italy, 2.2.2023}.

\bibitem{cardy_1996}
J.~Cardy, \href{http://dx.doi.org/10.1017/CBO9781316036440}{{\em {Scaling and
  Renormalization in Statistical Physics}}}.
\newblock Cambridge Lecture Notes in Physics. Cambridge University Press,
1996.
\newblock

\bibitem{CARDY1985123}
J.~L. Cardy, ``{Nonperturbative aspects of supersymmetry in statistical
  mechanics},''
  \href{http://dx.doi.org/https://doi.org/10.1016/0167-2789(85)90154-X}{{\em
  Physica D: Nonlinear Phenomena} {\bfseries 15} no.~1, (1985) 123 -- 128}.

\bibitem{de_dominicis_giardina_2006}
C.~De~Dominicis and I.~Giardina,
  \href{http://dx.doi.org/10.1017/CBO9780511534836}{{\em {Random Fields and
  Spin Glasses: A Field Theory Approach}}}.
\newblock Cambridge University Press, 2006.

\bibitem{Pelissetto:2000ek}
A.~Pelissetto and E.~Vicari, ``{Critical phenomena and renormalization-group
  theory},'' \href{http://dx.doi.org/10.1016/S0370-1573(02)00219-3}{{\em Phys.
  Rept.} {\bfseries 368} (2002) 549--727},
\href{http://arxiv.org/abs/cond-mat/0012164}{{\ttfamily
  arXiv:cond-mat/0012164}}.

\bibitem{Chester:2020iyt}
S.~M. Chester, W.~Landry, J.~Liu, D.~Poland, D.~Simmons-Duffin, N.~Su, and
  A.~Vichi, ``{Bootstrapping Heisenberg magnets and their cubic instability},''
  \href{http://dx.doi.org/10.1103/PhysRevD.104.105013}{{\em Phys. Rev. D}
  {\bfseries 104} no.~10, (2021) 105013},
  \href{http://arxiv.org/abs/2011.14647}{{\ttfamily arXiv:2011.14647
  [hep-th]}}.

\bibitem{Hasenbusch:2022zur}
M.~Hasenbusch, ``{Cubic fixed point in three dimensions: Monte Carlo
  simulations of the \ensuremath{\phi}4 model on the simple cubic lattice},''
  \href{http://dx.doi.org/10.1103/PhysRevB.107.024409}{{\em Phys. Rev. B}
  {\bfseries 107} no.~2, (2023) 024409},
  \href{http://arxiv.org/abs/2211.16170}{{\ttfamily arXiv:2211.16170
  [cond-mat.stat-mech]}}.

\bibitem{Badel:2019oxl}
G.~Badel, G.~Cuomo, A.~Monin, and R.~Rattazzi, ``{The Epsilon Expansion Meets
  Semiclassics},'' \href{http://dx.doi.org/10.1007/JHEP11(2019)110}{{\em JHEP}
  {\bfseries 11} (2019) 110}, \href{http://arxiv.org/abs/1909.01269}{{\ttfamily
  arXiv:1909.01269 [hep-th]}}.

\bibitem{Peskin:1995ev}
M.~E. Peskin and D.~V. Schroeder, {\em {An Introduction to quantum field
  theory}}.
\newblock Addison-Wesley, Reading, USA, 1995.

\bibitem{vasil2020field}
A.~Vasil'ev, {\em The Field Theoretic Renormalization Group in Critical
  Behavior Theory and Stochastic Dynamics}.
\newblock CRC Press LLC, 2020.

\bibitem{Zinn-Justin:2002ecy}
J.~Zinn-Justin, ``{Quantum field theory and critical phenomena},'' {\em Int.
  Ser. Monogr. Phys.} {\bfseries 113} (2002) 1--1054.

\bibitem{Kleinert:2001ax}
H.~Kleinert and V.~Schulte-Frohlinde,
  \href{http://dx.doi.org/10.1142/4733}{{\em Critical properties of $\phi^4$
  theories}}.
\newblock World Scientiic, 2001.

\bibitem{Korchemsky:2015cyx}
G.~P. Korchemsky, ``{On level crossing in conformal field theories},''
  \href{http://dx.doi.org/10.1007/JHEP03(2016)212}{{\em JHEP} {\bfseries 03}
  (2016) 212}, \href{http://arxiv.org/abs/1512.05362}{{\ttfamily
  arXiv:1512.05362 [hep-th]}}.

\bibitem{Behan:2017mwi}
C.~Behan, ``{Conformal manifolds: ODEs from OPEs},''
  \href{http://dx.doi.org/10.1007/JHEP03(2018)127}{{\em JHEP} {\bfseries 03}
  (2018) 127}, \href{http://arxiv.org/abs/1709.03967}{{\ttfamily
  arXiv:1709.03967 [hep-th]}}.

\bibitem{Henriksson:2022gpa}
J.~Henriksson, S.~R. Kousvos, and M.~Reehorst, ``{Spectrum continuity and level
  repulsion: the Ising CFT from infinitesimal to finite
  \ensuremath{\varepsilon}},''
  \href{http://dx.doi.org/10.1007/JHEP02(2023)218}{{\em JHEP} {\bfseries 02}
  (2023) 218}, \href{http://arxiv.org/abs/2207.10118}{{\ttfamily
  arXiv:2207.10118 [hep-th]}}.

\bibitem{Hsu2005}
H.-P. Hsu, W.~Nadler, and P.~Grassberger, ``Simulations of lattice animals and
  trees,'' \href{http://dx.doi.org/10.1088/0305-4470/38/4/001}{{\em J. Phys. A}
  {\bfseries 38} no.~4, (2005) 775–806},
  \href{http://arxiv.org/abs/cond-mat/0408061}{{\ttfamily
  arXiv:cond-mat/0408061 [cond-mat.stat-mech]}}.

\bibitem{PhysRevLett.46.871}
G.~Parisi and N.~Sourlas, ``{Critical Behavior of Branched Polymers and the
  Lee-Yang Edge Singularity},''
  \href{http://dx.doi.org/10.1103/PhysRevLett.46.871}{{\em Phys. Rev. Lett.}
  {\bfseries 46} (1981) 871--874}.

\bibitem{PhysRevA.20.2130}
T.~C. Lubensky and J.~Isaacson, ``Statistics of lattice animals and dilute
  branched polymers,'' \href{http://dx.doi.org/10.1103/PhysRevA.20.2130}{{\em
  Phys. Rev. A} {\bfseries 20} (1979) 2130--2146}.

\bibitem{zbMATH02068689}
D.~C. {Brydges} and J.~Z. {Imbrie}, ``{{Branched polymers and dimensional
  reduction.}},'' {\em {Ann. Math.}} {\bfseries 158} no.~3, (2003) 1019--1039,
  \href{http://arxiv.org/abs/math-ph/0107005}{{\ttfamily
  arXiv:math-ph/0107005}}.

\bibitem{Miller:1992uv}
J.~D. Miller and K.~De'Bell, ``{Randomly branched polymers and conformal
  invariance},'' \href{http://arxiv.org/abs/hep-th/9211127}{{\ttfamily
  arXiv:hep-th/9211127}}.

\bibitem{Polchinski:1987dy}
J.~Polchinski, ``{Scale and Conformal Invariance in Quantum Field Theory},''
  \href{http://dx.doi.org/10.1016/0550-3213(88)90179-4}{{\em Nucl. Phys. B}
  {\bfseries 303} (1988) 226--236}.

\bibitem{Balog:2020sre}
I.~Balog, G.~Tarjus, and M.~Tissier, ``{Dimensional reduction breakdown and
  correction to scaling in the random-field Ising model},''
  \href{http://dx.doi.org/10.1103/PhysRevE.102.062154}{{\em Phys. Rev. E}
  {\bfseries 102} (2020) 062154},
  \href{http://arxiv.org/abs/2008.13650}{{\ttfamily arXiv:2008.13650
  [cond-mat.dis-nn]}}.

\bibitem{Gorbenko:2018ncu}
V.~Gorbenko, S.~Rychkov, and B.~Zan, ``{Walking, Weak first-order transitions,
  and Complex CFTs},'' \href{http://dx.doi.org/10.1007/JHEP10(2018)108}{{\em
  JHEP} {\bfseries 10} (2018) 108},
  \href{http://arxiv.org/abs/1807.11512}{{\ttfamily arXiv:1807.11512
  [hep-th]}}.

\bibitem{Wiese:2021qpk}
K.~J. Wiese, ``{Theory and experiments for disordered elastic manifolds,
  depinning, avalanches, and sandpiles},''
  \href{http://dx.doi.org/10.1088/1361-6633/ac4648}{{\em Rept. Prog. Phys.}
  {\bfseries 85} no.~8, (2022) 086502},
  \href{http://arxiv.org/abs/2102.01215}{{\ttfamily arXiv:2102.01215
  [cond-mat.dis-nn]}}.

\bibitem{Cortona}
S.~Rychkov, ``{Random Field Ising Model and Parisi-Sourlas Supersymmetry - some
  recent developments}.''. Mathematics Meets Physics on Disordered Systems, PhD
  School, Cortona, Italy, April 25-May 6, 2022.

\bibitem{Montreal}
S.~Rychkov, ``{The fate of Parisi-Sourlas supersymmetry in random-field
  models}.''. Conformal field theory and quantum many-body physics, August 22 -
  September 9, 2022, Centre de Recherches Mathematiques, Université de
  Montréal, Quebec, Canada.

\bibitem{IHES}
S.~Rychkov, ``{Random Field Ising Model and Parisi-Sourlas Supersymmetry}.''.
  \href{https://www.youtube.com/watch?v=6ZywSBtcEKg}{Lectures at the Institut
  des Hautes Études Scientifiques, Nov. 2022}.

\end{thebibliography}\endgroup
\end{document}